\documentclass{aa}  

\usepackage{graphicx}
\usepackage{hyperref}
\usepackage{xcolor}
\usepackage{rotating}
\usepackage{float}
\usepackage[normalem]{ulem}
\newcommand{\fnsref}[1]{\hyperref[#1]{\textsuperscript{\ref{#1}}}}
\newcommand{\fink}{{\sc Fink}}

\newcommand{\emi}[1]{\textcolor{black}{#1}}

\usepackage{txfonts}
\begin{document}

   \title{Lost and Found --- A gallery of overlooked optical nuclear transients from the ZTF archive}


   \author{E. Quintin
          \inst{1}
          \and
          E. Russeil
          \inst{2}
          \and
          M. Llamas Lanza
          \inst{3}
          \and
          S. Karpov
          \inst{4}
          \and
          E. E. O. Ishida
          \inst{5}
          \and
          J. Peloton
          \inst{6}
          \and
          M.~V.~Pruzhinskaya
          \inst{7}
          \and
          A. Möller
          \inst{8,9}
          \and
          M. Giustini
          \inst{10}
          \and
          G. Miniutti
          \inst{10}
          \and
          R.S. Saxton
          \inst{11}
          \and
          P. Sánchez-Sáez
          \inst{12}
          \and
          S. Zheltoukhov 
          \inst{7}
          \and
          A. Dodin
          \inst{7}
          \and
          A. Belinski 
          \inst{7}}

   \institute{European Space Agency (ESA), European Space Astronomy Centre (ESAC), Camino Bajo del Castillo s/n, 28692 Villanueva de la Cañada, Madrid, Spain\\
              \email{erwan.quintin@esa.int}
         \and
             The Oskar Klein Centre, Department of Astronomy, Stockholm University, AlbaNova, SE-10691 Stockholm, Sweden
         \and
            IRAP, Université de Toulouse, CNRS, CNES, UPS, Toulouse, France.
         \and
             Institute of Physics of the Czech Academy of Sciences, Na Slovance 1999/2, 182 00 Prague 8, Czech Republic
         \and
             Université Clermont Auvergne, CNRS, LPCA, Clermont-Ferrand, F-63000, France
         \and
             Université Paris-Saclay, CNRS/IN2P3, IJCLab, Orsay, France
         \and
             Lomonosov Moscow State University, Sternberg Astronomical Institute, Universitetsky 13, Moscow 119234, Russia
         \and
             Centre for Astrophysics and Supercomputing, Swinburne University of Technology, Mail Number H29, PO Box 218, 31122 Hawthorn, VIC, Australia
         \and
             ARC Centre of Excellence for Gravitational Wave Discovery (OzGrav), John St, Hawthorn, VIC 3122, Australia
         \and
             Centro de Astrobiología (CAB), CSIC-INTA, Camino Bajo del Castillo s/n, 28692 Villanueva de la Cañada, Madrid, Spain
         \and
             Telespazio UK for the European Space Agency (ESA), European Space Astronomy Centre (ESAC), Camino Bajo del Castillo s/n, 28692 Villanueva de la Cañada, Madrid, Spain
         \and
             European Southern Observatory, Karl-Schwarzschild-Strasse 2, 85748 Garching bei München, Germany
            }

   \date{}

 
  \abstract
   {Tidal disruption events (TDEs) correspond to the destruction of a star by the tidal forces around a black hole, leading to outbursts which can last from months to years. These transients are rare, and increasing the current sample is paramount to understand them. As part of the \fink\ alert broker, we have developed an early detection system for TDEs for the Zwicky Transient Facility (ZTF) data. 
   }
   {In this paper, we report on the optical transients we found either during the development of this tool, or when applying the classifier to the existing archive. We use this sample to anticipate what improvements to the TDE detection systems will need to be implemented for future surveys.}
   {For all the transients, we present optical and infrared archival photometry from ZTF, WISE, and Catalina, and assess the previous nuclear activity of the host. We fit the ZTF lightcurves with both a phenomenological and a physically-motivated model.} 
   {We report on a total of 19 optical nuclear transients, out of which nine are in passive galaxies, eight in active galaxies, and two for which the activity of the host is uncertain. Two transients are newly discovered repeated TDE candidates, and we compare them to the current sample of repeated optical nuclear transients. One transient is exceptionally long-lived ($>$5 years), in an until-now passive galaxy. Three of the TDE-like flares in active galaxies have absolute $g-$band magnitudes brighter than -24, making them new Extreme Nuclear Transient (ENT) candidates. One seemingly repeated object was revealed to be two independent supernovae in the same galaxy.}
   {This sample shows both the potential of our detection system for future discovery, and the relevance of archival searches to reveal overlooked transients. It also raises several points of concern and avenues of improvement for current and future classifiers.}

   \keywords{Methods: observational, data analysis --        
                Black hole physics --
                Surveys
               }

   \maketitle
%


\section{Introduction}

\begin{figure}
    \centering
    \includegraphics[width=\linewidth]{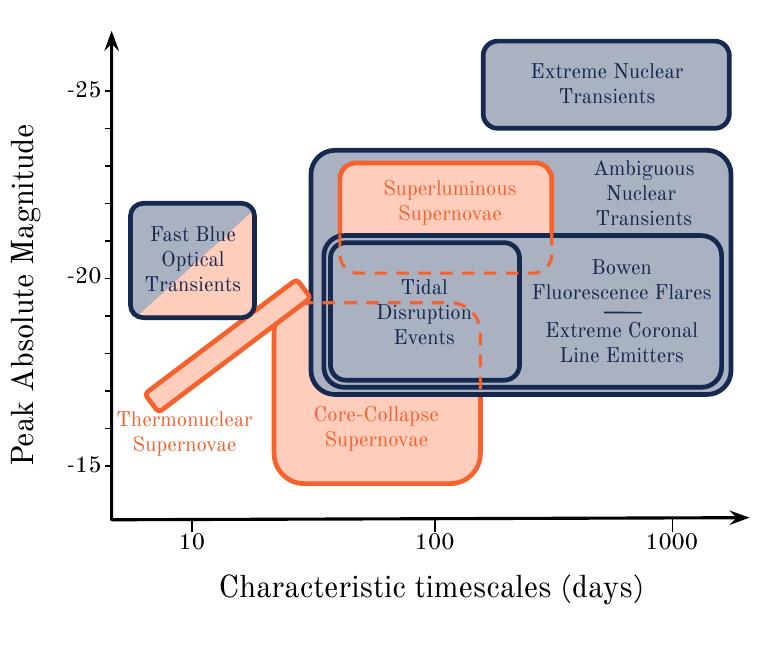}
    \caption{Representation of the timescale and peak absolute magnitude of the various optical transients considered in this paper. The exact borders of each category are generally not well-defined -- in particular, there might be a continuum between ANTs and ENTs, the latter being for now crudely defined as brighter than absolute magnitude -24. The orange boxes correspond to supernovae events, and the blue ones to accretion-powered events (the interpretation of FBOTs is still open). The clear overlaps between categories indicate the difficulty of classifying these events based on photometric data only. This figure  is inspired from Fig. 4 of \citet{hinkle_extreme_2024}.}
    \label{fig:OpticalTransients}
\end{figure}

Due to the advent of large scale surveys, the diversity of transient astrophysical sources has increased significantly in the last few decades. Successive observatories and surveys, and increasingly more intricate detection and classification schemes, have allowed to paint a diverse picture of the transient sky, covering orders of magnitude in terms of luminosity and timescale. In particular, as the samples grow and outliers are discovered, the boundaries between different classes can be challenged -- and so can be our understanding of these objects \citep[e.g.][]{zabludoff_distinguishing_2021}. In this work, we focus in particular on optical nuclear transients, i.e. objects lying in the nucleus of their host galaxy that show an episode of dramatic variability, way beyond their usual behavior. A schematic representation of a selection of these transients can be seen in Fig.~\ref{fig:OpticalTransients}, where the overlap between the classes is clearly visible. In this work, our goal is to show the value of using archival datasets in discovering optical nuclear transients, especially outliers which might have been rejected by already-existing classifiers.

As the galactic nuclei have a dense stellar population, one can expect to observe nuclear supernovae (SNe -- nuclear here indicating their location within the galaxy, not their triggering mechanism), which typically last a few weeks, showing slightly different behaviors depending on their type, but consistently displaying a fast cooling in their decay phase. SNe also have a slower and brighter cousin, aptly named superluminous supernovae \citep[SLSNe, e.g.][]{moriya_superluminous_2018}, which can last for several months. Another category of transients which is yet to receive a definitive interpretation are the Fast Blue Optical Transients \citep[FBOTs, e.g.][]{ho_search_2023}, which can reach luminosities similar to those of SLSNe but on timescales comparable to typical, faster SNe. One category of nuclear transients that is the subject of significant effort from the community at the moment are Tidal Disruption Events \citep[TDEs, e.g.][]{gezari_tidal_2021}. These correspond to the outburst resulting from the destruction of an inbound star by the tidal forces around the central super-massive black hole (SMBH) of a galaxy. They can be seen in various wavelengths, typically lasting a few months, and with optical brightness at the frontier between SNe and SLSNe. While most of the $\sim$150 known TDEs happened in passive galaxies (i.e. where the central SMBH was not accreting matter before the stellar disruption), in theory, they can also happen in comparable rates in Active Galactic Nuclei (AGNs) as well \citep[e.g.][]{chan_tidal_2019}. Recently, a number of nuclear transient events have been suggested to be linked to such TDEs in AGNs. The first, quite broad, category are Ambiguous Nuclear Transients \citep[ANTs, e.g.][]{kankare_population_2017,wiseman_systematically_2025}, and correspond to bursts that photometrically roughly look like TDEs (with a smooth rise and slower decay), but for which the spectra do not match expected TDE emission. The exact mechanism behind these objects is still to be understood. The most luminous of them (absolute magnitudes above -24) are called Extreme Nuclear Transients, with only a handful of instances \citep[ENTs, e.g.][]{hinkle_extreme_2024, russeil_identification_2025, graham_extremely_2025}. Somewhat perhaps linked to ANTs and TDEs are the so-called Bowen Fluorescence Flares \citep[BFFs, e.g.][]{trakhtenbrot_new_2019} and the Extreme Coronal Line Emitters \citep[ECLEs, e.g.][]{komossa_discovery_2008}, which are both characterized by smooth, long TDE-like lightcurves and optical spectra showing specific transient emission line features. For BFFs and ECLEs, the exact triggering mechanism is unclear, and they could arise from flares in AGN accretion flow or from TDEs alike. A final type of nuclear optical transient are the Changing Look Active Galactic Nuclei \citep[CL-AGNs, e.g.][]{trakhtenbrot_1es_2019, ricci_changing-look_2023}. These correspond to AGNs where the optical classification as type 1 or 2 changes over time -- this is usually associated with flaring behaviour in the optical, and sometimes a dip in the X-ray lightcurve. Such a change should not be possible in the unified AGN model, where the type is a function of the inclination angle only \citep[e.g.][]{antonucci_unified_1993}. Several scenarios can be invoked to explain these transients, including intrinsic changes in the accretion flow due to a TDE embedded within the AGN \citep[e.g.][]{merloni_tidal_2015}.

One final note on the zoo of optical nuclear transients is that some of them have been shown to repeat. These repeated bursts can take the form of rebrightening during an otherwise smooth decay, or a fully fledged second peak at later times. For some objects, this peculiar behavior can put constraints on the corresponding models. For instance, a repeated TDE can indicate that the inbound star survived the initial encounter, with a disruption that was partial only. \emi{However, in order to survive} a significant number of such partial TDEs, a main sequence star requires strong fine-tuning of the parameters of the problem \citep[e.g.][]{payne_asassn-14ko_2021}. If this fine-tuning becomes too constraining, it might indicate issues in the models. As such, collecting repeated transients and assessing their fraction within the overall population is of particular interest to constrain the models and their limits. 

For all those events, the current sample is limited and prevents us from strongly constraining the existing models. As such, any new detection should ideally lead to a multi-wavelength photometric and spectroscopic monitoring. But the first step is to ensure a timely detection of these events. This was the idea behind the development of an early TDE detection system \citep{llamas_lanza_early_2025} in the Zwicky Transient Facility \citep[ZTF,][]{bellm_zwicky_2014} data, for the \fink\ broker \citep{moller_fink_2021}. We purposely built it for high completeness rather than high purity, in order to detect the outlier events that will push the existing models of each class. While developing this tool, we have encountered several objects of interest. These objects were either picked up directly by the TDE detection system itself, or found serendipitously in the archive during its development. The goal of this paper is to present these objects in a single place, and put them into perspective in order to anticipate the efforts that should be done for future, larger transient surveys such as LSST \citep{2019ApJ...873..111I}.


\emi{
Section {\ref{sec:methods}} gives a general idea of the methods which lead to the identification of this sample, and how we analysed them to reveal their nature. Section \ref{sec:results} presents detailed information about the sample and highlights a few candidates of particular interest. Finally, we discuss the broader impact of these findings in Section \ref{sec:discussion}, especially from the point of view of automatic classifiers, and present our conclusions in Section \ref{sec:conclusions}.}
    
\section{Methods}
\label{sec:methods}
\subsection{Finding the candidates}
\label{subsec:finding}

As a part of the \fink\ alert broker \citep{moller_fink_2021} we created a dedicated science module aimed at detecting TDEs during their initial rising phase \citep{llamas_lanza_early_2025}. It is based on applying a physically-motivated parametric fit \citep[the Rainbow model,][]{russeil_rainbow_2023} to the rising parts of the lightcurves, as a feature-extraction step, followed by a machine learning classifier applied on these features, without relying on additional external information.
The module’s design choices, which prioritize completeness over purity and result in a relatively low precision but a recall as high as 76\%, make it particularly well suited for identifying off-nuclear TDE-like events and those superimposed on AGN-like variability, alongside standard nuclear TDEs.

We extracted and analyzed the data for all ZTF observed objects having significant rising parts in their lightcurves between late 2019 and mid 2025, and applied different versions of the classifier to them. Visual inspection of the objects flagged as prominent TDE candidates but lacking proper classification in Transient Name Server\footnote{\url{https://www.wis-tns.org/}} (TNS) revealed 89 objects of interest -- out of those, about 40 were interpreted as AGNs, about 40 as SNe, and the remaining dozen objects were interesting TDE/ANT sources, that are presented below. We additionally found serendipitously a number of transients of interest, which were added to the sample presented here, leading to the total of 19 optical nuclear transients. It is important to note that the selection biases of our sample are thus relatively unknown, due to the diverse ways \emi{in which} these transients \emi{were found.}

\subsection{Analysing the candidates}
\label{subsec:analysing}
\subsubsection{Photometric data}
\label{subsubsec:photo}
This study made use of various archival data for each object. Once the candidates were found in alert photometry, we retrieved the detailed ZTF lightcurves through the forced photometry pipeline \citep{masci_zwicky_2018, masci_new_2023}, making use of the $g$, $r$ and $i$ bands. Where needed, we converted these differential fluxes into apparent magnitudes following the standard instructions in \citet{masci_new_2023}. 

We also collected any available $V$-band data from the Catalina Real-time Transient Survey \citep[CRTS,][]{drake_first_2009}. Infrared photometric data is obtained from the NeoWISE catalog data \citep{mainzer_preliminary_2011}, in order to reveal reprocessing of the light from the nuclear transient by dust in the form a delayed infrared echo \citep[e.g.][]{masterson_new_2024}. We use both the $W1$ and $W2$ data, with a 30 days binning. When it was available, we retrieved the Swift/XRT 0.3-10 keV detections and $3\sigma$ upper limits using the online products \citep{evans_real-time_2023}.


\subsection{Discriminating between active and passive host galaxies}
\label{subsec:discriminating}
Before studying the burst in each of the objects presented here, we try to determine whether the host galaxy was previously active (in the sense of having an active galactic nucleus) or passive. This allows to separate the transients between candidate TDEs (in passive galaxies) and candidate ANTs (in active galaxies). We emphasize that this is here mostly a phenomenological consideration, and we make this distinction mostly to support the point that known AGN hosts should not be blindly excluded from transient search algorithms. This classification is ideally performed using spectroscopic data, for instance through the use of BPT diagrams \citep{baldwin_classification_1981}. Whenever such a spectroscopic classification is available \citep[e.g. for objects with SDSS or DESI archival spectra,][]{york_sloan_2000,desi_collaboration_data_2025}, we use it. 

In the absence of previous spectroscopic classification, the task of discriminating between active and passive galaxies is more difficult. 
We make use of Gaia DR3 photometric classification when available \citep{gaia_collaboration_gaia_2023}. We assess the presence of significant AGN-like archival variability in the optical and infrared lightcurves, as an indication of the presence of an AGN. In the absence of any of this information, we use the infrared color criterion $W1-W2>0.8$ \citep{stern_mid-infrared_2012} to assess the presence of an AGN.

We take this opportunity to raise the caveat that, in the context of variability studies, the definition of AGN might be somewhat ambiguous. Indeed, an AGN diagnostics based on the relative amplitudes of narrow lines might indicate that the SMBH has been accreting in a relatively recent past, leading to non-stellar ionization of a narrow line region (and thus be typically identified as Type 2 AGN). However, it does not necessarily indicate that the SMBH is currently accreting -- which is the most important question in our context, both to exclude standard AGN flaring variability, and differentiate between TDEs in vacuum or embedded withing an accretion disk. Fortunately, these concerns should be lifted in the presence of broad emission lines or photometric variability, which require current accretion.

\subsection{Phenomenological fitting}
\label{subsec:phenomenological}
In order to obtain quantified estimates of the various timing parameters of the TDE candidates we present here, we fitted their $g-$band flux lightcurve with a standard phenomenological model to describe TDE evolution, a Gaussian rise followed by a power-law decay \citep[e.g.][]{van_velzen_seventeen_2021}:
\begin{equation}
    F_{g}(t) = F_{g}^{\rm host} + F_{g}^{\rm peak}\times\left\{
    \begin{array}{ll}
        e^{-(t-t_{peak})^{2}/2\tau_{rise}^{2}} & \mbox{if }t\leq t_{peak} \\
        \big((t-t_{peak}+\tau_{decay})/\tau_{decay}\big)^{p} & \mbox{if }t> t_{peak} 
    \end{array}
    \right. 
\end{equation}
The free parameters are thus six: $F_{g}^{\rm host}$ and $F_{g}^{\rm peak}$ the $g$-band fluxes of the host and the (host-subtracted) transient at the peak, $t_{peak}$ the time of the peak, $\tau_{rise}$ the rising timescale, $\tau_{decay}$ the decay timescale, and $p$ the decay power-law index, which we fix at $p=-5/3$. 
A late-time (> one year) plateau is known to sometimes happen in TDEs, at least in optical and ultraviolet wavelengths \citep[e.g.][]{van_velzen_late-time_2019}. This is not accounted for in this model, and for some of our objects this plateau is detected as an excess from the model. 

In practice, the fit is performed on the ZTF g-band lightcurve, using the function \texttt{curve\_fit} from the \texttt{scipy} Python package. Where needed, a late-time component was added (either a late-time plateau emission, or a negative component arising from a quiescent emission lower after the transient compared to before it). In these cases, the components are simply added to the '$\mbox{if }t> t_{peak}$' part of the equation.

\subsection{Physically-motivated fitting}
\label{subsec:physical}

The selection in the \fink\ early TDE module is based on Rainbow \citep{russeil_rainbow_2023}, a phenomenological model well suited to describe the evolution of the lightcurve. However, such approach is not enough to constrain the physics and build a finer understanding of the event. In order to obtain physically informative parameters, we turn to analytical models. For this purpose, we use the \textsc{Redback} python package v.1.0.31 \citep{sarin_redback_2024,ashton_bilby_2019} that provides a collection of models able to fit transients of various natures. It uses Bayesian inference to build a full posterior distribution of the physical parameters, offering highly informative fits. Note that although the parameters carry valuable information regarding the physics of the event, fitting semi-analytical models to every lightcurve is computationally impossible. Indeed, a single lightcurve fit typically takes hours to tens of hours to converge, and therefore can only be applied to carefully selected objects of particular interest. \\

We use the \texttt{tde\_fallback} model, which is a direct adaptation of the model proposed by \citet{mockler_weighing_2019}. In the latter, the bolometric flux of the TDE is powered by the fallback rate of the matter coming from the disrupted star onto the black hole, given an efficiency parameter. It also assumes a photosphere around the black hole, allowing the reprocessing of high energy electromagnetic radiations to visible/UV emissions. Within \textsc{Redback}, the model consists of 9 parameters: $z$ the redshift, $M_{BH}$ the mass of the black hole, $M_{star}$ the mass of the star, $t_{visc}$ the time delay due to viscosity effects (which comprise circularization and accretion processes), $\beta^{*}$ a proxy for the impact parameter, $\eta$ the efficiency of the conversion of the fallback rate of material into electromagnetic emissions, $R_{ph-0}$ a normalizing parameter for the radius of the photosphere, $L^*_{photo}$ the exponent in the power-law describing the dependence of the radius of the photosphere to the luminosity, and $t_0$ the time of the TDE start. The priors for the parameters were chosen following the analysis of \citet{nicholl_systematic_2022}, that provides a systematic modeling of a sample of TDE lightcurves\footnote{Given the extreme behavior of AT2020ukj (see Section \ref{subsec:passive}), the $t_{visc}$ prior upper bound for this particular event was increased by one order of magnitude}. Only the prior on the mass of the star differs. In our case, it was approximated by a gaussian truncated between 0.01 and 100 $M_{\odot}$, with $\mu =0.01$ and $\sigma =0.5$. The Markov chain Monte Carlo (MCMC) is performed using the dynesty sampler \citep{speagle_dynesty_2020}, a robust and standard nested sampling method. We set the number of live points to 1000. 

 
\section{Results}
\label{sec:results}
We present here a sample of 23 optical transients of interest that were found in the ZTF archive. Out of those, 9 are TDE-like flares in galaxies that were previously passive, and as such are good TDE candidates. This includes one very long-duration burst, and one repeated TDE candidate. 8 flares are TDE-like in galaxies that were previously active, and so are ANT candidates. For 2 other transients, the diagnostic methods to assess the host activity were not decisive. Finally, we present 4 supernova candidates, including a seemingly repeated event which is in fact two different SNe in the same galaxy, a few years apart.

A summary of the sources can be found in Table \ref{tab:SummarySources}, with their respective properties and an assessment of their state of activity at the time of writing, to motivate follow-up studies where relevant. For more details, the properties of the bursts are shown in Tab. \ref{tab:TDE_TimingProperties} for the phenomenological fit, and Tab. \ref{tab:Redback_values} for the physically motivated fit. Fig. \ref{fig:ImageTDEs} provides an optical sky view of the host of the transients, Fig. \ref{fig:ZTF_TDEs1} and Fig. \ref{fig:ZTF_TDEs2} provide a detailed view of their ZTF lightcurves, and Fig. \ref{fig:MultiL_TDEs1} and Fig. \ref{fig:MultiL_TDEs2} gives the broader multi-wavelength, multi-epoch context, by showing their CRTS and WISE lightcurves as well. 

\begin{table*}
    \centering
    \resizebox{\textwidth}{!}{\begin{tabular}{lcccccccc}
        \hline\hline \\[-0.2cm]
        Name       & RA          & Dec  & Redshift & $M^{\rm peak}_{\rm{abs,}g}$ [mag] & IR echo & Comments &  Still active\\[0.2cm] \hline\hline \\[-0.1cm]

        \textbf{Candidates TDE}    & & & & & & & \\[0.2cm]
        AT2020ukj / ZTF20accxwrk   & 07:31:34.20 & +57:16:29.04& $0.089\pm0.016$ (P) &$-19.0\pm0.2$&No &Very slow flare in passive galaxy& Yes\\
        AT2023adr / ZTF22abzajwl   & 14:36:19.83 & +32:23:16.46& $0.131$ (S) &-20.7$\pm$0.1/-19.3$\pm0.1$ &Hints&Repeated flares in passive galaxy& Maybe\\
        AT2023npi / ZTF18acfhomp   & 00:12:15.73 & +22:19:18.34& $0.0256$ (S) &$-16.8\pm0.1$&Yes&Very faint flare&No\\
        AT2020pno / ZTF20abjwqqq   & 16:46:37.04 & +55:36:26.81& 0.279 (S) &$-21.5\pm0.1$&Yes&Hints of late-time plateau&No\\
        AT2020aexc / ZTF21aabgjcz  & 12:07:13.45 & +44:10:15.60& $0.43\pm0.03$ (P) &$-22.6\pm0.3$&Hints&&No\\
        ZTF20aatpzog & 11:18:54.34 & +21:49:42.44& $0.38 \pm 0.06$ (P) &$-21.8\pm0.5$&Yes&&No\\
        AT2020afap / ZTF18aasvknh  & 11:36:33.82 & +61:43:39.00& 0.099 (S) &$-18.5\pm0.1$&Yes &&No\\
        AT2023jag / ZTF23aajsmul   & 16:09:12.09 & +46:33:42.4& $0.13\pm0.05$ (P) &$-19.2\pm0.1$&No&&No\\
        AT2021ovg / ZTF21aazenvp   & 17:48:05.01 & +40:14:17.42& $0.24\pm0.06$ (P) &$-20.4\pm0.1$&No&&No\\[0.2cm] \hline \\[-0.1cm] 

        \textbf{Candidates ANT}    & & & & & & & \\[0.2cm]
        ZTF23abjvojy & 02:34:39.04 & +01:07:40.14& 0.27 (S) &$-21.2\pm0.1$&Yes &Repeated flares in active galaxy&Unclear\\
        AT2022yhf / ZTF19aayijkh   & 21:46:06.85 & -11:58:01.02& $0.56\pm0.03$ (P) &$-24.2\pm0.2$& Yes &ENT candidate& Yes\\
        AT2024nxp / ZTF24aamfius   & 17:31:27.23 & +57:36:58.09& $1.7\pm0.8$ (P) &-25.5$\pm1.5$&No&ENT candidate -- Rebrightening?&Yes\\
        AT2023rav / ZTF23aaxaehw   & 23:51:14.69 & +25:41:50.65& $0.504$ (S) &$-24.2\pm0.05$&Yes&ENT candidate&Below quiescence\\
        AT2020actc / ZTF20acxtaau  & 10:03:16.37  & +21:05:45.85& $0.145$ (S) &$-20.2\pm0.1$&Yes&&No\\
        AT2021wxd / ZTF21abvydim   & 15:58:19.21  & +07:28:24.59& 0.231 (S) &$-21.0\pm0.3$&No&&No\\
        AT2024hhj / ZTF24aaimfrw   & 08:48:37.12 & +33:50:15.01& $0.217$ (S) &$-21.5\pm0.5$&No&&No\\
        AT2024gzn / ZTF24aafvgzk   & 17:55:16.11 & +38:58:41.44& $0.23\pm0.03$ (P) &$-20.5\pm0.2$&No&Likely AGN burst& Rebrightening?\\[0.2cm] \hline \\[-0.1cm] 

        \textbf{Ambiguous}    & & & & & & & \\[0.2cm]
        AT2023zaj / ZTF23abowyjf   & 00:37:34.44 & -13:33:51.5& $0.4\pm0.1$ (P) &-22.5$\pm$0.6&Yes&ANT or TDE&Yes\\
        AT2023szj / ZTF23aajnfna   & 14:46:28.16   & +08:41:36.7& $0.57\pm0.16$ (P) &-23$\pm$0.75& Yes &Likely AGN burst&Rebrightening?\\[0.2cm] \hline \\[-0.1cm] 
 
        \textbf{Candidates SNe}    & & & & & & & \\[0.2cm]
        AT2021lnu / ZTF21aazrgtw   & 15:32:18.68 & -00:00:00.74& $0.15\pm0.07$ (P) &-20.2$\pm$1.2&No&SLSN candidate&No\\
        AT2020mvg / ZTF20aazyvre   & 12:05:26.53 & +28:54:34.36& $0.31\pm0.11$ (P) &-22.4$\pm$0.9&Unclear&SLSN candidate&No\\
        AT2019agc / ZTF19aafmytc   & 14:06:45.32 & +13:55:04.7&  $0.15\pm0.05$ (P) & -19.7$\pm$0.7/-19.7$\pm$0.7& No&Multiple SNe&No\\
        AT2024ljd / ZTF20aacedmi   & 21:13:39.44   & +02:29:36.59& $0.046$ (S) &$-19\pm0.05$&No&Confirmed SN Ia&No\\[0.2cm] \hline \\[-0.1cm]        

    \end{tabular}}
    \caption{Summary information about the various transient objects present11ed here. The table is divided in four sections, corresponding to flares in passive galaxies (secure TDE candidates), flares in active galaxies (TDE or flaring AGN candidates), ambiguous sources where classification is uncertain, and the most interesting SN candidates. For the redshift, we indicate with (S) those for which it is spectroscopic, and with (P) when it is photometric. The absolute magnitude, given for the $g$ band, comes from the fitted phenomenological burst profile and takes into account the redshift uncertainties.}
    \label{tab:SummarySources}
\end{table*}

\subsection{TDE candidates in passive galaxies}
\label{subsec:passive}

\subsubsection{\href{https://ztf.fink-portal.org/ZTF20accxwrk}{AT2020ukj}: A long-lived optical nuclear transient}
Because of the selection biases of our early detection system, aiming specifically at slow and hot nuclear transients to distinguish SNe candidates from TDEs, we tend to select objects with long rising timescales. AT2020ukj was selected because of this, but its decay timescale is even more outstanding. Its host is WISEA J073134.20+571628.9, a galaxy with a photo-z=$0.089\pm0.016$ \citep{duncan_all-purpose_2022}, which is likely passive \citep[no sign of variability in the past ZTF, CRTS or WISE lightcurves, WISE color W1-W2$\sim$0, and classified as passive galaxy by Gaia DR3, ][]{gaia_collaboration_gaia_2022}. Between June and December 2020, it underwent a slow blue brightening (rise timescale of $68.5\pm8.8$ days, peak difference color of $g-r\sim-0.3$), peaking at differences magnitudes of 19.0 (19.2), or absolute magnitudes of -19.0 (-18.8). It then started an exceptionally slow decay, at constant color, with a decay timescale of $4984.9\pm497.2$ days -- since its peak in December 2020, it has not yet reached the quiescent level. 

\begin{figure}[h]
    \centering
    \includegraphics[width=\columnwidth]{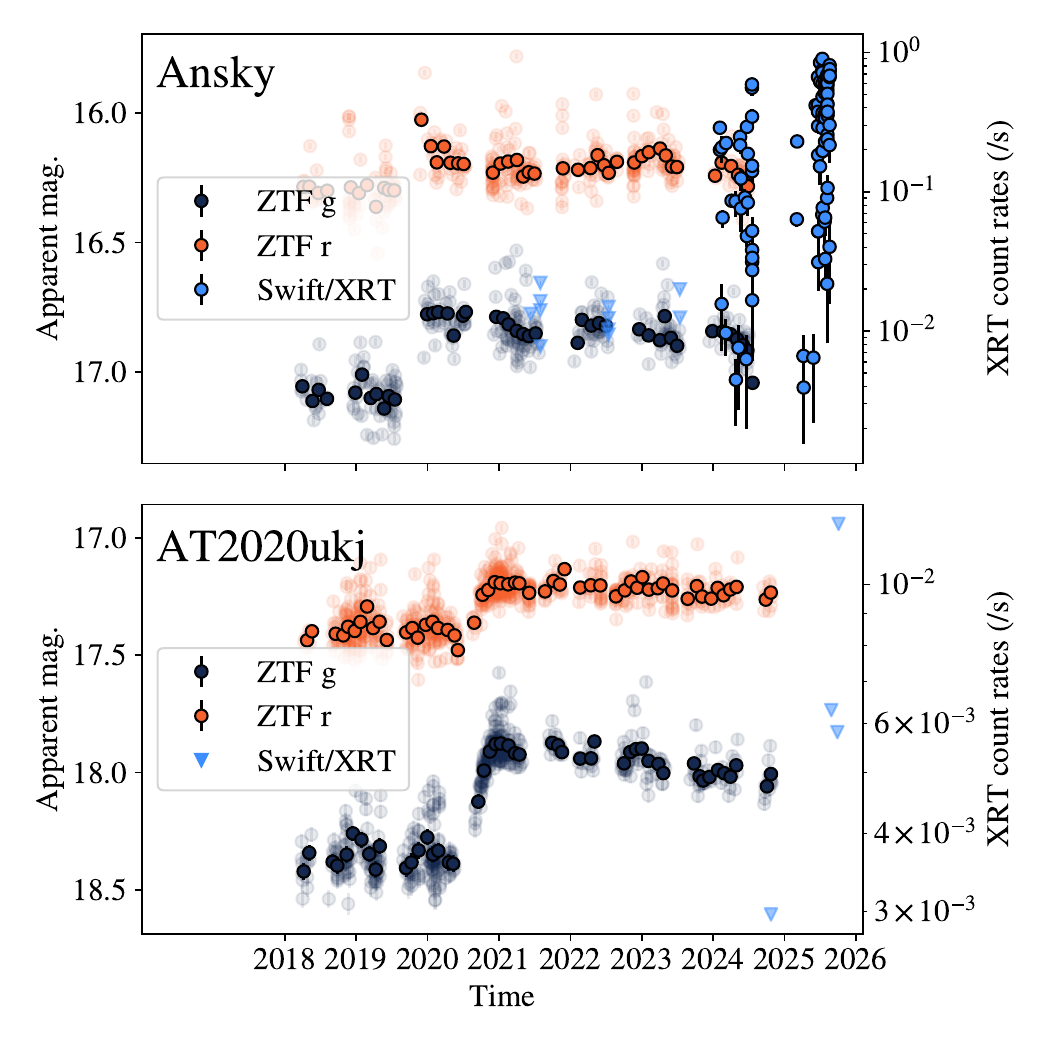}
    \caption{Comparison of the behaviors of ZTF19acnskyy ("Ansky", top panel) and AT2020ukj (bottom panel), showing their ZTF r and g bands (orange and dark blue), and the Swift/XRT detections and upper limits (light blue circles and downwards triangles). This shows their similarities in terms of long decay timescales, as well as the complex late-time X-ray behavior of Ansky.}
    \label{fig:AT2020ukj_Ansky}
\end{figure}


Such long timescales exclude SNe contaminants. In fact, this source shows striking similarities with ZTF19acnskyy \citep[hereafter Ansky,][]{sanchez-saez_sdss13350728_2024} in terms of timescales, as can be seen in Fig. \ref{fig:AT2020ukj_Ansky}. Ansky has a comparable decay timescale of $6600.5\pm1484.6$ days, which is very large compared for instance to that of the slowest TDE in the ZTF sample from \citet{hammerstein_final_2022}, AT2018iih, which is about 200 days. In particular, one way to obtain such long bursts in a standard TDE model is through a combination of high viscosity and low efficiency of the accretion flow, which is confirmed by the \texttt{tde\_fallback} fit, showing AT2020ukj as an outlier in both parameters (respectively log($t_{\rm visc})=3.21^{+0.02}_{-0.02}$ and log($\eta)=-3.98^{+0.03}_{-0.01}$). We fitted the same TDE model to Ansky's ZTF data, leading to similar very low viscosity (log($t_{\rm visc})=3.28\pm0.01$), although the lower absolute magnitude led to the other parameters being different between the objects.

As a consequence of their similarity, AT2020ukj and Ansky have the same possible interpretations: either an extreme long-duration TDE, a TDE in an AGN with a low inclination angle, or an AGN in the process of activation. The main difference between these models would be the physical origin of the materials being accreted, either a tidally disrupted star \citep[likely very massive, e.g.][]{zhu_ultraviolet_2025}, or a gas overdensity. However, compared to Ansky, AT2020ukj is significantly brighter (absolute g-band magnitudes of -19 for AT2020ukj compared to -16.5 for Ansky). Such a brightness over such timescales leads to a massive energy budget, which could be too important for the disruption of a single star, and the interpretation of a starting AGN would seem to be favored here. A smoking gun evidence of this interpretation would be the late-time appearance of typical AGN observables, such as emission lines. We have obtained two spectra from the Caucasian Mountain Observatory (CMO, Russia), taken on December 14$^{\rm th}$, 2024 and January 26$^{\rm th}$, 2025, which showed no sign of clear emission feature, although the signal to noise ratio was too low to perform any definitive conclusion \citep{russeil_identification_2024}. As such, we have requested an optical spectral monitoring of this source, and will report on its evolution. Ansky is also interesting for its multi-wavelength behaviour, through the late-time appearance of X-ray Quasi-Periodic Eruptions (QPEs) in its decay \citep{hernandez-garcia_discovery_2025}, which are rare repeated bursts of thermal X-rays, with a dozen known sources to this date. The QPEs in Ansky are also the first one to show an evolution of their periodicity \citep{hernandez-garcia_nicer_2025}. We have thus requested an X-ray monitoring of this source, in order to constrain the late-time X-ray behaviour of this peculiar object.



\subsubsection{\href{https://ztf.fink-portal.org/ZTF22abzajwl}{AT2023adr}: a repeated pTDE candidate}
Our pipeline detected two transient objects displaying repeated flares: AT2018mac/TDE2022dbl and AT2023adr. The former has been reported recently  \citep[e.g.][]{chen_fate_2024,lin_unluckiest_2024,liu_repeating_2025,makrygianni_double_2025}; however the rebrightening of the latter had gone unnoticed until we detected it \citep{llamas_lanza_identification_2024}.

\begin{figure}
    \centering
    \includegraphics[width=\columnwidth]{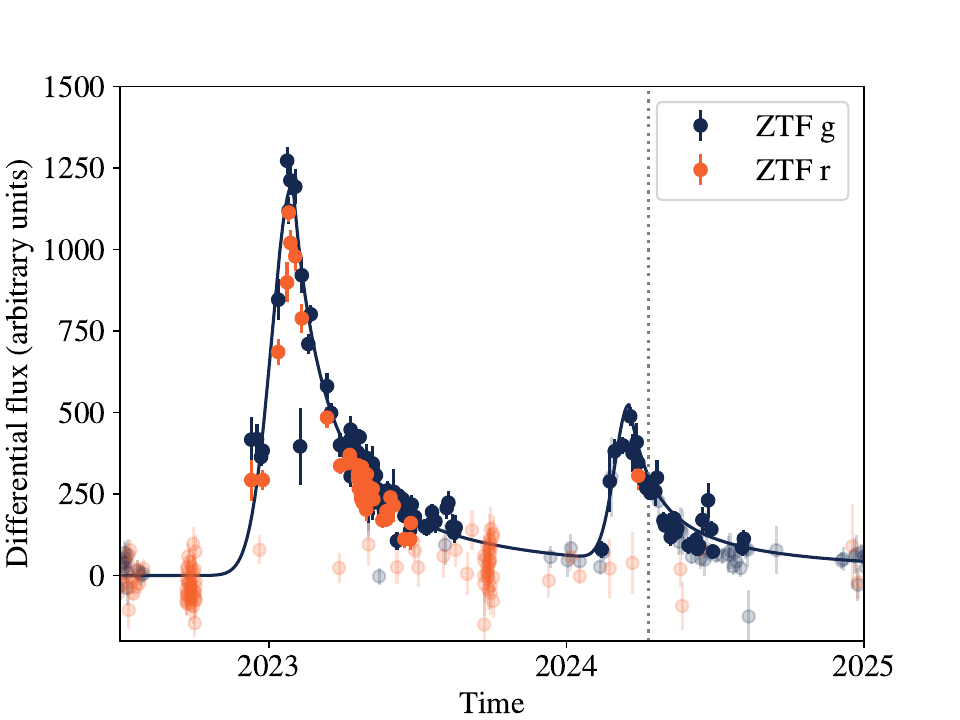}
    \caption{ZTF forced photometry lightcurve of AT2023adr, a candidate repeated TDE. Points with a signal to noise ratio below 3 have been plotted in transparence. The full line corresponds to a double TDE profile (gaussian rise, $\propto t^{-5/3}$ decay) fitted to the $g$-band lightcurve. The dotted vertical line shows the timing of the ePESSTO+ optical spectrum, during the second peak. }
    \label{fig:AT2023adr_ZTF_zoom}
\end{figure}

AT2023adr was first detected in early 2023. Its host is a bright relatively red galaxy ($g-r\simeq19.6-18.8=0.8$) at a spectroscopic redshift $z=0.131$, that displayed no variability before the first peak in either ZTF, WISE or CRTS data. Moreover, its WISE color is $W1-W2\sim0.1$. As such, the host can be safely classified as passive. In January 2023, it underwent a slow (rise time of $25.5\pm0.9$ days and decay of $64.3\pm1.6$ days) and blue (difference color of $g-r\sim-0.2$, constant during decay) brightening, peaking at difference magnitudes $18.15\pm0.1$ ($18.0\pm0.1$) and absolute magnitudes $-20.7\pm0.1$ ($-20.9\pm0.1$) in the ZTF $g$-band ($r$-band respectively). This is too bright for standard supernovae -- without a spectral follow-up, this object could be either a TDE or SLSN candidate, with the constant temperature being indicative of TDE. Over the course of 2023, it then dropped back down to the quiescence level (with a $3\sigma$ upper limit of the absolute $g$-band magnitude around -18).

However, about a year after the first peak, the source underwent a strong rebrightening (see Fig. \ref{fig:AT2023adr_ZTF_zoom}). This second peak had relatively faster timing properties (rise of $16.6\pm1.5$ days and decay of $40.5\pm3.7$ days). It was mostly covered by the ZTF g-band only, with a single good detection in the ZTF r-band, so no color evolution can be strongly inferred from the photometric data, although this single point indicates similar color and thus temperature as the first peak. It peaked about one magnitude fainter than the first event, at difference magnitude of 18.8, and absolute magnitude of -19.3 in the $g$-band (consistent with either SN or TDE). The separation between the peaks and the return to quiescence are not consistent with a single, not-smoothly-evolving SLSN. Additonnally, a follow-up optical spectrum was obtained during the decay of the second peak by the ePESSTO+ collaboration \citep[][see vertical dotted line in Fig.~\ref{fig:AT2023adr_ZTF_zoom}]{dalen_epessto_2024}. This spectrum, while having a relatively low signal to noise ratio, does not display any line expected from SN decay. It shows [O II] and [O III] emission lines, and an emission feature around 4610\AA ~which could be ascribed to blueshifted N III 4640 or He II 4686, which would be consistent with a TDE. Additionally, hints of an infrared echo can be found in the NeoWISE lightcurve of the object, with a slight increase by 0.3 magnitudes of the W2 band only.

As such, the lack of cooling during the first peak, and the lack of typical SN decay-related emission features in the second peak, lead us to the conclusion that a repeated TDE would be a strong interpretation of this event. By fitting the two distinct peaks with \textsc{Redback} separately, we obtained different physical parameters for the two events - in particular, the mass of the disrupted star is lower in the second peak ($0.76^{+0.14}_{-0.14} M_{\odot}$ compared to $0.32^{+0.23}_{-0.12} M_{\odot}$), which is physically consistent if the first disruption is only partial. However there are a handful of caveats, the main one being that the \textsc{Redback} model \texttt{tde\_fallback} was built for full disruptions, and might as such not provide proper estimates for partial disruptions. Moreover, the inferred mass of the central black hole is not consistent between the two peaks (log($M_{SMBH}/M_{\odot})=6.79^{+0.06}_{-0.08}$ compared to $5.84^{+0.32}_{-0.31}$) -- this could be explained if the two disruptions happened for one star in a relatively asymmetric binary SMBH system \citep[as has been suggested for instance for AT2018fyk, a similar repeated event, with similarly asymmetric SMBH mass estimates, ][]{wen_at2018fyk_2024}. This inconsistency could also arise from \texttt{tde\_fallback}'s assumption of full disruption, which would not be respected here, or from the absence of $r$-band detections in the second peak preventing a proper fit. 






\subsubsection{\href{https://ztf.fink-portal.org/ZTF18acfhomp}{AT2023npi}: an extremely faint TDE candidate}
The host of AT2023npi is UGC 113, a nearby galaxy at a spectroscopic redshift of $z=0.0256$. It is not bright in radio or X-rays despite being very close, its UV emission is not peaked at the nucleus, and its infrared color is $W1-W2\sim0$. As such, it is likely a passive galaxy.

In 2023, its nucleus underwent an outburst, rising with a timescale of $46.6\pm6.6$ days. It peaked at difference magnitudes of $\sim18.5$ in both $r$ and $g$ bands, corresponding to absolute magnitudes of $\sim-16.8$. The outburst was quite blue, with difference color $g-r\sim0$. It then decayed very slowly, with a timescale of $246.5\pm13.0$ days, with no clear cooling in the decay, and some hints of structure in the lightcurve. It showed one of the most significant infrared echo of our sample (owing to the nearby nature of the host), with a brightening of the $W1$ and $W2$ bands by 0.4 and 0.7 respectively, associated with a significant change in the infrared color of the galaxy (shifting from $W1-W2=-0.05$ to $W1-W2=0.25$). 

Such an absolute magnitude excludes SLSN, which should be at least three magnitudes brighter. Additionally, the very long timescales (active emission during more than 2 years) exclude standard supernovae. As such, AT2023npi is an excellent faint, nearby TDE candidate. The physical fit provides us with extreme parameters to account for the faintness of the transient, seemingly favoring a very low mass star (the fit hits the border of the prior, at a stellar mass of $0.01M_{\odot}$) and a small photosphere radius log($R_{ph-0})=0.03^{+0.05}_{-0.09}$. This would make this source one of the faintest optical TDE candidate to date, along with for instance AT2020wey, Ansky or J0744, all peaking around absolute magnitude -17 \citep{yao_tidal_2023, sanchez-saez_sdss13350728_2024, malyali_erasst_2023}. A part of this faintness could be due to absorption by a particularly dusty galactic nucleus (which could also account for the bright infrared echo) -- optical spectra of the transient while it was active would have been needed to confirm this hypothesis, and estimate clearly the intrinsic brightness of the transient.




\subsubsection{\href{https://ztf.fink-portal.org/ZTF20abjwqqq}{AT2020pno}: hints of late-time plateau}
The host of AT2020pno is WISE J120713.44+441015.6, a red galaxy at a spectroscopic redshift of $z=0.279$, that showed no variability in the historical CRTS, ZTF or WISE data, with infrared color $W1-W2\sim0.1$, and no UV counterpart in GALEX. As such it is likely a passive galaxy. 

\begin{figure}
    \centering
    \includegraphics[width=\columnwidth]{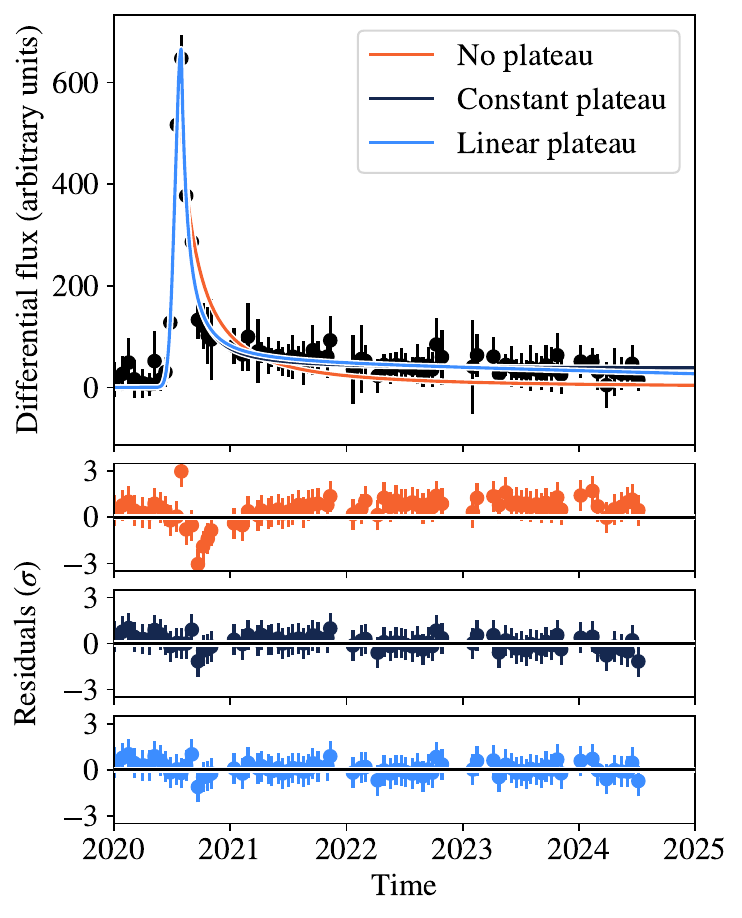}
    \caption{Evidence for late-time plateau in AT2020pno, with the data and the fitted models in the top panel, and the corresponding residuals in the bottom three panels. Here three models are compared: a simple $\propto t^{-5/3}$ decay (orange, significant negative then positive residuals), a constant plateau (dark blue, slight negative then positive residuals), and a linearly decreasing plateau (light blue, no strong residuals). Data here is \emi{shown} in bins of 15 days.}
    \label{fig:AT2020pnoPlateau}
\end{figure}

In 2020 it displayed a strong blue brightening, with a rise timescale of $19.0\pm0.7$ days, and a peak difference color of $g-r\sim-0.2$. The peak reached difference magnitudes of about 19.2, corresponding to absolute magnitudes of -21.5 in both bands. A spectrum was obtained during the burst from the Keck observatory \citep{pessi_spectroscopic_2024}, showing among other things a broad emission complex around 4600\AA. This could correspond to blueshifted HeII, which is a common feature in TDEs. It then decayed with a typical timescale of $50.5\pm1.9$ days. However, using a simple $\propto t^{-5/3}$ decay leads to significant late-time residuals. This can be interpreted as a late-time plateau emission, which is a typical TDE behaviour \citep[e.g.][]{mummery_fundamental_2024}. In particular, by fitting models to the photometric data, the absence of plateau is significantly worse than a constant plateau, which is marginally worse than a linearly decreasing plateau (Reduced $\chi^{2}$ of 0.48, 0.24 and 0.19 respectively, see residuals in Fig. \ref{fig:AT2020pnoPlateau}). Finally, in late 2024, this source showed a significant infrared echo, with an increase of NeoWISE magnitudes by 0.6 and 0.4 for the W1 and W2 bands respectively. 

This source has been noticed in \citet{pessi_sample_2025} and \citet{pavez-herrera_alerce_2025} as an ambiguous transient event. The broad He II emission complex, the lack of significant color evolution during the decay, and the late-time plateau and infrared echo, all lead us to classify this source as a strong TDE candidate. We obtained a monitoring of this source by Swift, which did not lead to any X-ray or UV detection.

\subsubsection{\href{https://ztf.fink-portal.org/ZTF23aajsmul}{AT2023jag}}
The host of AT2023jag is SDSS J160912.07+463342.4, which has a photo-z of 0.13$\pm0.05$. This galaxy has no detected infrared counterpart or radio counterpart, did not show any significant variability in CRTS, and is classified as passive by Gaia -- as such, it is likely a passive galaxy. 

In early 2023, it underwent a blue brightening, with rise timescale of $28.1\pm1.8$ days and decay timescale of $89.8\pm7.8$, and a peak difference color of $g-r\sim0$. It peaked at difference magnitudes of 19.75 in both bands, corresponding to absolute magnitudes of -19.15. The decay showed constant temperatures. All those properties strongly favor the TDE interpretation. 

\subsubsection{\href{https://ztf.fink-portal.org/ZTF21aabgjcz}{AT2020aexc}}
The host of AT2020aexc is SDSS J120713.45+441015.5, with a SDSS photo-$z$ of $0.43\pm0.03$. It is a relatively red galaxy with no UV counterpart in GALEX, whose infrared color is W1-W2$\sim0.1$, and that showed no significant variability in the ZTF and CRTS data -- as such, it is likely a passive galaxy. 

In the beginning of 2021, it showed a slow and blue brightening, with rise timescale of $38.2\pm2.8$ days and peak difference color of $g-r\sim-0.5$. It peaked at difference magnitudes of about 19.25 (19.75), corresponding to absolute magnitudes of -22.6 (-22.1). Such luminosities could only correspond to a SLSN or a (quite bright) TDE. It then decayed   slowly, following a timescale of $198.1\pm11.0$ days, with no signs of cooling during the decay. In 2023, the NeoWISE lightcurve showed hints of brightening, especially in the W1 band, by $\sim0.25$ magnitudes, although it is not extremely significant (see Fig. \ref{fig:MultiL_TDEs1}). 


AT2020aexc was noticed by another detection tool, FLEET \citep{gomez_search_2023}, with a 58\% chance of being a TDE, and flagged as a potential lensed supernova in \citet{magee_search_2023}. The long timescales, large amplitude, blue color, and constant temperature over the decay make this source a solid TDE candidate. 

\subsubsection{\href{https://ztf.fink-portal.org/ZTF18aasvknh}{AT2020afap}}

The host of AT2020afap is SDSS J113633.84+614338.9, a spectrally confirmed passive galaxy at a redshift of $z=0.099$. In the end of 2018, it showed a blue outburst, rising with a timescale of $59.2\pm5.0$ days, and reaching at the peak difference magnitudes of 19.75, corresponding to absolute magnitudes of -18.5. Its apparent color shifted from $g-r\sim0.9$ to $g-r\sim0.6$. It then decayed following a timescale of $92.2\pm10.6$ days, with no clear color evolution. Its NeoWISE lightcurve showed a clear infrared echo in both bands, with the $W1$ and $W2$ bands brightening by 0.25 and 0.5 magnitudes respectively.

The timescales are too long for standard supernovae, while the absolute magnitude is too low for a SLSN. Moreover, the amplitude, timescale, and color evolution all fit what would be expected from a TDE. As such, AT2020afap is a good TDE candidate. It was picked up as such in \citet{gomez_search_2023}.

\subsubsection{\href{https://ztf.fink-portal.org/ZTF20aatpzog}{ZTF20aatpzog}}
The host of ZTF20aatpzog is SDSS J111854.33+214942.4, a faint galaxy with a photometric redshift of $z=0.38 \pm 0.06$. It showed no strong variability in historical CRTS, ZTF or WISE data, and has an infrared color of $W1-W2\sim0.2$, making it a likely passive galaxy. 

In early 2020 it showed a strong brightening, with a typical rise time of $27.9\pm1.8$ days, and peak difference colors of $g-r\sim0$. 
It peaked at difference magnitudes of about 19.75 in both bands, corresponding to absolute magnitudes of $-21.8\pm0.4$ in both bands. It then decayed relatively quickly, following a typical timescale of $31.7\pm5.5$ days, with a constant color during the decay. The NeoWISE lightcurve of the host showed signs of an infrared echo in late 2020 and early 2021, with a brightening in both $W1$ and $W2$ bands by about 0.25 magnitudes.

The timing and amplitude of the burst point at either the TDE or SLSN interpretation, and the constant color in the decay tends to favor the former. As such, we conclude that ZTF20aatpzog is a good TDE candidate.

\subsubsection{\href{https://ztf.fink-portal.org/ZTF21aazenvp}{AT2021ovg}}
The host of AT2021ovg is SDSS J174805.00+401417.4, a faint, red galaxy at a SDSS photometric redshift of $z=0.24\pm0.06$. It is undetected in UV and radio, its CRTS, ZTF and NeoWISE historical lightcurves show no significant variability, and the WISE infrared color is $W1-W2\sim0.2$. As such, it is likely a passive galaxy.

In 2021, it underwent a blue outburst, following a timescale of $36.9\pm2.2$ days. It reached difference magnitudes of about 20, or absolute magnitudes of -20.4 in both bands. It also became significantly bluer, with the apparent color shifting from $g-r\sim1$ to $g-r\sim0.25$. It then slowly decayed, following a timescale of $59.5\pm5.7$ days, with no significant color change in the decay. Its infrared lightcurve showed no significant echo. While the peak absolute magnitude is consistent with either the TDE or SLSN interpretation, the lack of cooling in the decay allows to conclude that AT2021ovg is a good TDE candidate.

\subsection{Flaring events in active galaxies}
\label{subsec:flaring}

\subsubsection{\href{https://ztf.fink-portal.org/ZTF23abjvojy}{ZTF23abjvojy}}
The host of ZTF23abjvojy is SDSS J023439.04+010740.1, a spectrally-confirmed Narrow-Line Seyfert 1 \citep[NLSy1, e.g.][]{paliya_narrow-line_2024} at z=0.27. As an active galaxy, the quiescent ZTF level showed hints of variability (e.g. in early 2022 by 0.25 mag) -- however, the transient in late 2023 is in clear excess of this expected AGN-induced variability level. 

The source rose relatively quickly, with a typical timescale of $25.5\pm2.8$ days, up to difference magnitudes of about 19.5 in both bands, corresponding to absolute magnitudes of -21.2. It became significantly bluer, with the apparent color shifting from $g-r\sim0.6$ in quiescence to $g-r\sim 0.2$ at the peak. It then decayed slowly, following a timescale of $186.1\pm24.4$ days, with no significant color change in the decay. The NeoWISE lightcurve shows a clear echo to this 2023 transient, with a brightening of both the $W1$ and $W2$ bands by 0.6 magnitudes each.  

Interestingly, it was not the first time this galaxy displayed such a dramatic brightening: in January 2008, an even brighter event was detected by CRTS, where the source rose quickly (timescale of $16.1\pm8.4$ days) from a V-band apparent magnitude of $\sim20$ up to $\sim18.25$, corresponding to absolute magnitude of -22.5. It was then followed by a very slow decay over several years (decay timescale $523.5\pm58.2$ days).

\begin{figure}
    \centering
    \includegraphics[width=\columnwidth]{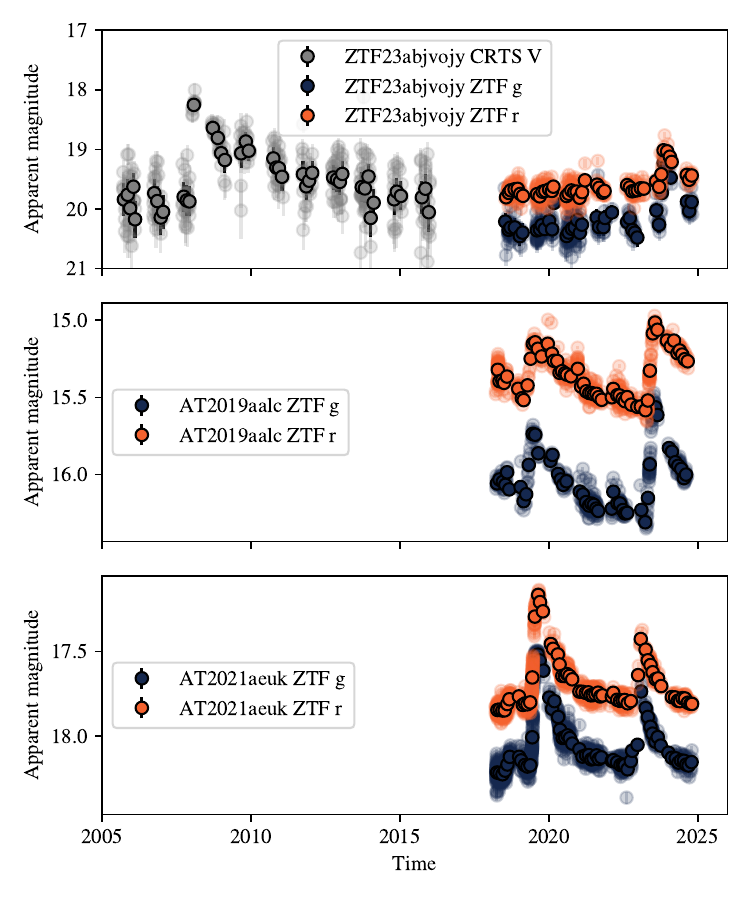}
    \caption{Evidence for two TDE-like flares 16 years apart in ZTF23abjvojy. Its lightcurve (top panel) is shown for both CRTS V-band (light blue) and ZTF g-band (dark blue) and r-band (orange) data. We also provide comparison with two other known repeated AGN flares: AT2019aalc (middle panel) and AT2021aeuk (botton panel). The transparent points correspond to unbinned data, and the opaque points to data binned in bins of 15 days. We draw attention to the fact that the bands for CRTS and ZTF here are not the same and not cross-calibrated, so their respective magnitudes should not be directly compared.}
    \label{fig:ZTF23abjvojy_RepeatedBurst}
\end{figure}

Both events are consistent in terms of timing and amplitude properties with a TDE or an AGN flare. However, the shape of the peaks (in particular their noticeable asymmetry) is not consistent with standard AGN red-noise variability. Unfortunately, it is not possible to conclude on the exact cause of the outburts in the absence of spectral data. This source is a good candidate to join the small group of repeated TDE-like flares in AGN, along with AT2019aalc \citep{milan_veres_back_2024}, AT2021aeuk \citep{sun_at2021aeuk_2025}, and IRAS F01004-2237 \citep{sun_recurring_2024}. The timing properties of the individual flares are roughly consistent between these sources, and they all showed infrared echoes. For AT2019aalc and AT2021aeuk, the time between flares was 3-4 years (which can be understood as a selection bias, as both flares had to be seen by ZTF over a baseline of seven years). However, ZTF23abjvojy repeats on a much longer timescale, with a time between peaks of $5785\pm10$ days, i.e. almost 16 years apart (see Fig. \ref{fig:ZTF23abjvojy_RepeatedBurst}) -- such a long baseline was made possible by the use of CRTS data. In that sense it is most similar to IRAS F01004-2237, which was revealed as well through a combined use of CRTS and ZTF (time between peaks of $4175\pm100$ days). Interestingly, this source is a NLSy1, which is a class of AGN that has been shown to be associated with similar flaring events \citep[e.g.][]{frederick_family_2021}


\subsubsection{\href{https://ztf.fink-portal.org/ZTF19aayijkh}{AT2022yhf}}

The host of AT2022yhf is WISEA J214606.86-115801.1, which is a blue galaxy with two different photo-$z$ estimates, $0.56\pm0.03$ or 0.769 \citep{gaia_collaboration_gaia_2023, fu_catnorth_2024}. The quiescent state, as seen by ZTF and CRTS, shows variations on timescales of $\sim6$ months that are consistent with nuclear activity - it is also classified as QSO by Gaia DR3. It is thus likely an active galaxy.

On top of the expected AGN-like variability, a significant brightening happened in September 2022, way beyond the amplitudes previously seen. The source slowly rose with a timescale of $64.0\pm2.0$  days, reaching its peak four month later in December 2022, at apparent magnitudes around 18.25 (18.5) in the $g$-band ($r$-band), corresponding to absolute magnitudes of -24.3 (-24.0) for the first redshift estimate, -25.1 (-24.8) for the second. It also became significantly bluer ($g-r$ decreasing from $\sim$0.1 in quiescence to $\sim$-0.2 at the peak). Since then, it has decayed slowly and steadily, with a typical decay time of $668.2\pm34.3$ days, although some gaps in exposures prevent a consistent sampling of the decay. The color has not reddened noticeably. It is still yet to reach its quiescent state, after more than three years. The NeoWISE lightcurve shows very clear signs of an infrared echo to the transient event, with a brightening of both $W1$ and $W2$ by almost one magnitude. All those properties make this object a good ANT, and perhaps ENT, candidate (depending on the actual redshift of the host).



\subsubsection{\href{https://ztf.fink-portal.org/ZTF23aaxaehw}{AT2023rav}}
This source is a known broad-line QSO, SDSS J235114.69+254150.6, at $z$=0.504. It is radio and ultraviolet bright \citep[NVSS J235114+254152 and GALEXASC J235114.69+254151.1,][]{condon_nrao_1998, bianchi_revised_2017}, further confirming the AGN nature of the object. 
It was seen by CRTS, showing long-term variability roughly consistent with AGN behavior.

The source underwent a fast brightening by about one magnitude over August 2023, reaching peak difference magnitudes of around 18 in both bands. This corresponds to a peak absolute difference magnitude of $\sim-24.2$ in both bands, which would make it an ENT candidate. The transient has a significantly larger amplitude than any previous variability, which tends to exclude standard AGN variability as an explanation. The transient is bluer than the host during its rise, however cools down during the $\sim6$ months of decay. There are hints of a slight infrared echo, especially in the W1 band (by $\sim0.25$ mag). 

One of the most striking features of this object is its post-peak behaviour: indeed, once the source reached a new steady state in late 2024, this new quiescence level was about 0.4 (0.3 respectively) magnitudes lower than it was before the transient for the $g$-band ($r$-band respectively). This could be indicative that the ENT, whatever its exact physical cause might be, depleted some of the materials from the region directly surrounding the SMBH, and thus slowed down the ulterior standard accretion once the transient is over. This is also supported by the fact that this decreased quiescence is more pronounced in the bluer $g$-band. This depletion of the accretion flow is akin to what was seen for instance in X-rays in 1ES 1927+654 \citep{trakhtenbrot_1es_2019}, where the X-ray activity of the source was consistent with an AGN before the transient event and subsequently significantly decreased, indicative of a possible depletion of the accretion flow. In particular, this depletion is expected from a retrograde TDE in an AGN disk \citep{mckernan_starfall_2022}. One could expect that, as the surroundings of the SMBH recover their previous content, the source should regain its quiescence level over the upcoming months / years - as such, we encourage a further monitoring of this source.

\subsubsection{\href{https://ztf.fink-portal.org/ZTF24aamfius}{AT2024nxp}}
The host of AT2024nxp is SDSS J173127.23+573658.1. It is faint ($g\sim20.7$), blue ($g-r\sim0.3$), UV-bright (GALEX J173127.2+573658), and its infrared color by NeoWISE ($W_1-W_2 \sim0.9$) is consistent with that of an AGN. As such, the host is likely a quasar \citep[e.g. $P_{QSO}=96\%$ for Gaia DR3][]{gaia_collaboration_gaia_2023}. It has a number of photometric redshift estimates, depending on the method and the wavelengths, ranging from 0.6 to 2.5.

In April 2024, the source underwent a significant brightening by more than two magnitudes, following a rising timescale of $30.4\pm1.0$ days. It became bluer ($g-r\sim0.1$) and peaked at magnitudes of $\sim18.75$ in both bands, corresponding to absolute magnitudes in the range -24 to -27, depending on the redshift value, qualifying it as an ENT candidate even for the most conservative estimates. It then started to decay at constant color, following a relatively long timescale of $137.1\pm5.9$ days. There is no sign of an infrared counterpart to the optical transient. The decay followed well a $t^{-5/3}$ shape for the first 6 months, after which a slight excess to this shape can be seen (see right panel of Fig. \ref{fig:AT2024gzn}, or Fig. \ref{fig:ZTF_TDEs2}). It is unclear whether this corresponds to a change of the power-law index of the decay, or a low-amplitude rebrightening. The current sample of ENTs is currently too limited to assess whether such irregular decay is to be expected. Two other potential ANTs from our sample present similar late rebrightenings, but with higher amplitudes (AT2023szj and AT2024gzn).

\subsubsection{\href{https://ztf.fink-portal.org/ZTF24aaimfrw}{AT2024hhj}}
The host of AT2024hhj is SDSS J084837.11+335014.9, a spectrally confirmed QSO at a redshift of $z=0.217$. In particular, it is another NLSy1 \citep{paliya_narrow-line_2024}. The CRTS, ZTF and NeoWISE lightcurves show levels of variability consistent with nuclear activity, with around 0.5-1 magnitude amplitude.

However, in 2024 it underwent a significant brightening, rising with a timescale of $18.8\pm2.6$ days, peaking at difference magnitudes of about 18.5, corresponding to absolute magnitudes of -21.5 (although the peak itself was missed due to poor sampling by ZTF). It also became significantly bluer, from apparent color of $g-r\sim0.5$ to $g-r\sim0$. It then decayed slowly following a timescale of $82.8\pm11.1$ days, with no significant color change in the decay. There doesn't seem to be an infrared counterpart to the optical transient, beyond some variability with comparable level to what was observed before the optical peak.

\subsubsection{\href{https://ztf.fink-portal.org/ZTF21abvydim}{AT2021wxd}}
The host of AT2021wxd is SDSS J155819.21+072824.5, a spectrally confirmed AGN at a redshift of $z=0.231$. Its CRTS, ZTF and NeoWISE archival lightcurves displayed small-amplitude variability consistent with an AGN. 

In late 2021, it underwent a slow blue brightening, with a timescale of $24.8\pm3.4$ days, peaking at difference magnitudes 19.25 or absolute magnitudes of -21. Its color went significantly bluer, from $g-r\sim0.5$ to $g-r\sim0$. It then decayed following a timescale of $160.6\pm16.6$ days, with no sign of color change in the decay. The infrared lightcurve shows no sign of variability beyond the historical levels of AGN-induced variability.

\subsubsection{\href{https://ztf.fink-portal.org/ZTF20acxtaau}{AT2020actc}}
The host of AT2020actc is SDSS J100316.36+210545.8, a spectrally confirmed Seyfert 1 AGN at a redshift of $z=0.145$. The historical ZTF, CRTS and NeoWISE lightcurves showed very slight levels of variability (around 0.1 magnitude), which is consistent with a steady AGN accretion flow. 

In late 2020 the source underwent a brightening over a timescale of $29.5\pm2.8$ days, peaking at difference magnitudes of about 19, corresponding to absolute magnitudes of -20.2 in both bands. It then decayed steadily over a timescale of $136.6\pm15.4$ days, with no signs of color change in the decay. The infrared lightcurve showed signs of an echo just after the optical burst, brightening by about 0.2 magnitudes in both $W1$ and $W2$ bands.

\subsubsection{\href{https://ztf.fink-portal.org/ZTF24aafvgzk}{AT2024gzn}}
The host of AT2024gzn is SDSS J175516.11+385841.3, a spectroscopically confirmed QSO at redshift of $z=0.201$. It showed no variability in historical CRTS or ZTF data. The infrared lightcurve shows hints of a slow and steady increase ($\sim0.25$ magnitudes over a decade).

In late 2023, the nucleus displayed a slow blue outburst, with a rise timescale of $64.9\pm4.9$ days and a peak difference color of $g-r\sim0$. It reached peak difference magnitudes of about 19.75 in both bands, corresponding to absolute magnitudes of $-20.5\pm0.2$ (taking into account the redshift uncertainty) - this could be consistent with a bright SN, a faint SLSN, or a TDE/ANT. It then slowly decayed over the course of 2024, with a typical timescale of $91.1\pm7.8$ days. There was no color change in the decay, which would rather be indicative of a TDE. There is no clear infrared counterpart to the optical transient, beyond the aforementioned slow rise. The timescale of this transient thus tends to exclude SNe; the color evolution favors a TDE/ANT rather than a SLSN.

However, around the beginning of 2025, the transient started to rebrighten, showing a significant excess compared to the expected $\propto t^{-5/3}$ decay (see Fig. \ref{fig:AT2024gzn}). The main interpretations of the behaviour of AT2024gzn are a low-accretion AGN that underwent an episode of increased accretion, or a peculiar TDE/ANT with a bumpy decay. The former is the most likely interpretation. Additional monitoring of the source is needed to confirm either hypothesis.

\subsection{Flares in uncertain hosts}
\label{subsec:uncertain}

\subsubsection{\href{https://ztf.fink-portal.org/ZTF23abowyjf}{AT2023zaj}}
The host of AT2023zaj is WISEA J003734.48-133352.0, a relatively faint galaxy with a photometric redshift of $z=0.4\pm0.1$. Its NeoWISE lightcurve showed signs of low-amplitude variability, it is detected but faint in GALEX, and its infrared color is $W1-W2\sim0.8$, meaning that it might be an active galaxy, but we cannot definitely confirm it. This uncertainty is further confirmed by the Gaia classification, which gives $P_{gal}=0.32$ and $P_{QSO}=0.2$. 

The source brightened in late 2023 up to magnitudes 19, corresponding to absolute magnitudes between -21.9 and -23.2, depending on the redshift value. Its rise was relatively fast, with a typical timescale of $23.1\pm2.6$ days, and the apparent color evolved from $g-r\sim0.6$ to $g-r\sim0$. It then decayed very slowly, following a timescale of $887.5\pm137.0$ days. It showed a clear infrared echo to the transient, with a strong brightening of the $W1$ band by 1.5 magnitudes, and of the $W2$ band by 1 magnitude.

The brightness of the objects means that it is either a SLSN or a TDE/ANT candidate (depending on the actual underlying activity of the host). However, the decay timescale is very long, way too long for standard SLSNe, and as such this source is a good candidate ANT.

\subsubsection{\href{https://ztf.fink-portal.org/ZTF23aajnfna}{AT2023szj}}
The host of AT2023szj is SDSS J144628.14+084136.6, a faint red galaxy with a photometric redshift of $z=0.57 \pm 0.16$. It is undetected in UV and radio, has an infrared color of $W1-W2=0.67$, so it could be passive - however, its historical ZTF and NeoWISE lightcurves showed hints of low-amplitude variability, hinting at nuclear activity. As such, it is difficult to confirm on the AGN nature of the host -- if it is an AGN, its activity level has been relatively low. 

In 2023, it underwent a very slow brightening, over a timescale of $162.3\pm11.5$ days, peaking at difference magnitudes of 19.5, corresponding to absolute magnitudes of $-23.0\pm0.75$, depending on the redshift estimate. It also became bluer, with the apparent color shifting from $g-r\sim0.25$ to $g-r\sim-0.2$ at the peak. It then decayed over a long timescale as well, $375.4\pm50.6$ days, with no signs of color change. Its NeoWISE lightcurve showed a clear echo, both the $W1$ and $W2$ brightening by one magnitude. 

\begin{figure*}
    \centering
    \includegraphics[width=\textwidth]{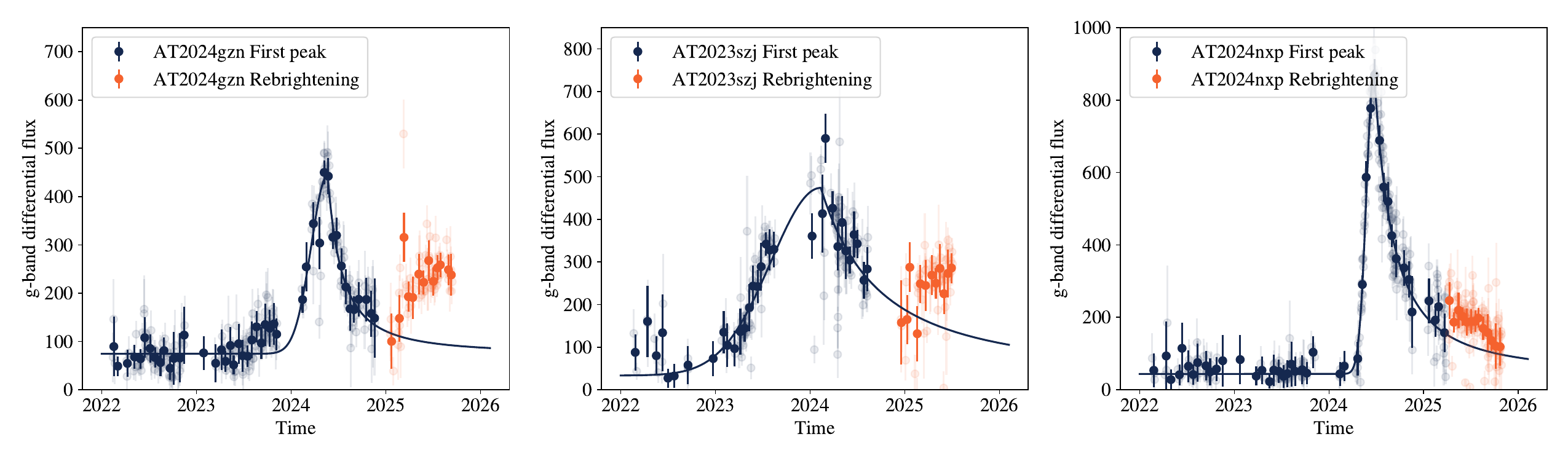}
    \caption{Evidence for late-time features in AT2024gzn (left), AT2023szj (centre), and AT2024nxp (right). The line displays the best fit phenomenological $t^{-5/3}$ profile to the $g$-band lightcurve, with the fitting performed on the blue data points (before rebrightening, displayed in orange). Transparent points are unbinned data, opaque points are binned in bins of 15 days.}
    \label{fig:AT2024gzn}
\end{figure*}

While the amplitude is consistent with either a SLSN, a TDE, or an ANT, the very long timescales are rather indicative of either an extreme TDE or of an ANT. In 2025, during its decay the source underwent a rebrightening, with a clear excess to the rough $\propto t^{-5/3}$ expected decay (see Fig. \ref{fig:AT2024gzn}). There are two main interpretations of this behaviour. First, and most likely, the past slight optical and infrared variability were indicative of a low-accretion AGN, and this outburst is a sudden episode of higher accretion -- or, this was a long-duration TDE/ANT with a complex decaying lightcurve \citep[as was seen for instance in AT2022exr, e.g.][]{langis_repeating_2025}. Further monitoring of this source's behaviour will allow to discriminate between these interpretations.

\subsection{Physical parameters analysis}
\label{subsec:parameter}

\begin{figure}
    \centering
    \includegraphics[width=1\linewidth]{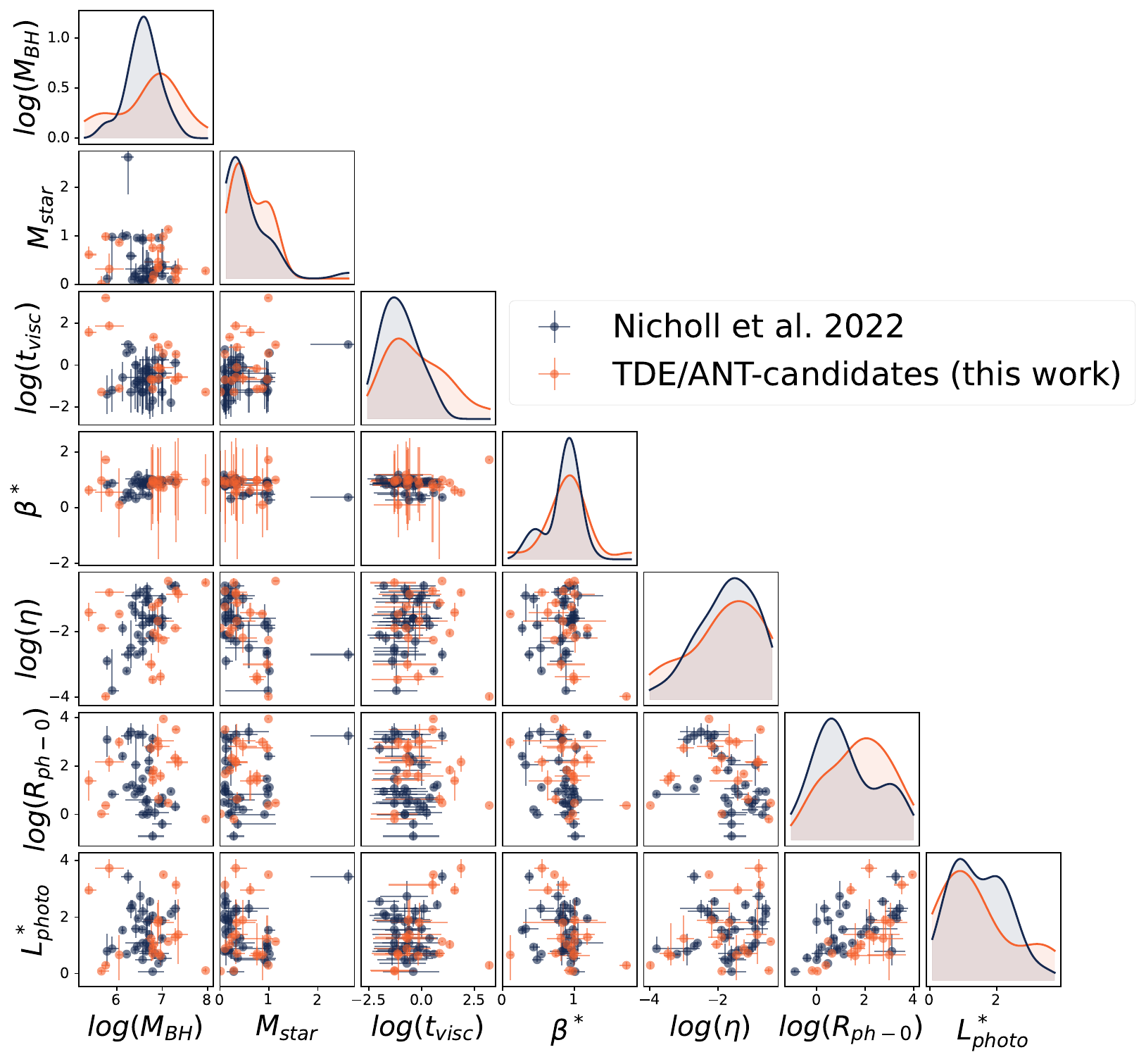}
    \caption{Distribution of the posterior parameters from the \textit{tde\_fallback} \textsc{Redback} model. The values correspond to the median of the posterior distribution. Upper and lower error bars indicate the 84th and 16th percentiles. The orange points are events studied in this work, and blue points are TDE from a similar analysis \citep{nicholl_systematic_2022} performed with the same \textit{tde\_fallback} model but using MOSFiT. Exact values are available in Table \ref{tab:Redback_values}. Because the fit failed to converge, AT2020actc and AT2023rav are missing from this sample.}
    \label{fig:Redback_distrib}
\end{figure}

The results from the \textsc{Redback} bayesian fits are displayed in Fig.~\ref{fig:Redback_distrib}. They are useful to constrain the physical parameters of each source, assuming the model is suited to describe the events. In particular, the fit of AT2020ukj reveals that its exceptionally slow decay can occur in a TDE scenario given that: a massive star is disrupted, the time delay due to viscosity is high, and the efficiency of the emission is very low. We also compare our global distribution with the large sample presented by \citet{nicholl_systematic_2022}. It consists of 32 confirmed TDEs, that were modeled using the same physical model and methodology as this work but using the MOSFiT software \citep{guillochon_mosfit_2018}. We find that the two distributions are highly consistent with each other, supporting the conclusion that our classifier has identified a sample of objects \emi{hosting similar events.} 

We also attempted to distinguish between the TDEs and ANTs within our sample, but did not observe any significant differences between the two distributions. However, given the limited sample size and the unquantified selection biases inherent to our analysis, further investigations will be necessary to better understand their statistical behaviors, by incorporating a larger number of systematically selected TDE and ANT events.

\subsection{Supernova candidates}
\label{subsec:sn_candidates}



We identified a series of SN candidates while searching for TDEs. Here, we considered as such any lightcurve which visually resembles the expected behavior of a SN (consistent burst with relatively fast rise and different rates of decay). Following \citet{pruzhinskaya2023}, selected sources were fitted \emi{with SN templates provided by Peter Nugent\footnote{\url{https://c3.lbl.gov/nugent/nugent_templates.html}\label{foot:sn_models}} and the SALT2 model} (the procedure is described in Appendix \ref{app:sn}). It was not possible to perform a good fit for most of the candidates since many of them had \emi{insufficient number of photometric points}. From an initial set of 40 objects, 11 were compatible with Type Ia supernova \citep[SALT2,][]{guy2007}, 8 were better modeled as Type~IIP and 7 as Type IIn. We also identified two long-lasting lightcurves which were not compatible with any of the available SN models and which can be better explained as superluminous supernova candidates. We detail the most interesting SN cases below.

\subsubsection{SLSN candidate: \href{https://ztf.fink-portal.org/ZTF21aazrgtw}{AT2021lnu}}
The host of AT2021lnu is SDSS J153218.78+000000.0, a faint irregular dwarf galaxy at a spectroscopic redshift of $z=0.15\pm0.07$. In 2021, it underwent a fast brightening over a month, up to difference magnitudes of 18.75, corresponding to absolute magnitudes of $-20.2\pm1.2$, depending on the redshift estimate. It stayed in a plateau phase about a month, and then decayed over about a month, with no strong sign of cooling. It finally showed a secondary, lower-amplitude peak, with redder colors.

The shape, timescale and amplitude of this outburst all point towards the SLSN interpretation. In particular, its lightcurve shows strong similarities to that of another known SLSN, SN2024afav.



\subsubsection{SLSN candidate: \href{https://ztf.fink-portal.org/ZTF20aazyvre}{AT2020mvg}}

The host of AT2020mvg is SDSS J120526.52+285434.6, a faint galaxy at a photometric redshift of $z=0.31\pm0.11$. It is blue, UV-bright, and its CRTS lightcurve showed hints of decade-long low-amplitude variability. The NeoWISE data is dominated by the nearby (5") source, WISEA J120526.74+285429.8 - as such, its own infrared emission is difficult to assess. Overall, this source is consistent with being active, although no definitive evidence is available.

In 2020, it underwent a significant relatively fast outburst, with a rise time of $12.9\pm0.5$ days. It reached difference magnitudes of $18.5\pm0.05$ at its peak, corresponding to absolute magnitudes $-22.4\pm0.9$ depending on the redshift value. The transient was relatively blue (difference color of $g-r\sim-0.3$ at the peak). It then decayed following a $31.5\pm2.4$ days timescale, with signs of cooling in the decay.

The absolute magnitude, timescale, and color evolution all point towards a good SLSN candidate.

\subsubsection{Confirmed SN Ia: \href{https://ztf.fink-portal.org/ZTF20aacedmi}{AT2024ljd}}
The host of AT2024ljd is SDSS J211339.40+022937.3, at a redshift of $z=0.046$. In 2024, its nucleus showed a fast outburst, rising over 15 days and decaying over a month. It reached difference magnitudes of $17.5\pm0.05$, corresponding to absolute magnitude of $19.0\pm0.05$. There is evidence of cooling during the decay, manifest in the color evolution. 

We detected this object initially at an early stage of development of our TDE pipeline. We obtained a spectral follow-up by the CMO-2.5m telescope \citep{2020gbar.conf..127S}, allowing to classify AT2024ljd as a SN~Ia \citep{quintin_type_2024}.

\subsubsection{Supernova siblings: \href{https://ztf.fink-portal.org/ZTF19aafmytc}{AT2019agc}}
\label{sec:AT2019agc}
The host of AT2019agc is SDSS J140645.34+135505.1, a passive galaxy at a photometric redshift $z=0.15\pm0.05$. It showed a brightening by about 1.5 magnitudes in both ZTF bands in early 2019, and a second one in June 2021. Both peak at around absolute magnitude of $-19.7\pm0.7$ (taking into account redshift uncertainties) and cool down during the decay. The two bursts are associated to the same ZTF / TNS object.

\begin{figure}
    \centering
    \includegraphics[width=\columnwidth]{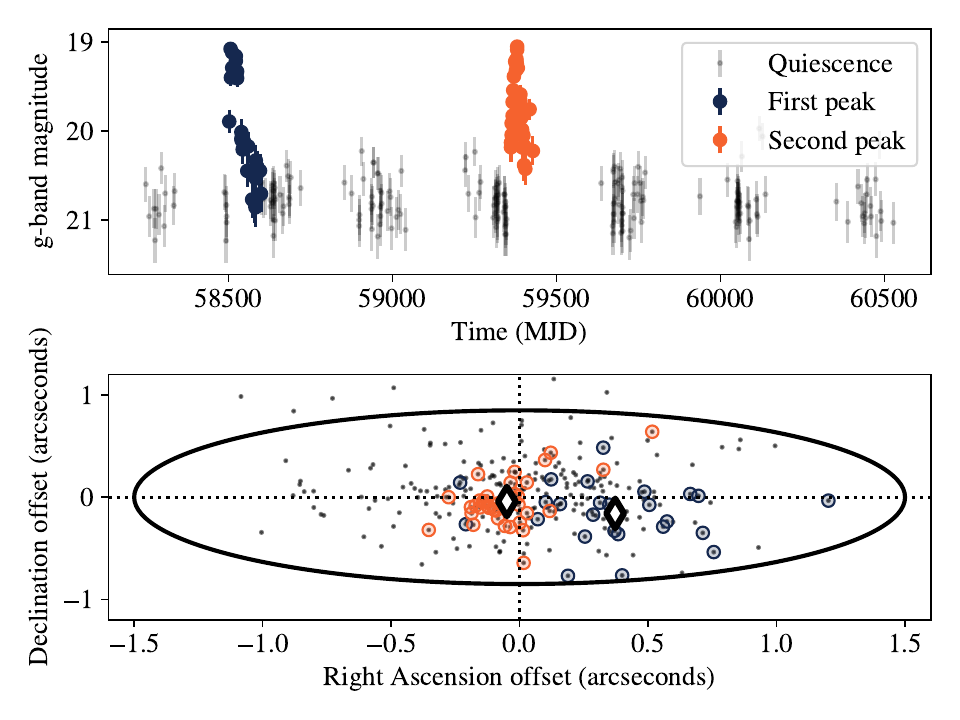}
    \caption{Comparison of the position of the two bursts associated with AT2019agc, with color-codes corresponding to a temporal clustering of the bursts. \textit{Top panel:} Apparent magnitudes in the $g$ band of the two transients (blue and orange), with black dots for the quiescent emission from the galaxy. \textit{Bottom panel:} positional offset of the corresponding detections from the center of the host galaxy, with the same color code (dark blue for the first burst, orange for the second). The black dotted lines represent the center of the host galaxy, and the black ellipse corresponds to its approximate optical extent. The two black diamonds show the average position of the first (blue) peak and the second (orange) peak. This shows that the transients do not originate from the same position, and neither exactly from the galactic center.}
    \label{fig:AT2019agc}
\end{figure}

Interestingly, by retrieving the positions of each individual detection from the ZTF DR23 data, and plotting them compared to the visual extent of the galaxy, one can see that these objects come from distinct regions (see the $g$-band points in Fig. \ref{fig:AT2019agc}, where the blue and orange points corresponding to the two peaks display a clear offset). A double-sided KS-test on the right ascension and declination distributions of both bursts lead to $p$-values of $8.3\times10^{-10}$ and $2.0\times10^{-1}$ respectively for the g-band, and $3.2\times10^{-6}$ and $3.7\times10^{-4}$ for the r-band. These $p$-values in both bands (especially for right ascension) allow to safely ($>6\sigma$) exclude the possibility that the two bursts come from the same position. This is further confirmed by the fact that, after averaging the positions of the alerts during the first and then the second peak, the offset between these average positions is about 0.4", which is significantly larger than the typical position accuracy of ZTF \citep[45 to 85 milliarcsec,][]{masci_zwicky_2019}. \emi{It is thus unlikely that they are} physically related. However, they are both associated to the same ZTF/TNS object, ZTF19aafmytc/AT2019agc. This highlights one limitation of the current position-based identification scheme for transients. In the current paradigm, the association of a new alert with an existing transient is done individually for each new alert -- the individual position error of each alert is quite large, and it would be difficult to separate two spatially close but separate transients this way\footnote{A simple visual representation would be trying to find two populations in the bottom panel of Fig. \ref{fig:AT2019agc} if the points were not color-coded.}. However, for this specific object the clearly distinct bursts allowed a temporal clustering, that revealed the two underlying populations in the individual alerts. 


Comparing the two bursts individually with the \emi{ \textsc{sncosmo}  built-in supernova models}, both were consistent with Type Ia supernova. The 2021 event is in clear agreement with SALT2 model ($x_1 = 0.98 \pm 0.33$, $c = 0.094 \pm 0.033$, $z = 0.149 \pm 0.026$), while the lightcurve format of the event from 2019 does not allow us to reliably discard other supernova types. The study of supernovae siblings (occuring in the same galaxy), has been advocated  in the literature as a possible path to further investigate the influence of local environment in supernova properties. A study of spectroscopically confirmed siblings in ZTF \citep{graham2022} reported 5 such pairs, highlighting the importance and difficulty in identifying such events. Another study using an active anomaly detection algorithm discovered an additional SN sibling, SN2018fcg \citep{majumder_superluminous_2024}. A recent work, focusing exclusively in Type Ias for optimizing standardization, found 25 such pairs \citep{dhawan2024}. The serendipitous discovery of this pair in our analysis points to the potential of ZTF, as well as other large scale sky surveys, in also enabling purely photometric siblings studies, as it was already performed in the Dark Energy Survey \citep{scolnic2020}. As a final note, we cannot definitely exclude the possibility of contamination by a background galaxy (e.g. two distinct galaxies as hosts for the two peaks, aligned along the line of sight and reasonably close to each other), although the available data shows no clear sign of such a background contribution. This situation would not change our conclusion that the two peaks are physically independent.

\section{Discussion}
\label{sec:discussion}
\subsection{Rate considerations}
\label{subsec:rate}
The selections biases of our study are uncertain, as the sample has been built inhomogeneously, not only through the systematic application of the early TDE detection module to the part of the ZTF archive that was not classified in TNS, but also through serendipitous discoveries during the developement of said module. As such, converting the number of TDE or ANT candidates to physical rates \citep[as was done for instance in][]{yao_tidal_2023} is not directly possible. What can be done however is to compare these numbers to those of other studies of a roughly similar sample. 

For instance, \citet{yao_tidal_2023} presents a sample of 33 spectroscopically-confirmed optical TDEs, obtained by studying 3 years worth of ZTF data. Our sample covers roughly 5 years worth of ZTF data, and presents 9 TDE candidates and 7 ANT candidates. These studies lead to a respective yearly detection rate of 11 TDEs per year compared to 2.2 in our case (or 3.2 if one includes ANT candidates). This indicates that a small, but not insignificant, number of nuclear transients might still be lying in the archives, and justifies future work to ensure an exhaustive search. In particular, as will be discussed in the later points in more details, such exhaustive searches will require the community to be careful about the assumptions put into the classifiers (e.g. hard cut-off values for some parameters such as the duration or the amplitude, or excluding any host that has sign of prior nuclear activity).  

\subsection{Insights on repeated nuclear transients}
\label{subse:repeated}
We presented here two newly-revealed repeated nuclear optical transients, one in a passive galaxy (AT2023adr) and one in an active galaxy (ZTF23abjvojy). Additionally, two other objects show hints of late-time rebrightenings during their decay after the initial burst (AT2023szj and AT2024gzn), and AT2024nxp could be a third candidate (although with a much lower amplitude). Conservatively keeping only the first two secure repeated transients, this amounts to 10\% of the sample presented in this paper. This is a good opportunity to assess the current state of our knowledge of repeated optical nuclear transients, and what conclusions we might draw from it.

We present in Tab. \ref{tab:repeated} an up-to-date inventory of the currently known repeated optical nuclear transients. This table was inspired from Tab. 2 of \citet{lin_unluckiest_2024}, but we updated it and also added sources that show a strong rebrightening in their decay -- for these, it is unclear whether they are repeated or simply show a more complex behaviour than the usual monotonous decay. We also limit this sample to optical transients, otherwise for instance X-ray repeated nuclear transients would include numerous other sources, such as all quasi-periodic eruptions sources \citep[e.g.][]{miniutti_nine-hour_2019}. We include information about the host (whether it contains an AGN or not, i.e. whether the transient is a TDE or an ANT candidate), and information about the flaring pattern (number of flares and approximate interval between them, and the amplitude evolution of the flares). Inspired by a recent discussion about the repeated TDE AT2022dbl at a conference\footnote{See "Predicting the future" in \href{https://www.cosmos.esa.int/web/x-ray-quasi-periodic-eruptions/programme}{https://www.cosmos.esa.int/web/x-ray-quasi-periodic-eruptions/programme}}, we add a column corresponding to the estimated time of the anticipated next peak, assuming the interval between peaks is constant. While some are in the future, some anticipated third peaks were supposed to have happened in the past. For these, we indicate whether the ZTF photometry covered the expected epoch of the peak or if it fell in a gap, and in the former case whether the lightcurve showed a clear rebrightening or not. Several conclusions can be drawn from this table only.

First, in the last few years the existing sample has grown noticeably, reaching a significant fraction of the total TDE sample. \emi{Considering} only events with clearly separated peaks, and excluding OJ 287 as it is not TDE-related \citep{sillanpaa_oj_1988}, we are left with 16 sources out of the 25 in the table. These 16 sources amounts to about 10\% of the total sample of known optical TDEs \citep[e.g. 134 confirmed optical TDEs in][]{langis_repeating_2025}. This fraction is comparable to the one obtained from our sample only, with two clearly repeated events out of 20 nuclear transients. While the poorly constrained biases of our selection process prevent us from drawing any clear conclusion as to the precise rate of repeated TDEs, it appears clear that they represent a non-negligible \emi{portion of} the total population, and should be accounted for in both the rate estimates and the physical models of (partial) disruption \citep[e.g. see considerations in][]{makrygianni_double_2025}. This fraction is also bound to increase over time, as the probed temporal baseline increases linearly (e.g. about 20 years now separate the earliest CRTS data and the latest ZTF data). It is however important to note that peaks separated by decades could reasonably arise from independent TDEs instead of repeated partial TDEs of the same initial star,  notably in the presence of a locally increase TDE rate \citep[e.g. for a particularly dense nuclear star cluster, see for instance][]{stone_rates_2016}. One can also see that among the clearly separated events, there is no strong pattern in the relative amplitudes of the peaks (out of the 15 repeated events, 6 have comparable luminosities between peaks, 7 have a smaller second peak, and 2 have a higher second peak). In order to continue building up this sample, we strongly encourage the community to keep an up-to-date inventory of repeated nuclear transients, similar to that presented in Tab. \ref{tab:repeated}. 

Second, a natural approach to confirm the repeated partial TDE nature of a repeated event (as opposed to two independent TDEs) is to assess the presence of a third peak, which would strenghten the former interpretation. This in turn requires to anticipate an observational strategy, and thus the expected timing of the supposed third peak needs to be estimated. The simplest way to do this is to assume that the time between peaks is not modified by the partial disruptions, and extrapolate the date of the third peak from the date of the second peak and the delay between the first and second. This corresponds to the \textit{"Next" Peak} column in Tab. \ref{tab:repeated}. For recent events, this date is in the future, and could be used to design observational strategies. However, one can also look back at objects for which this third peak should have happened in the past. For some, this third peak happened during observational gaps (e.g. AT2023adr), or a faint hint of a peak could perhaps be detected (e.g. AT2019ehz). But for others, a third peak was not detected at all (e.g. AT2019avd). In fact, as of now no known optical TDE has three strong separated peaks in its lightcurve, and only ASASSN-14ko and AT2023uqm show multiple TDE-like peaks. It is unclear whether this means that the second disruption is always total, or that the delay between peaks is modified by the partial disruptions, or even that these events were possibly two independent TDEs -- it does warn us though against using this simplified method to anticipate observational strategies.  

Third, the impostor repeated event AT2019agc sheds light on some limitations of the current transient identification system. Indeed, while AT2019agc is identified as a singular object showing a rebrightening in both TNS and ZTF catalogs, the study of the position of individual detections we have shown in Sec. \ref{sec:AT2019agc}, and in Fig. \ref{fig:AT2019agc} in particular, allows to conclude that the rebrightening \emi{is likely to have happened in a} distinct location within the galaxy, and as such the two peaks are actually physically independent. More generally, while it could be impossible to distinguish the positions of two independent supernovae arising from the nuclear cluster, objects like AT2019agc where one of the peaks lies in the periphery of the galaxy are bound to happen, and their numbers will increase with the new upcoming surveys and the increasing temporal baseline. This situation indicates that proper systematic checks for positional overlaps should be performed in the case of repeated transients. This is technically different from the standard automatic cross-match of alerts that leads to the existing transient identification, which is performed on a detection-by-detection basis. What we suggest is to (ideally automatically) perform a temporal clustering of the alerts of similar \emi{ events} 
(repeated transients that are clearly separated in time), and assess whether the alerts from these temporally distinct peaks are spatially consistent or not (e.g. through a simple KS-test on the right ascensions and declinations of the alerts, as we have done here). This should, at a relatively low computational cost, provide an effective solution to filter out most of the impostors repeated events.




\begin{table*}
    \centering
    \resizebox{\textwidth}{!}{
    \begin{tabular}{ccccccc}
    \hline\hline \\[-0.1cm]
        Name & Host & Approx. Interval (d.) & Flares & Peak Evolution &"Next" Peak & Reference\\[0.2cm] \hline\hline \\[-0.1cm]
        OJ 287 &Active&4000&$\sim$10&Similar&$\sim$2030&\citet{sillanpaa_oj_1988}\\
        IC 3599 & Active & $\sim$3470?& 2/3 &Similar&2019 (No)&\citet{campana_multiple_2015}\\
        ASASSN-14ko  & Active & 115.2 &$\sim$30 &Similar&Aug. 2025&\citet{payne_asassn-14ko_2021}\\
        AT 2019avd & Passive & $\sim$500& 2& Lower&May 2022 (No)&\citet{malyali_at_2021}\\
        AT 2019azh & Passive&$\sim$4800&2&Similar&2032 &\citet{hinkle_discovery_2021}\\
        AT 2018fyk & Passive &$\sim$1200 &2 &Lower&Apr. 2025 (No)&\citet{wevers_rebrightening_2022}\\
        $^\dagger$AT 2019baf &Passive&250&2&Lower&Jul. 2020 (No)& \citet{yao_tidal_2023}\\
        AT 2019ehz&Passive&$\sim$300&2&Lower&Dec. 2020 (Hints)&\citet{yao_tidal_2023}\\
        $^\dagger$AT 2020acka&Passive&300&2&Lower&Aug. 2022 (No)&\citet{yao_tidal_2023}\\
        $^\dagger$AT 2021uqv&Passive&225&2&Lower&Jan 2023 (Hints)&\citet{yao_tidal_2023}\\
        AT 2020vdq & Passive & $\sim$870& 2& Higher&Oct. 2025&\citet{somalwar_first_2023}\\
        $^\dagger$AT 2021loi & Active &$\sim$400 &2&Lower&Aug. 2023 (No)&\citet{makrygianni_at_2023}\\
        IRAS F01004-2237 & Passive &$\sim$4175  &2&Similar&2032&\citet{makrygianni_at_2023,sun_recurring_2024}\\
        AT 2022dbl & Passive&  $\sim$710& 2 &Lower&Jan. 2026&\citet{lin_unluckiest_2024}\\
        AT 2019aalc &Active&1500&2&Higher&Aug. 2028&\citet{milan_veres_back_2024}\\
        AT 2024pvu&Passive&$\sim$6500&2&Similar&2042&\citet{langis_repeating_2025}\\
        $^\dagger$AT 2022exr&Passive&120-50&2/3&Lower&Nov. 2022 (Gap)&\citet{langis_repeating_2025}\\
        $^\dagger$AT 2021uvz&Passive&175&2&Lower&Dec. 2021 (Gap)&\citet{langis_repeating_2025}\\
        AT 2019teq&Passive&375&2&Similar&Nov. 2020 (No)&\citet{langis_repeating_2025}\\
        AT 2021aeuk &Active&1300&2/3&Lower (+precursor)&Oct. 2026&\citet{sun_at2021aeuk_2025}\\
        AT 2023uqm & Passive & 525 & 5 & Higher & Aug. 2026& \citet{wang_stars_2025}\\ 
        $^\dagger$AT 2024gzn&Active&400&2&Lower&Sep. 2026 &This work\\
        $^\dagger$AT 2023szj&Active&600&2&Lower&Feb. 2027&This work\\
        ZTF23abjvojy&Active&$\sim5785$&2&Lower&2039&This work\\
        AT 2023adr&Passive&400&2&Lower&Apr. 2025 (Gap)&This work\\
        
    \end{tabular}
    }
    \caption{Summary of the currently known optical repeated nuclear transient events, sorted by date of reporting. Similar to Tab. 2 from \citet{lin_unluckiest_2024}, simplified and updated. Note this only includes optically repeated objects, otherwise X-ray repeated flares would include numerous other objects, including repeated X-ray TDEs but also quasi-periodic eruption sources. Sources marked with the $\dagger$ symbol correspond to a lower amplitude rebrightening during the decay of the first peak, while for the others the peaks are clearly separated in time. For completeness, we included OJ 287, which is interpreted not as a TDE but a binary blazar with periodic increases of accretion rates due to interactions between the disks. We did not include AT 2021mhg \citep{somalwar_first_2023}, for which the second peak was photometrically identified as likely SN due to its timescale and red color, or AT 2024nxp, as the amplitude of the second peak is so low that it could be a simple flattening of the decay. The 'Next Peak' column corresponds to an estimation of the time of the next peak, assuming the delay between peaks is the same as the last delay. For values that are in the past, we indicated whether this additional peak fell in ZTF observational gaps, whether some hints of such a new peak were seen, or whether it was observed and not seen. }
    \label{tab:repeated}
\end{table*}

\subsection{Insights on TDE-like flares in AGNs}
\label{subsec:agn}
Our selection of TDE-like flares in ZTF data lead to a number of transients hosted in previously active galaxies - we classify them as candidates ANTs. We kept only those where the flare itself is clearly in excess to the previous accretion-driven variability witnessed in the host (see their lightcurves in Fig. \ref{fig:MultiL_TDEs1} and Fig. \ref{fig:MultiL_TDEs2}). Discriminating between fluctuations of accretion in a standard AGN and the flaring pattern of an ANT from photometric data only is a non-trivial problem in general.
The selection in our study is mostly qualitative, but one could reasonably design a more quantitative approach \citep[e.g. as was done in][]{wiseman_systematically_2025}. For instance, the temporal evolution of photometric properties of standard AGNs is well described by stochastic processes, in particular Damped Random Walks (DRWs). By estimating the parameters of the process in the historical lightcurve, one can then quantify the deviation from this process observed in the flare. Another alternative would be to use anomaly detection methods \citep[e.g.][]{sanchez-saez_searching_2021}. Such methods will need to be refined and applied to the upcoming LSST data streams, in order to increase the existing sample of ANTs and e.g. perform population studies. Interestingly, in our sample the number of ANT candidates is very comparable to that of TDE candidates, indicative of the fact that these events might not be exceedingly rare. \emi{However,}  our inhomogeneous selection biases prevent us from definitive rate estimates.


AT2020ukj, along with Ansky, hint at another possibly new class of nuclear transients. These two events show a previously passive galaxy display a TDE-like flare, but on much longer timescales. The main possible interpretations are that, similar to Ansky, AT2020ukj is either a turn-on AGN, a tidal disruption in an AGN with a very low angle, or the tidal disruption of a massive star or massive cloud of gas. We invite further study and monitoring of this object. More generally, the existence of these long-lived objects indicates that we should be careful about including hard cut-offs on the timing properties of the bursts into TDE classifiers, which would lead to rejecting this sort of very long events. In a similar manner, the faintness of AT2023npi despite its long duration indicates that the parameter spaces of different types of optical transients overlap significantly, and hard cut-offs should be avoided if possible, at risk of losing these physically interesting outliers.

Two AGN-related transients in our sample are narrow-line Seyfert 1 sources (ZTF23abjvojy and AT2024hhj). This class has been linked to flaring events in the past \citep{frederick_family_2021}. While our unquantified selection biases prevent us from definitely confirming whether they are over-represented in our variability-selected sample, it is still of note that this class is clearly present. A pre-existing catalog of NLSy1 could for instance be used as a tool to prioritize early flare detection alerts, in order to organise monitoring campagins of a flaring event.

Finally, AT2023szj and AT2024gzn are both nuclear transients with clear signs of a late-time rebrightening. AT2024gzn is a spectroscopically confirmed QSO, while AT2023szj could be an AGN but the diagnostics methods do not lead to a definitive answer. A likely interpretation of both these events that would account for their observed properties are that they are turn-on AGNs, with very low activity in the past that suddenly increased, leading to the optical burst. It is reasonable to assume that future optical transient surveys will detect more of these objects, which are contaminants for TDE classifiers but could still provide very useful physical insights about the onset of accretion in AGNs. 

\subsection{Potential of multi-wavelength synergy for transient detection} 
\label{subsec:synergy}

The main purpose of the real-time early TDE detection tool developed by the \fink\  collaboration \citep{llamas_lanza_early_2025} is to provide early warning on optical nuclear transients matching TDE-expected properties, for the ZTF data stream and for upcoming LSST. The sample presented here is archival in nature, and was obtained as a secondary result of the development of this pipeline. As most of these transients were already over when we found them, it was not possible to obtain spectroscopic follow-ups to confirm their nature. For the few objects that are still active, we requested such follow-ups whenever possible, being often limited by their relative faintness. This however highlights a general difficulty with photometric transient detection, which is the availability of timely spectroscopic follow-ups for classification. Despite significant efforts in upcoming spectroscopic survey capabilities \citep[e.g. the TiDES survey on the 4MOST spectrometer][]{frohmaier_tides_2025}, this issue will be prevalent among LSST transients, due both to the large number of transients and their faintness compared e.g. to ZTF transients, putting them out of reach of the more modest spectrometers. 

One solution to help classification in the absence of spectroscopic data is to inform photometric classification methods by using multi-wavelength data. In particular, using variability information and transient alerts from multiple other wavelengths can help discriminate between different classifications. As such, multi-wavelength synergies of transient-detection systems should be encouraged. There are numerous existing transient detection projects in almost all wavelengths: in X-rays \citep[e.g.][]{evans_real-time_2023, quintin_stonks_2024}, ultraviolet \citep[e.g.][]{modiano_tuvopipe_2022}, radio \citep[e.g.][]{murphy_vast_2013}, infrared \citep[e.g.][]{rose_hourglass_2025}, and even other messengers, with neutrinos \citep[e.g.][]{adrian-martinez_letter_2016} and of course gravitational waves \citep[e.g.][]{abbott_open_2023}. Some efforts have been made to centralize alerts from all those channels \citep[e.g.][]{reichherzer_astro-colibri-coincidence_2021}. In the case of nuclear transients, comparing for instance the X-ray, radio and infrared variability to the optical one could in some situations lift uncertainties (e.g. the presence of quiescent X-ray emission, that could perhaps present a dip during the optical burst, is indicative of a pre-existing AGN rather than a simple TDE).

\section{Conclusions}
\label{sec:conclusions}
In this paper, we presented a sample of optical nuclear transient events that were found during the development of an early TDE detection system for ZTF data, as part of the \fink\ broker. We report on the discovery of 19 optical nuclear transients, among which 9 in previously passive galaxies:
\begin{itemize}
    \item \textbf{AT2020ukj} is a long-lived transient that has been active for the last 5 years, with a constant blue color, and shows strong similarities to ZTF19acnskyy \citep["Ansky",][]{sanchez-saez_sdss13350728_2024} - it could be an exotic TDE with extreme timescales, or an AGN in the process of turning on;
    \item \textbf{AT2023adr} is a TDE candidate that displayed a clear second peak about 400 days after the first one, with lower amplitude but an optical spectrum consistent with that of a TDE;
    \item \textbf{AT2023npi} is a TDE candidate in a nearby galaxy, and would be the faintest optical TDE to date, despite its very strong infrared echo;
    \item \textbf{AT2020pno} is a TDE candidate that displays a clear optical plateau phase in its decay;
    \item \textbf{AT2020aexc}, \textbf{AT2020afap}, \textbf{AT2023jag}, \textbf{AT2021ovg}, \textbf{ZTF20aaptzog} are all TDE candidates, with no particular distinctive features 
\end{itemize}

We also presented 8 in previously active galaxies:
\begin{itemize}
    \item \textbf{ZTF23abjvojy} is a blue flare in a confirmed NLSy1 that had already displayed a similar behaviour almost 16 years before, witnessed at the time by CRTS. It is the repeated ANT with the longest timeline among the limited sample of such objects;
    \item \textbf{AT2022yhf}, \textbf{AT2024nxp}, and \textbf{AT2023rav} are bright transients in previously active galaxies, that reach absolute magnitudes above -24, classifying them as likely ENTs. AT2024nxp shows hints of late-time rebrightening (or at least flattening of the decay), and AT2023rav shows a lower quiescent flux after the transient, indicative of a possible depletion of the initial AGN-like accretion flow;
    \item \textbf{AT2020actc}, \textbf{AT2021wxd}, and \textbf{AT2024hhj} are all ANT candidates with no particular distinctive features, apart from the latter one being a NLSy1.
    \item \textbf{AT2024gzn} is a flare in a confirmed QSO, with a late-time rebrightening.
\end{itemize}

We presented 2 transients in galaxies where we were not able to confirm or infirm previous nuclear activity:
\begin{itemize}
    \item \textbf{AT2023szj} is a bright flare that displays clear late-time rebrightening, which could be turn-on AGN;
    \item \textbf{AT2023zaj} is a slow blue transient with no particular distinctive features
\end{itemize}

We also have a significant number of supernovae candidates, among which:
\begin{itemize}
    \item \textbf{AT2024ljd} was spectrally confirmed as SN Ia \citep{quintin_type_2024};
    \item \textbf{AT2019agc} is a seemingly repeated optical nuclear transients, but further inspection revealed that it is consistent with two independent supernovae in the same galaxy, two and a half years apart;
    \item \textbf{AT2021lnu} and \textbf{AT2020mvg} are both SLSN candidates;
\end{itemize}

We encourage monitoring of these objects, and spectroscopic confirmation of their redshift when it is only photometric. 

All these objects, and the peculiar features some of them display, lead us to a certain number of considerations for future optical transient detection systems, especially with the perspective of upcoming LSST data :
\begin{itemize}
    \item A significant fraction of the optical nuclear transients we present here (about half of them) happened in a galaxy that showed signs of previous nuclear activity. Future transient detection systems should devise clear methods regarding these nuclear transients in active galaxies (ANTs and ENTs), in particular to discriminate between a simple AGN flare and an exceptional brightening. 
    \item The discovery of AT2020ukj, a long-lived transient in a previously passive host that could be a turn-on AGN, indicates that ZTF19acnskyy was not the only instance of its class, and upcoming classifiers should keep it in mind (e.g. by not discarding nuclear transients with timescales larger than expected);
    \item A fraction of the presented nuclear transients are repeated objects (about 10\% of them). Even though our selection is biased and as such it is difficult to estimate exact rates, it is clear that repeated events are not exceptionally rare, and as such classifiers should anticipate this situation. This means for instance monitoring previous events, or not excluding events with past variability in TDE classifiers;
    \item We attempted to create an inventory of the current sample of optical repeated nuclear transients (see Tab. \ref{tab:repeated}). We invite the community to keep it up to date. In particular, this allowed us to assess that one should not use the timing of a two-peaks event to anticipate the date of an hypothetical third peak;
    \item The discovery of AT2019agc, an impostor repeated event, indicates that we should be mindful of the possibility of independent events being associated with a single object. A systematic check of the consistency of positions between flares could be one way to tackle this issue.
\end{itemize}


\emi{In summary, these results highlight the scientific potential still enclosed in already available data, and demonstrate how the sheer volume of observations being generated by large scale surveys impose a completely different approach to data mining in astronomy. There has been extensive discussion on both sides of this topic: how this change of paradigm impacts research goals focused on well established classes \citep[e.g. SNIa,][]{ishida2019}, and how it opens the possibility to completely unexpected discoveries using unsupervised anomaly detection \citep[e.g. ][]{volnova2024}.} 

In this work, we have presented several examples of optical nuclear transients that appear as outliers from the typically expected behaviours (e.g. the long duration of AT2020ukj, the low amplitude of AT2023npi, or the repeated natures of AT2023adr and ZTF23abjvojy). Their peculiar properties make them valuable to constrain models and understand the physics at play, but might also lead to them being excluded by purity-focused classifiers.

\emi{Especially in the case of LSST, which will deliver a large number of all possible sources, it is crucial that automatic classifiers and filters are designed for the specific goal of showing to the user a large diversity of examples which could still be generated by the same underlying astrophysical process. While some might argue that the overwhelming number of alerts it will generate can allow us to strive for high purity at the price of low completeness, we argue that doing so would prevent us from detecting 
the extremely interesting and rare events populating the borders of each classes. This will require adaptation of tools and cultural practices, which are certain to result in the expansion of the current astronomical taxonomy, as hinted by the results presented in this work.}

\begin{acknowledgements}
Softwares: \texttt{numpy} \citep{harris_array_2020}, \texttt{matplotlib} \citep{hunter_matplotlib_2007}, \texttt{astropy} 
\citep{astropy_collaboration_astropy_2013,astropy_collaboration_astropy_2018,astropy_collaboration_astropy_2022}, \texttt{astroquery} \citep{ginsburg_astroquery_2019}, \texttt{Redback} \citep{sarin_redback_2024}, \texttt{scipy} \citep{virtanen_scipy_2020}, \texttt{SNAD} \citep{malanchev_snad_2023}, \texttt{SNCosmo} \citep{barbary_sncosmo_2025}. This research has made use of \texttt{hips2fits}, a tool developed at CDS, Strasbourg, France aiming at extracting FITS images from HiPS sky maps with respect to a WCS.

EQ acknowledges support from the European Space Agency, through the Internal Research Fellowship programme. EQ acknowledges funding from the European
Union’s Horizon 2020 research and innovation programme under grant agreement number 101004168, the XMM2ATHENA project \citep{webb_xmm2athena_2023}. MVP contribution was carried out under the state assignment of Lomonosov Moscow State University. ER was Funded by the European Union (ERC, project number 101042299, TransPIre). Views and opinions expressed are however those of the author(s) only and do not necessarily reflect those of the European Union or the European Research Council Executive Agency. Neither the European Union nor the granting authority can be held responsible for them. This work was developed within the \fink\ community and made use of the \fink\ community broker resources. \fink\ is supported by LSST-France and CNRS/IN2P3. 
AM is supported by the Australian Research Council Discovery Early Research Award (DE230100055). Parts of this research were conducted by the Australian Research Council Centre of Excellence for Gravitational Wave Discovery (OzGrav), through project number CE230100016.
This research has made use of the SIMBAD database, operated at CDS, Strasbourg, France.
This work was co-funded by the European Union and supported by the Czech Ministry of Education, Youth and Sports (Project No. CZ.02.01.01/00/22\_008/0004632 -- FORTE).
GM acknowledges support from grants n. PID2020-115325GB-C31 and n. PID2023-147338NB-C21 funded by MICIU/AEI/10.13039/501100011033 and ERDF/EU.
\end{acknowledgements}

%
%

\bibliographystyle{aa} 
\bibliography{references_erwan, references_emille}

@ARTICLE{pruzhinskaya2023,
       author = {{Pruzhinskaya}, M.~V. and {Ishida}, E.~E.~O. and {Novinskaya}, A.~K. and {Russeil}, E. and {Volnova}, A.~A. and {Malanchev}, K.~L. and {Kornilov}, M.~V. and {Aleo}, P.~D. and {Korolev}, V.~S. and {Krushinsky}, V.~V. and {Sreejith}, S. and {Gangler}, E.},
        title = "{Supernova search with active learning in ZTF DR3}",
      journal = {\aap},
         year = 2023,
        month = apr,
       volume = {672},
          eid = {A111},
        pages = {A111},
          doi = {10.1051/0004-6361/202245172},
archivePrefix = {arXiv},
       eprint = {2208.09053},
 primaryClass = {astro-ph.HE},
       adsurl = {https://ui.adsabs.harvard.edu/abs/2023A&A...672A.111P}
}

@INPROCEEDINGS{2020gbar.conf..127S,
       author = {{Shatsky}, N. and {Belinski}, A. and {Dodin}, A. and {Zheltoukhov}, S. and {Kornilov}, V. and {Postnov}, K. and {Potanin}, S. and {Safonov}, B. and {Tatarnikov}, A. and {Cherepashchuk}, A.},
        title = "{The Caucasian Mountain Observatory of the Sternberg Astronomical Institute: First Six Years of Operation}",
    booktitle = {Ground-Based Astronomy in Russia. 21st Century},
         year = 2020,
       editor = {{Romanyuk}, I.~I. and {Yakunin}, I.~A. and {Valeev}, A.~F. and {Kudryavtsev}, D.~O.},
        month = dec,
        pages = {127-132},
          doi = {10.26119/978-5-6045062-0-2_2020_127},
archivePrefix = {arXiv},
       eprint = {2010.10850},
 primaryClass = {astro-ph.IM},
       adsurl = {https://ui.adsabs.harvard.edu/abs/2020gbar.conf..127S}
}

@ARTICLE{guy2007,
       author = {{Guy}, J. and {Astier}, P. and {Baumont}, S. and {Hardin}, D. and {Pain}, R. and {Regnault}, N. and {Basa}, S. and {Carlberg}, R.~G. and {Conley}, A. and {Fabbro}, S. and {Fouchez}, D. and {Hook}, I.~M. and {Howell}, D.~A. and {Perrett}, K. and {Pritchet}, C.~J. and {Rich}, J. and {Sullivan}, M. and {Antilogus}, P. and {Aubourg}, E. and {Bazin}, G. and {Bronder}, J. and {Filiol}, M. and {Palanque-Delabrouille}, N. and {Ripoche}, P. and {Ruhlmann-Kleider}, V.},
        title = "{SALT2: using distant supernovae to improve the use of type Ia supernovae as distance indicators}",
      journal = {\aap},
         year = 2007,
        month = apr,
       volume = {466},
       number = {1},
        pages = {11-21},
          doi = {10.1051/0004-6361:20066930},
archivePrefix = {arXiv},
       eprint = {astro-ph/0701828},
 primaryClass = {astro-ph},
       adsurl = {https://ui.adsabs.harvard.edu/abs/2007A&A...466...11G}
}

@ARTICLE{graham2022,
       author = {{Graham}, Melissa L. and {Fremling}, Christoffer and {Perley}, Daniel A. and {Biswas}, Rahul and {Phillips}, Christopher A. and {Sollerman}, Jesper and {Nugent}, Peter E. and {Nance}, Sarafina and {Dhawan}, Suhail and {Nordin}, Jakob and {Goobar}, Ariel and {Miller}, Adam and {Neill}, James D. and {Hall}, Xander J. and {Hankins}, Matthew J. and {Duev}, Dmitry A. and {Kasliwal}, Mansi M. and {Rigault}, Mickael and {Bellm}, Eric C. and {Hale}, David and {Mr{\'o}z}, Przemek and {Kulkarni}, S.~R.},
        title = "{Supernova siblings and their parent galaxies in the Zwicky Transient Facility Bright Transient Survey}",
      journal = {\mnras},
         year = 2022,
        month = mar,
       volume = {511},
       number = {1},
        pages = {241-254},
          doi = {10.1093/mnras/stab3802},
archivePrefix = {arXiv},
       eprint = {2112.14819},
 primaryClass = {astro-ph.HE},
       adsurl = {https://ui.adsabs.harvard.edu/abs/2022MNRAS.511..241G}
}

@ARTICLE{scolnic2020,
       author = {{Scolnic}, D. and {Smith}, M. and {Massiah}, A. and {Wiseman}, P. and {Brout}, D. and {Kessler}, R. and {Davis}, T.~M. and {Foley}, R.~J. and {Galbany}, L. and {Hinton}, S.~R. and {Hounsell}, R. and {Kelsey}, L. and {Lidman}, C. and {Macaulay}, E. and {Morgan}, R. and {Nichol}, R.~C. and {M{\"o}ller}, A. and {Popovic}, B. and {Sako}, M. and {Sullivan}, M. and {Thomas}, B.~P. and {Tucker}, B.~E. and {Abbott}, T.~M.~C. and {Aguena}, M. and {Allam}, S. and {Annis}, J. and {Avila}, S. and {Bechtol}, K. and {Bertin}, E. and {Brooks}, D. and {Burke}, D.~L. and {Rosell}, A. Carnero and {Carollo}, D. and {Kind}, M. Carrasco and {Carretero}, J. and {Costanzi}, M. and {da Costa}, L.~N. and {De Vicente}, J. and {Desai}, S. and {Diehl}, H.~T. and {Doel}, P. and {Drlica-Wagner}, A. and {Eckert}, K. and {Eifler}, T.~F. and {Everett}, S. and {Flaugher}, B. and {Fosalba}, P. and {Frieman}, J. and {Garc{\'\i}a-Bellido}, J. and {Gaztanaga}, E. and {Gerdes}, D.~W. and {Glazebrook}, K. and {Gruen}, D. and {Gruendl}, R.~A. and {Gschwend}, J. and {Gutierrez}, G. and {Hartley}, W.~G. and {Hollowood}, D.~L. and {Honscheid}, K. and {James}, D.~J. and {Kuehn}, K. and {Kuropatkin}, N. and {Lewis}, G.~F. and {Li}, T.~S. and {Lima}, M. and {Maia}, M.~A.~G. and {Marshall}, J.~L. and {Menanteau}, F. and {Miquel}, R. and {Palmese}, A. and {Paz-Chinch{\'o}n}, F. and {Plazas}, A.~A. and {Pursiainen}, M. and {Sanchez}, E. and {Scarpine}, V. and {Schubnell}, M. and {Serrano}, S. and {Sevilla-Noarbe}, I. and {Sommer}, N.~E. and {Suchyta}, E. and {Swanson}, M.~E.~C. and {Tarle}, G. and {Varga}, T.~N. and {Walker}, A.~R. and {Wilkinson}, R. and {DES Collaboration}},
        title = "{Supernova Siblings: Assessing the Consistency of Properties of Type Ia Supernovae that Share the Same Parent Galaxies}",
      journal = {\apjl},
         year = 2020,
        month = jun,
       volume = {896},
       number = {1},
          eid = {L13},
        pages = {L13},
          doi = {10.3847/2041-8213/ab8735},
archivePrefix = {arXiv},
       eprint = {2002.00974},
 primaryClass = {astro-ph.GA},
       adsurl = {https://ui.adsabs.harvard.edu/abs/2020ApJ...896L..13S}
}

@ARTICLE{dhawan2024,
       author = {{Dhawan}, S. and {Mortsell}, E. and {Johansson}, J. and {Goobar}, A. and {Rigault}, M. and {Smith}, M. and {Maguire}, K. and {Nordin}, J. and {Dimitriadis}, G. and {Nugent}, P.~E. and {Galbany}, L. and {Sollerman}, J. and {de Jaeger}, T. and {Terwel}, J.~H. and {Kim}, Y. -L. and {Burgaz}, Umut and {Helou}, G. and {Purdum}, J. and {Groom}, S.~L. and {Laher}, R. and {Healy}, B.},
        title = "{ZTF SN\raisebox{-0.5ex}\textasciitildeIa DR2: Cosmology-independent constraints on Type Ia supernova standardisation from supernova siblings}",
      journal = {arXiv e-prints},
         year = 2024,
        month = jun,
          eid = {arXiv:2406.01434},
        pages = {arXiv:2406.01434},
          doi = {10.48550/arXiv.2406.01434},
archivePrefix = {arXiv},
       eprint = {2406.01434},
 primaryClass = {astro-ph.CO},
       adsurl = {https://ui.adsabs.harvard.edu/abs/2024arXiv240601434D}
}

@ARTICLE{malanchev2023,
       author = {{Malanchev}, Konstantin and {Kornilov}, Matwey V. and {Pruzhinskaya}, Maria V. and {Ishida}, Emille E.~O. and {Aleo}, Patrick D. and {Korolev}, Vladimir S. and {Lavrukhina}, Anastasia and {Russeil}, Etienne and {Sreejith}, Sreevarsha and {Volnova}, Alina A. and {Voloshina}, Anastasiya and {Krone-Martins}, Alberto},
        title = "{The SNAD Viewer: Everything You Want to Know about Your Favorite ZTF Object}",
      journal = {\pasp},
         year = 2023,
        month = feb,
       volume = {135},
       number = {1044},
          eid = {024503},
        pages = {024503},
          doi = {10.1088/1538-3873/acb292},
archivePrefix = {arXiv},
       eprint = {2211.07605},
 primaryClass = {astro-ph.IM},
       adsurl = {https://ui.adsabs.harvard.edu/abs/2023PASP..135b4503M}
}

@ARTICLE{volnova2024,
       author = {{Volnova}, Alina A. and {Aleo}, Patrick D. and {Lavrukhina}, Anastasia and {Russeil}, Etienne and {Semenikhin}, Timofey and {Gangler}, Emmanuel and {Ishida}, Emille E.~O. and {Kornilov}, Matwey V. and {Korolev}, Vladimir and {Malanchev}, Konstantin and {Pruzhinskaya}, Maria V. and {Sreejith}, Sreevarsha},
        title = "{Exploring the Universe with SNAD: Anomaly Detection in Astronomy}",
      journal = {arXiv e-prints},
         year = 2024,
        month = oct,
          eid = {arXiv:2410.18875},
        pages = {arXiv:2410.18875},
          doi = {10.48550/arXiv.2410.18875},
archivePrefix = {arXiv},
       eprint = {2410.18875},
 primaryClass = {astro-ph.IM},
       adsurl = {https://ui.adsabs.harvard.edu/abs/2024arXiv241018875V}
}

@ARTICLE{ishida2019,
       author = {{Ishida}, Emille E.~O.},
        title = "{Machine learning and the future of supernova cosmology}",
      journal = {Nature Astronomy},
         year = 2019,
        month = aug,
       volume = {3},
        pages = {680-682},
          doi = {10.1038/s41550-019-0860-6},
archivePrefix = {arXiv},
       eprint = {1908.02315},
 primaryClass = {astro-ph.IM},
       adsurl = {https://ui.adsabs.harvard.edu/abs/2019NatAs...3..680I}
}

@ARTICLE{2011ApJ...737..103S,
       author = {{Schlafly}, Edward F. and {Finkbeiner}, Douglas P.},
        title = "{Measuring Reddening with Sloan Digital Sky Survey Stellar Spectra and Recalibrating SFD}",
      journal = {\apj},
         year = 2011,
        month = aug,
       volume = {737},
       number = {2},
          eid = {103},
        pages = {103},
          doi = {10.1088/0004-637X/737/2/103},
archivePrefix = {arXiv},
       eprint = {1012.4804},
 primaryClass = {astro-ph.GA},
       adsurl = {https://ui.adsabs.harvard.edu/abs/2011ApJ...737..103S}
}

@ARTICLE{2014AJ....147..118R,
       author = {{Richardson}, Dean and {Jenkins}, III, Robert L. and {Wright}, John and {Maddox}, Larry},
        title = "{Absolute-magnitude Distributions of Supernovae}",
      journal = {\aj},
         year = 2014,
        month = may,
       volume = {147},
       number = {5},
          eid = {118},
        pages = {118},
          doi = {10.1088/0004-6256/147/5/118},
archivePrefix = {arXiv},
       eprint = {1403.5755},
 primaryClass = {astro-ph.SR},
       adsurl = {https://ui.adsabs.harvard.edu/abs/2014AJ....147..118R}
}

@misc{barbary_sncosmo_2025,
	title = {{SNCosmo}},
	url = {https://zenodo.org/records/15019859},
	urldate = {2025-11-20},
	publisher = {Zenodo},
	author = {Barbary, Kyle and Bailey, Stephen and Barentsen, Geert and Barclay, Tom and Biswas, Rahul and Boone, Kyle and Craig, Matt and Feindt, Ulrich and Friesen, Brian and Goldstein, Danny and Jha, Saurabh W. and Jones, David O. and Mondon, Florian and Papadogiannakis, Seméli and Perrefort, Daniel and Pierel, Justin and Rodney, Steve and Rose, Benjamin and Saunders, Clare and Sipőcz, Brigitta and Sofiatti, Caroline and Thomas, Rollin C. and van Santen, Jakob and Vincenzi, Maria and Wang, David and Wood-Vasey, Michael},
	month = mar,
	year = {2025},
	doi = {10.5281/zenodo.15019859},
}

@article{malanchev_snad_2023,
	title = {The {SNAD} {Viewer}: {Everything} {You} {Want} to {Know} about {Your} {Favorite} {ZTF} {Object}},
	volume = {135},
	issn = {0004-6280},
	shorttitle = {The {SNAD} {Viewer}},
	url = {https://ui.adsabs.harvard.edu/abs/2023PASP..135b4503M/abstract},
	doi = {10.1088/1538-3873/acb292},
	language = {en},
	number = {1044},
	urldate = {2025-11-20},
	journal = {Publications of the Astronomical Society of the Pacific},
	author = {Malanchev, Konstantin and Kornilov, Matwey V. and Pruzhinskaya, Maria V. and Ishida, Emille E. O. and Aleo, Patrick D. and Korolev, Vladimir S. and Lavrukhina, Anastasia and Russeil, Etienne and Sreejith, Sreevarsha and Volnova, Alina A. and Voloshina, Anastasiya and Krone-Martins, Alberto},
	month = feb,
	year = {2023},
	pages = {024503},
}

@misc{majumder_superluminous_2024,
	title = {Superluminous supernova search with {PineForest}},
	url = {http://arxiv.org/abs/2410.21077},
	doi = {10.48550/arXiv.2410.21077},
	urldate = {2025-11-19},
	publisher = {arXiv},
	author = {Majumder, T. and Pruzhinskaya, M. V. and Ishida, E. E. O. and Malanchev, K. L. and Semenikhin, T. A.},
	month = oct,
	year = {2024},
	note = {arXiv:2410.21077},
}

@misc{masci_zwicky_2019,
	title = {The {Zwicky} {Transient} {Facility}: {Data} {Processing}, {Products}, and {Archive}},
	shorttitle = {The {Zwicky} {Transient} {Facility}},
	url = {http://arxiv.org/abs/1902.01872},
	doi = {10.48550/arXiv.1902.01872},
	urldate = {2025-11-18},
	publisher = {arXiv},
	author = {Masci, Frank J. and Laher, Russ R. and Rusholme, Ben and Shupe, David L. and Groom, Steven and Surace, Jason and Jackson, Edward and Monkewitz, Serge and Beck, Ron and Flynn, David and Terek, Scott and Landry, Walter and Hacopians, Eugean and Desai, Vandana and Howell, Justin and Brooke, Tim and Imel, David and Wachter, Stefanie and Ye, Quan-Zhi and Lin, Hsing-Wen and Cenko, S. Bradley and Cunningham, Virginia and Rebbapragada, Umaa and Bue, Brian and Miller, Adam A. and Mahabal, Ashish and Bellm, Eric C. and Patterson, Maria T. and Jurić, Mario and Golkhou, V. Zach and Ofek, Eran O. and Walters, Richard and Graham, Matthew and Kasliwal, Mansi M. and Dekany, Richard G. and Kupfer, Thomas and Burdge, Kevin and Cannella, Christopher B. and Barlow, Tom and Sistine, Angela Van and Giomi, Matteo and Fremling, Christoffer and Blagorodnova, Nadejda and Levitan, David and Riddle, Reed and Smith, Roger M. and Helou, George and Prince, Thomas A. and Kulkarni, Shrinivas R.},
	month = feb,
	year = {2019},
	note = {arXiv:1902.01872},
}

@article{mckernan_starfall_2022,
	title = {Starfall: a heavy rain of stars in 'turning on' {AGN}},
	volume = {514},
	issn = {0035-8711},
	shorttitle = {Starfall},
	url = {https://ui.adsabs.harvard.edu/abs/2022MNRAS.514.4102M/abstract},
	doi = {10.1093/mnras/stac1310},
	language = {en},
	number = {3},
	urldate = {2025-11-11},
	journal = {Monthly Notices of the Royal Astronomical Society},
	author = {McKernan, B. and Ford, K. E. S. and Cantiello, M. and Graham, M. and Jermyn, A. S. and Leigh, N. W. C. and Ryu, T. and Stern, D.},
	month = aug,
	year = {2022},
	pages = {4102--4110},
}

@article{paliya_narrow-line_2024,
	title = {Narrow-line {Seyfert} 1 galaxies in {Sloan} {Digital} {Sky} {Survey}: a new optical spectroscopic catalogue},
	volume = {527},
	issn = {0035-8711},
	shorttitle = {Narrow-line {Seyfert} 1 galaxies in {Sloan} {Digital} {Sky} {Survey}},
	url = {https://ui.adsabs.harvard.edu/abs/2024MNRAS.527.7055P/abstract},
	doi = {10.1093/mnras/stad3650},
	language = {en},
	number = {3},
	urldate = {2025-11-11},
	journal = {Monthly Notices of the Royal Astronomical Society},
	author = {Paliya, Vaidehi S. and Stalin, C. S. and Domínguez, Alberto and Saikia, D. J.},
	month = jan,
	year = {2024},
	pages = {7055--7069},
}

@article{frederick_family_2021,
	title = {A {Family} {Tree} of {Optical} {Transients} from {Narrow}-line {Seyfert} 1 {Galaxies}},
	volume = {920},
	issn = {0004-637X},
	url = {https://ui.adsabs.harvard.edu/abs/2021ApJ...920...56F/abstract},
	doi = {10.3847/1538-4357/ac110f},
	language = {en},
	number = {1},
	urldate = {2025-11-11},
	journal = {The Astrophysical Journal},
	author = {Frederick, Sara and Gezari, Suvi and Graham, Matthew J. and Sollerman, Jesper and van Velzen, Sjoert and Perley, Daniel A. and Stern, Daniel and Ward, Charlotte and Hammerstein, Erica and Hung, Tiara and Yan, Lin and Andreoni, Igor and Bellm, Eric C. and Duev, Dmitry A. and Kowalski, Marek and Mahabal, Ashish A. and Masci, Frank J. and Medford, Michael and Rusholme, Ben and Smith, Roger and Walters, Richard},
	month = oct,
	year = {2021},
	pages = {56},
}

@misc{desi_collaboration_data_2025,
	title = {Data {Release} 1 of the {Dark} {Energy} {Spectroscopic} {Instrument}},
	url = {https://arxiv.org/abs/2503.14745v1},
	language = {en},
	urldate = {2025-11-11},
	journal = {arXiv.org},
	author = {DESI Collaboration and Abdul-Karim, M. and Adame, A. G. and Aguado, D. and Aguilar, J. and Ahlen, S. and Alam, S. and Aldering, G. and Alexander, D. M. and Alfarsy, R. and Allen, L. and Prieto, C. Allende and Alves, O. and Anand, A. and Andrade, U. and Armengaud, E. and Avila, S. and Aviles, A. and Awan, H. and Bailey, S. and Lizancos, A. Baleato and Ballester, O. and Bault, A. and Bautista, J. and BenZvi, S. and Silva, L. Beraldo e and Bermejo-Climent, J. R. and Beutler, F. and Bianchi, D. and Blake, C. and Blum, R. and Bolton, A. S. and Bonici, M. and Brieden, S. and Brodzeller, A. and Brooks, D. and Buckley-Geer, E. and Burtin, E. and Canning, R. and Rosell, A. Carnero and Carr, A. and Carrilho, P. and Casas, L. and Castander, F. J. and Cereskaite, R. and Cervantes-Cota, J. L. and Chaussidon, E. and Chaves-Montero, J. and Chen, S. and Chen, X. and Claybaugh, T. and Cole, S. and Cooper, A. P. and Cousinou, M.-C. and Cuceu, A. and Davis, T. M. and Dawson, K. S. and de Belsunce, R. and de la Cruz, R. and de la Macorra, A. and de Mattia, A. and Deiosso, N. and Della Costa, J. and Demina, R. and Demirbozan, U. and DeRose, J. and Dey, A. and Dey, B. and Ding, J. and Ding, Z. and Doel, P. and Douglass, K. and Dowicz, M. and Ebina, H. and Edelstein, J. and Eisenstein, D. J. and Elbers, W. and Emas, N. and Escoffier, S. and Fagrelius, P. and Fan, X. and Fanning, K. and Fawcett, V. A. and Fernández-García, E. and Ferraro, S. and Findlay, N. and Font-Ribera, A. and Forero-Romero, J. E. and Forero-Sánchez, D. and Frenk, C. S. and Gänsicke, B. T. and Galbany, L. and García-Bellido, J. and Garcia-Quintero, C. and Garrison, L. H. and Gaztañaga, E. and Gil-Marín, H. and Gnedin, O. Y. and Gontcho, S. Gontcho A. and Gonzalez-Morales, A. X. and Gonzalez-Perez, V. and Gordon, C. and Graur, O. and Green, D. and Gruen, D. and Gsponer, R. and Guandalin, C. and Gutierrez, G. and Guy, J. and Hahn, C. and Han, J. J. and Han, J. and He, S. and Herrera-Alcantar, H. K. and Honscheid, K. and Hou, J. and Howlett, C. and Huterer, D. and Iršič, V. and Ishak, M. and Jacques, A. and Jimenez, J. and Jing, Y. P. and Joachimi, B. and Joudaki, S. and Joyce, R. and Jullo, E. and Juneau, S. and Karaçaylı, N. G. and Karim, T. and Kehoe, R. and Kent, S. and Khederlarian, A. and Kirkby, D. and Kisner, T. and Kitaura, F.-S. and Kizhuprakkat, N. and Kong, H. and Koposov, S. E. and Kremin, A. and Krolewski, A. and Lahav, O. and Lai, Y. and Lamman, C. and Lan, T.-W. and Landriau, M. and Lang, D. and Lange, J. U. and Lasker, J. and Goff, J. M. Le and Guillou, L. Le and Leauthaud, A. and Levi, M. E. and Li, S. and Li, T. S. and Lodha, K. and Lokken, M. and Luo, Y. and Magneville, C. and Manera, M. and Manser, C. J. and Margala, D. and Martini, P. and Maus, M. and McCullough, J. and McDonald, P. and Medina, G. E. and Medina-Varela, L. and Meisner, A. and Mena-Fernández, J. and Menegas, A. and Mezcua, M. and Miquel, R. and Montero-Camacho, P. and Moon, J. and Moustakas, J. and Muñoz-Gutiérrez, A. and Muñoz-Santos, D. and Myers, A. D. and Myles, J. and Nadathur, S. and Najita, J. and Napolitano, L. and Newman, J. A. and Nikakhtar, F. and Nikutta, R. and Niz, G. and Noriega, H. E. and Padmanabhan, N. and Paillas, E. and Palanque-Delabrouille, N. and Palmese, A. and Pan, J. and Pan, Z. and Parkinson, D. and Peacock, J. and Percival, W. J. and Pérez-Fernández, A. and Pérez-Ràfols, I. and Peterson, P. and Piat, J. and Pieri, M. M. and Pinon, M. and Poppett, C. and Porredon, A. and Prada, F. and Pucha, R. and Qin, F. and Rabinowitz, D. and Raichoor, A. and Ramírez-Pérez, C. and Ramirez-Solano, S. and Rashkovetskyi, M. and Ravoux, C. and Riley, A. H. and Rocher, A. and Rockosi, C. and Rohlf, J. and Ross, A. J. and Rossi, G. and Ruggeri, R. and Ruhlmann-Kleider, V. and Sabiu, C. G. and Said, K. and Saintonge, A. and Samushia, L. and Sanchez, E. and Sanders, N. and Saulder, C. and Schlafly, E. F. and Schlegel, D. and Scholte, D. and Schubnell, M. and Seo, H. and Shafieloo, A. and Sharples, R. and Silber, J. and Siudek, M. and Smith, A. and Sprayberry, D. and Suárez-Pérez, J. and Swanson, J. and Tan, T. and Tarlé, G. and Taylor, P. and Thomas, G. and Tojeiro, R. and Turner, R. J. and Turner, W. and Ureña-López, L. A. and Vaisakh, R. and Valluri, M. and Vargas-Magaña, M. and Verde, L. and Walther, M. and Wang, B. and Wang, M. S. and Wang, W. and Weaver, B. A. and Weaverdyck, N. and Wechsler, R. H. and White, M. and Wolfson, M. and Yang, J. and Yèche, C. and Youles, S. and Yu, J. and Yuan, S. and Zaborowski, E. A. and Zarrouk, P. and Zhang, H. and Zhao, C. and Zhao, R. and Zheng, Z. and Zhou, R. and Zou, H. and Zou, S. and Zu, Y.},
	month = mar,
	year = {2025},
}

@misc{graham_extremely_2025,
	title = {An {Extremely} {Luminous} {Flare} {Recorded} from a {Supermassive} {Black} {Hole}},
	url = {http://arxiv.org/abs/2511.02178},
	doi = {10.48550/arXiv.2511.02178},
	urldate = {2025-11-05},
	publisher = {arXiv},
	author = {Graham, Matthew J. and McKernan, Barry and Ford, K. E. Saavik and Stern, Daniel and Cantiello, Matteo and Drake, Andrew J. and Ding, Yuanze and Kasliwal, Mansi and Koss, Mike and Margutti, Raffaella and Rose, Sam and Somalwar, Jean and Wiseman, Phil and Djorgovski, S. G. and Veres, Patrik M. and Bellm, Eric C. and Chen, Tracy X. and Groom, Steven L. and Kulkarni, Shrinivas R. and Mahabal, Ashish},
	month = nov,
	year = {2025},
	note = {arXiv:2511.02178},
}

@misc{wang_stars_2025,
	title = {A {Star}'s {Death} by a {Thousand} {Cuts}: {The} {Runaway} {Periodic} {Eruptions} of {AT2023uqm}},
	shorttitle = {A {Star}'s {Death} by a {Thousand} {Cuts}},
	url = {http://arxiv.org/abs/2510.26561},
	doi = {10.48550/arXiv.2510.26561},
	urldate = {2025-10-31},
	publisher = {arXiv},
	author = {Wang, Yibo and Wang, Tingui and Huang, Shifeng and Zhu, Jiazheng and Jiang, Ning and Lu, Wenbin and Shen, Rongfeng and Zhong, Shiyan and Lai, Dong and Yang, Yi and Shu, Xinwen and Xia, Tianyu and Luo, Di and Lyu, Jianwei and Brink, Thomas and Filippenko, Alex and Zheng, Weikang and Cai, Minxuan and Xu, Zelin and Wu, Mingxin and Zhang, Xiaer and Wu, Weiyu and Fan, Lulu and Kong, Ji-an Jiang Xu and Li, Bin and Lin, Feng and Liang, Ming and Luo, Wentao and Tang, Jinlong and Wan, Zhen and Wang, Hairen and Wang, Jian and Xue, Yongquan and Yao, Dazhi and Zhang, Hongfei and Zhao, Wen and Zheng, Xianzhong and Zhu, Qingfeng and Zuo, Yingxi},
	month = oct,
	year = {2025},
	note = {arXiv:2510.26561},
}

@misc{zhu_ultraviolet_2025,
	title = {Ultraviolet {Spectral} {Evidence} for {Ansky} as a {Slowly} {Evolving} {Featureless} {Tidal} {Disruption} {Event} with {Quasi}-periodic {Eruptions}},
	url = {http://arxiv.org/abs/2510.22211},
	doi = {10.48550/arXiv.2510.22211},
	urldate = {2025-10-28},
	publisher = {arXiv},
	author = {Zhu, Jiazheng and Jiang, Ning and Wang, Yibo and Wang, Tinggui and Sun, Luming and Zhong, Shiyan and Yao, Yuhan and Chornock, Ryan and Dai, Lixin and Lyu, Jianwei and Shu, Xinwen and Fremling, Christoffer and Hammerstein, Erica and Huang, Shifeng and Li, Wenkai and You, Bei},
	month = oct,
	year = {2025},
	note = {arXiv:2510.22211},
}

@article{speagle_dynesty_2020,
	title = {{DYNESTY}: a dynamic nested sampling package for estimating {Bayesian} posteriors and evidences},
	volume = {493},
	issn = {0035-8711},
	shorttitle = {{DYNESTY}},
	url = {https://ui.adsabs.harvard.edu/abs/2020MNRAS.493.3132S/abstract},
	doi = {10.1093/mnras/staa278},
	language = {en},
	number = {3},
	urldate = {2025-10-23},
	journal = {Monthly Notices of the Royal Astronomical Society},
	author = {Speagle, Joshua S.},
	month = apr,
	year = {2020},
	pages = {3132--3158},
}

@article{mockler_weighing_2019,
	title = {Weighing {Black} {Holes} {Using} {Tidal} {Disruption} {Events}},
	volume = {872},
	issn = {0004-637X},
	url = {https://ui.adsabs.harvard.edu/abs/2019ApJ...872..151M/abstract},
	doi = {10.3847/1538-4357/ab010f},
	language = {en},
	number = {2},
	urldate = {2025-10-23},
	journal = {The Astrophysical Journal},
	author = {Mockler, Brenna and Guillochon, James and Ramirez-Ruiz, Enrico},
	month = feb,
	year = {2019},
	pages = {151},
}

@article{ashton_bilby_2019,
	title = {{BILBY}: {A} {User}-friendly {Bayesian} {Inference} {Library} for {Gravitational}-wave {Astronomy}},
	volume = {241},
	issn = {0067-0049},
	shorttitle = {{BILBY}},
	url = {https://ui.adsabs.harvard.edu/abs/2019ApJS..241...27A/abstract},
	doi = {10.3847/1538-4365/ab06fc},
	language = {en},
	number = {2},
	urldate = {2025-10-23},
	journal = {The Astrophysical Journal Supplement Series},
	author = {Ashton, Gregory and Hübner, Moritz and Lasky, Paul D. and Talbot, Colm and Ackley, Kendall and Biscoveanu, Sylvia and Chu, Qi and Divakarla, Atul and Easter, Paul J. and Goncharov, Boris and Hernandez Vivanco, Francisco and Harms, Jan and Lower, Marcus E. and Meadors, Grant D. and Melchor, Denyz and Payne, Ethan and Pitkin, Matthew D. and Powell, Jade and Sarin, Nikhil and Smith, Rory J. E. and Thrane, Eric},
	month = apr,
	year = {2019},
	pages = {27},
}

@article{sarin_redback_2024,
	title = {{REDBACK}: a {Bayesian} inference software package for electromagnetic transients},
	volume = {531},
	issn = {0035-8711},
	shorttitle = {{REDBACK}},
	url = {https://ui.adsabs.harvard.edu/abs/2024MNRAS.531.1203S/abstract},
	doi = {10.1093/mnras/stae1238},
	language = {en},
	number = {1},
	urldate = {2025-10-23},
	journal = {Monthly Notices of the Royal Astronomical Society},
	author = {Sarin, Nikhil and Hübner, Moritz and Omand, Conor M. B. and Setzer, Christian N. and Schulze, Steve and Adhikari, Naresh and Sagués-Carracedo, Ana and Galaudage, Shanika and Wallace, Wendy F. and Lamb, Gavin P. and Lin, En-Tzu},
	month = jun,
	year = {2024},
	pages = {1203--1227},
}

@article{fu_catnorth_2024,
	title = {{CatNorth}: {An} {Improved} {Gaia} {DR3} {Quasar} {Candidate} {Catalog} with {Pan}-{STARRS1} and {CatWISE}},
	volume = {271},
	issn = {0067-0049},
	shorttitle = {{CatNorth}},
	url = {https://ui.adsabs.harvard.edu/abs/2024ApJS..271...54F/abstract},
	doi = {10.3847/1538-4365/ad2ae6},
	language = {en},
	number = {2},
	urldate = {2025-10-22},
	journal = {The Astrophysical Journal Supplement Series},
	author = {Fu, Yuming and Wu, Xue-Bing and Li, Yifan and Pang, Yuxuan and Joshi, Ravi and Zhang, Shuo and Wang, Qiyue and Yang, Jing and Ng, FanLam and Liu, Xingjian and Qiu, Yu and Zhu, Rui and Wang, Huimei and Wolf, Christian and Zhang, Yanxia and Huo, Zhi-Ying and Ai, Y. L. and Ma, Qinchun and Feng, Xiaotong and Bouwens, R. J.},
	month = apr,
	year = {2024},
	pages = {54},
}

@article{wen_at2018fyk_2024,
	title = {{AT2018fyk}: {Candidate} {Tidal} {Disruption} {Event} by a ({Super}){Massive} {Black} {Hole} {Binary}},
	volume = {970},
	issn = {0004-637X},
	shorttitle = {{AT2018fyk}},
	url = {https://ui.adsabs.harvard.edu/abs/2024ApJ...970..116W/abstract},
	doi = {10.3847/1538-4357/ad4da3},
	language = {en},
	number = {2},
	urldate = {2025-10-21},
	journal = {The Astrophysical Journal},
	author = {Wen, S. and Jonker, P. G. and Levan, A. J. and Li, D. and Stone, N. C. and Zabludoff, A. I. and Cao, Z. and Wevers, T. and Pasham, D. R. and Lewin, C. and Kara, E.},
	month = aug,
	year = {2024},
	pages = {116},
}

@misc{hernandez-garcia_nicer_2025,
	title = {{NICER} observations reveal doubled timescales in {Ansky}'s quasi-periodic eruptions ({QPEs})},
	url = {http://arxiv.org/abs/2509.16304},
	doi = {10.48550/arXiv.2509.16304},
	urldate = {2025-09-23},
	publisher = {arXiv},
	author = {Hernández-García, L. and Sánchez-Sáez, P. and Chakraborty, J. and Cuadra, J. and Miniutti, G. and Arcodia, R. and Arévalo, P. and Giustini, M. and Kara, E. and Ricci, C. and Pasham, D. R. and Arzoumanian, Z. and Gendreau, K. and Lira, P.},
	month = sep,
	year = {2025},
	note = {arXiv:2509.16304},
}

@article{stern_mid-infrared_2012,
	title = {Mid-infrared {Selection} of {Active} {Galactic} {Nuclei} with the {Wide}-{Field} {Infrared} {Survey} {Explorer}. {I}. {Characterizing} {WISE}-selected {Active} {Galactic} {Nuclei} in {COSMOS}},
	volume = {753},
	issn = {0004-637X},
	url = {https://ui.adsabs.harvard.edu/abs/2012ApJ...753...30S/abstract},
	doi = {10.1088/0004-637X/753/1/30},
	language = {en},
	number = {1},
	urldate = {2025-09-17},
	journal = {The Astrophysical Journal},
	author = {Stern, Daniel and Assef, Roberto J. and Benford, Dominic J. and Blain, Andrew and Cutri, Roc and Dey, Arjun and Eisenhardt, Peter and Griffith, Roger L. and Jarrett, T. H. and Lake, Sean and Masci, Frank and Petty, Sara and Stanford, S. A. and Tsai, Chao-Wei and Wright, E. L. and Yan, Lin and Harrison, Fiona and Madsen, Kristin},
	month = jul,
	year = {2012},
	pages = {30},
}

@article{york_sloan_2000,
	title = {The {Sloan} {Digital} {Sky} {Survey}: {Technical} {Summary}},
	volume = {120},
	issn = {0004-6256},
	shorttitle = {The {Sloan} {Digital} {Sky} {Survey}},
	url = {https://ui.adsabs.harvard.edu/abs/2000AJ....120.1579Y/abstract},
	doi = {10.1086/301513},
	language = {en},
	number = {3},
	urldate = {2025-09-17},
	journal = {The Astronomical Journal},
	author = {York, Donald G. and Adelman, J. and Anderson, John E. and Anderson, Scott F. and Annis, James and Bahcall, Neta A. and Bakken, J. A. and Barkhouser, Robert and Bastian, Steven and Berman, Eileen and Boroski, William N. and Bracker, Steve and Briegel, Charlie and Briggs, John W. and Brinkmann, J. and Brunner, Robert and Burles, Scott and Carey, Larry and Carr, Michael A. and Castander, Francisco J. and Chen, Bing and Colestock, Patrick L. and Connolly, A. J. and Crocker, J. H. and Csabai, István and Czarapata, Paul C. and Davis, John Eric and Doi, Mamoru and Dombeck, Tom and Eisenstein, Daniel and Ellman, Nancy and Elms, Brian R. and Evans, Michael L. and Fan, Xiaohui and Federwitz, Glenn R. and Fiscelli, Larry and Friedman, Scott and Frieman, Joshua A. and Fukugita, Masataka and Gillespie, Bruce and Gunn, James E. and Gurbani, Vijay K. and de Haas, Ernst and Haldeman, Merle and Harris, Frederick H. and Hayes, J. and Heckman, Timothy M. and Hennessy, G. S. and Hindsley, Robert B. and Holm, Scott and Holmgren, Donald J. and Huang, Chi-hao and Hull, Charles and Husby, Don and Ichikawa, Shin-Ichi and Ichikawa, Takashi and Ivezić, Željko and Kent, Stephen and Kim, Rita S. J. and Kinney, E. and Klaene, Mark and Kleinman, A. N. and Kleinman, S. and Knapp, G. R. and Korienek, John and Kron, Richard G. and Kunszt, Peter Z. and Lamb, D. Q. and Lee, B. and Leger, R. French and Limmongkol, Siriluk and Lindenmeyer, Carl and Long, Daniel C. and Loomis, Craig and Loveday, Jon and Lucinio, Rich and Lupton, Robert H. and MacKinnon, Bryan and Mannery, Edward J. and Mantsch, P. M. and Margon, Bruce and McGehee, Peregrine and McKay, Timothy A. and Meiksin, Avery and Merelli, Aronne and Monet, David G. and Munn, Jeffrey A. and Narayanan, Vijay K. and Nash, Thomas and Neilsen, Eric and Neswold, Rich and Newberg, Heidi Jo and Nichol, R. C. and Nicinski, Tom and Nonino, Mario and Okada, Norio and Okamura, Sadanori and Ostriker, Jeremiah P. and Owen, Russell and Pauls, A. George and Peoples, John and Peterson, R. L. and Petravick, Donald and Pier, Jeffrey R. and Pope, Adrian and Pordes, Ruth and Prosapio, Angela and Rechenmacher, Ron and Quinn, Thomas R. and Richards, Gordon T. and Richmond, Michael W. and Rivetta, Claudio H. and Rockosi, Constance M. and Ruthmansdorfer, Kurt and Sandford, Dale and Schlegel, David J. and Schneider, Donald P. and Sekiguchi, Maki and Sergey, Gary and Shimasaku, Kazuhiro and Siegmund, Walter A. and Smee, Stephen and Smith, J. Allyn and Snedden, S. and Stone, R. and Stoughton, Chris and Strauss, Michael A. and Stubbs, Christopher and SubbaRao, Mark and Szalay, Alexander S. and Szapudi, Istvan and Szokoly, Gyula P. and Thakar, Anirudda R. and Tremonti, Christy and Tucker, Douglas L. and Uomoto, Alan and Vanden Berk, Dan and Vogeley, Michael S. and Waddell, Patrick and Wang, Shu-i and Watanabe, Masaru and Weinberg, David H. and Yanny, Brian and Yasuda, Naoki and Collaboration, Sdss},
	month = sep,
	year = {2000},
	pages = {1579--1587},
}

@article{mainzer_preliminary_2011,
	title = {Preliminary {Results} from {NEOWISE}: {An} {Enhancement} to the {Wide}-field {Infrared} {Survey} {Explorer} for {Solar} {System} {Science}},
	volume = {731},
	issn = {0004-637X},
	shorttitle = {Preliminary {Results} from {NEOWISE}},
	url = {https://ui.adsabs.harvard.edu/abs/2011ApJ...731...53M/abstract},
	doi = {10.1088/0004-637X/731/1/53},
	language = {en},
	number = {1},
	urldate = {2025-09-17},
	journal = {The Astrophysical Journal},
	author = {Mainzer, A. and Bauer, J. and Grav, T. and Masiero, J. and Cutri, R. M. and Dailey, J. and Eisenhardt, P. and McMillan, R. S. and Wright, E. and Walker, R. and Jedicke, R. and Spahr, T. and Tholen, D. and Alles, R. and Beck, R. and Brandenburg, H. and Conrow, T. and Evans, T. and Fowler, J. and Jarrett, T. and Marsh, K. and Masci, F. and McCallon, H. and Wheelock, S. and Wittman, M. and Wyatt, P. and DeBaun, E. and Elliott, G. and Elsbury, D. and Gautier, T. and Gomillion, S. and Leisawitz, D. and Maleszewski, C. and Micheli, M. and Wilkins, A.},
	month = apr,
	year = {2011},
	pages = {53},
}

@article{masci_new_2023,
	title = {A {New} {Forced} {Photometry} {Service} for the {Zwicky} {Transient} {Facility}},
	url = {https://ui.adsabs.harvard.edu/abs/2023arXiv230516279M/abstract},
	doi = {10.48550/arXiv.2305.16279},
	language = {en},
	urldate = {2025-09-17},
	journal = {arXiv e-prints},
	author = {Masci, Frank J. and Laher, Russ R. and Rusholme, Benjamin and Shupe, David and Paladini, Roberta and Groom, Steve and Wold, Avery and Miller, Adam A. and Drake, Andrew},
	month = may,
	year = {2023},
	pages = {arXiv:2305.16279},
}

@article{drake_first_2009,
	title = {First {Results} from the {Catalina} {Real}-{Time} {Transient} {Survey}},
	volume = {696},
	issn = {0004-637X},
	url = {https://ui.adsabs.harvard.edu/abs/2009ApJ...696..870D/abstract},
	doi = {10.1088/0004-637X/696/1/870},
	language = {en},
	number = {1},
	urldate = {2025-09-17},
	journal = {The Astrophysical Journal},
	author = {Drake, A. J. and Djorgovski, S. G. and Mahabal, A. and Beshore, E. and Larson, S. and Graham, M. J. and Williams, R. and Christensen, E. and Catelan, M. and Boattini, A. and Gibbs, A. and Hill, R. and Kowalski, R.},
	month = may,
	year = {2009},
	pages = {870--884},
}

@article{rose_hourglass_2025,
	title = {The {Hourglass} {Simulation}: {A} {Catalog} for the {Roman} {High}-latitude {Time}-domain {Core} {Community} {Survey}},
	volume = {988},
	issn = {0004-637X},
	shorttitle = {The {Hourglass} {Simulation}},
	url = {https://ui.adsabs.harvard.edu/abs/2025ApJ...988...65R/abstract},
	doi = {10.3847/1538-4357/ade1d6},
	language = {en},
	number = {1},
	urldate = {2025-09-01},
	journal = {The Astrophysical Journal},
	author = {Rose, B. M. and Vincenzi, M. and Hounsell, R. and Qu, H. and Aldoroty, L. and Scolnic, D. and Kessler, R. and Macias, P. and Brout, D. and Acevedo, M. and Chen, R. C. and Gomez, S. and Peterson, E. and Rubin, D. and Sako, M. and Team, the Roman Supernova Project Infrastructure},
	month = jul,
	year = {2025},
	pages = {65},
}

@article{murphy_vast_2013,
	title = {{VAST}: {An} {ASKAP} {Survey} for {Variables} and {Slow} {Transients}},
	volume = {30},
	issn = {1323-3580},
	shorttitle = {{VAST}},
	url = {https://ui.adsabs.harvard.edu/abs/2013PASA...30....6M/abstract},
	doi = {10.1017/pasa.2012.006},
	language = {en},
	urldate = {2025-09-01},
	journal = {Publications of the Astronomical Society of Australia},
	author = {Murphy, Tara and Chatterjee, Shami and Kaplan, David L. and Banyer, Jay and Bell, Martin E. and Bignall, Hayley E. and Bower, Geoffrey C. and Cameron, Robert A. and Coward, David M. and Cordes, James M. and Croft, Steve and Curran, James R. and Djorgovski, S. G. and Farrell, Sean A. and Frail, Dale A. and Gaensler, B. M. and Galloway, Duncan K. and Gendre, Bruce and Green, Anne J. and Hancock, Paul J. and Johnston, Simon and Kamble, Atish and Law, Casey J. and Lazio, T. Joseph W. and Lo, Kitty K. and Macquart, Jean-Pierre and Rea, Nanda and Rebbapragada, Umaa and Reynolds, Cormac and Ryder, Stuart D. and Schmidt, Brian and Soria, Roberto and Stairs, Ingrid H. and Tingay, Steven J. and Torkelsson, Ulf and Wagstaff, Kiri and Walker, Mark and Wayth, Randall B. and Williams, Peter K. G.},
	month = feb,
	year = {2013},
	pages = {e006},
}

@article{reichherzer_astro-colibri-coincidence_2021,
	title = {Astro-{COLIBRI}-{The} {COincidence} {LIBrary} for {Real}-time {Inquiry} for {Multimessenger} {Astrophysics}},
	volume = {256},
	issn = {0067-0049},
	url = {https://ui.adsabs.harvard.edu/abs/2021ApJS..256....5R/abstract},
	doi = {10.3847/1538-4365/ac1517},
	language = {en},
	number = {1},
	urldate = {2025-09-01},
	journal = {The Astrophysical Journal Supplement Series},
	author = {Reichherzer, P. and Schüssler, F. and Lefranc, V. and Yusafzai, A. and Alkan, A. K. and Ashkar, H. and Becker Tjus, J.},
	month = sep,
	year = {2021},
	pages = {5},
}

@article{frohmaier_tides_2025,
	title = {{TiDES}: {The} {4MOST} {Time} {Domain} {Extragalactic} {Survey}},
	shorttitle = {{TiDES}},
	url = {https://ui.adsabs.harvard.edu/abs/2025arXiv250116311F/abstract},
	doi = {10.48550/arXiv.2501.16311},
	language = {en},
	urldate = {2025-09-01},
	journal = {arXiv e-prints},
	author = {Frohmaier, C. and Vincenzi, M. and Sullivan, M. and Hönig, S. F. and Smith, M. and Addison, H. and Collett, T. and Dimitriadis, G. and Ellis, R. S. and Gandhi, P. and Graur, O. and Hook, I. and Kelsey, L. and Kim, Y. L. and Lidman, C. and Maguire, K. and Makrygianni, L. and Martin, B. and Möller, A. and Nichol, R. C. and Nicholl, M. and Schady, P. and Simmons, B. D. and Smartt, S. J. and Tempel, E. and Wiseman, P. and Collaboration, the LSST Dark Energy Science},
	month = jan,
	year = {2025},
	pages = {arXiv:2501.16311},
}

@misc{makrygianni_at_2023,
	title = {{AT} 2021loi: {A} {Bowen} {Fluorescence} {Flare} with a {Rebrightening} {Episode}, {Occurring} in a {Previously}-{Known} {AGN}},
	shorttitle = {{AT} 2021loi},
	url = {http://arxiv.org/abs/2305.01694},
	doi = {10.48550/arXiv.2305.01694},
	urldate = {2025-08-27},
	publisher = {arXiv},
	author = {Makrygianni, Lydia and Trakhtenbrot, Benny and Arcavi, Iair and Ricci, Claudio and Lam, Marco C. and Horesh, Assaf and Sfaradi, Itai and Bostroem, K. Azalee and Hosseinzadeh, Griffin and Howell, D. Andrew and Pellegrino, Craig and Fender, Rob and Green, David A. and Williams, David R. A. and Bright, Joe},
	month = may,
	year = {2023},
	note = {arXiv:2305.01694},
}

@article{stone_rates_2016,
	title = {Rates of stellar tidal disruption as probes of the supermassive black hole mass function},
	volume = {455},
	issn = {0035-8711},
	url = {https://ui.adsabs.harvard.edu/abs/2016MNRAS.455..859S/abstract},
	doi = {10.1093/mnras/stv2281},
	language = {en},
	number = {1},
	urldate = {2025-08-27},
	journal = {Monthly Notices of the Royal Astronomical Society},
	author = {Stone, Nicholas C. and Metzger, Brian D.},
	month = jan,
	year = {2016},
	pages = {859--883},
}

@misc{llamas_lanza_early_2025,
	title = {Early {Identification} of {Optical} {Tidal} {Disruption} {Events}: {A} science module for the {Fink} broker},
	shorttitle = {Early {Identification} of {Optical} {Tidal} {Disruption} {Events}},
	url = {http://arxiv.org/abs/2507.17499},
	doi = {10.48550/arXiv.2507.17499},
	urldate = {2025-07-24},
	publisher = {arXiv},
	author = {Llamas Lanza, Miguel and Karpov, Sergey and Russeil, Etienne and Quintin, Erwan and Ishida, Emille and Peloton, Julien and Pruzhinskaya, Maria and Möller, Anais},
	month = jul,
	year = {2025},
	note = {arXiv:2507.17499},
}

@article{merloni_tidal_2015,
	title = {A tidal disruption flare in a massive galaxy? {Implications} for the fuelling mechanisms of nuclear black holes},
	volume = {452},
	issn = {0035-8711},
	shorttitle = {A tidal disruption flare in a massive galaxy?},
	url = {https://ui.adsabs.harvard.edu/abs/2015MNRAS.452...69M/abstract},
	doi = {10.1093/mnras/stv1095},
	language = {en},
	number = {1},
	urldate = {2025-07-22},
	journal = {Monthly Notices of the Royal Astronomical Society},
	author = {Merloni, A. and Dwelly, T. and Salvato, M. and Georgakakis, A. and Greiner, J. and Krumpe, M. and Nandra, K. and Ponti, G. and Rau, A.},
	month = sep,
	year = {2015},
	pages = {69--87},
}

@article{komossa_discovery_2008,
	title = {Discovery of {Superstrong}, {Fading}, {Iron} {Line} {Emission} and {Double}-peaked {Balmer} {Lines} of the {Galaxy} {SDSS} {J095209}.56+214313.3: {The} {Light} {Echo} of a {Huge} {Flare}},
	volume = {678},
	issn = {0004-637X},
	shorttitle = {Discovery of {Superstrong}, {Fading}, {Iron} {Line} {Emission} and {Double}-peaked {Balmer} {Lines} of the {Galaxy} {SDSS} {J095209}.56+214313.3},
	url = {https://ui.adsabs.harvard.edu/abs/2008ApJ...678L..13K/abstract},
	doi = {10.1086/588281},
	language = {en},
	number = {1},
	urldate = {2025-07-22},
	journal = {The Astrophysical Journal},
	author = {Komossa, S. and Zhou, H. and Wang, T. and Ajello, M. and Ge, J. and Greiner, J. and Lu, H. and Salvato, M. and Saxton, R. and Shan, H. and Xu, D. and Yuan, W.},
	month = may,
	year = {2008},
	pages = {L13},
}

@article{russeil_identification_2025,
	title = {Identification of {AT2021aeuj} as a likely {Extreme} {Nuclear} {Transients} ({ENT})},
	volume = {202},
	url = {https://ui.adsabs.harvard.edu/abs/2025TNSAN.202....1R/abstract},
	language = {en},
	urldate = {2025-07-21},
	journal = {Transient Name Server AstroNote},
	author = {Russeil, E. and Quintin, E.},
	month = jul,
	year = {2025},
	pages = {1},
}

@article{kankare_population_2017,
	title = {A population of highly energetic transient events in the centres of active galaxies},
	volume = {1},
	copyright = {2017 The Author(s)},
	issn = {2397-3366},
	url = {https://www.nature.com/articles/s41550-017-0290-2},
	doi = {10.1038/s41550-017-0290-2},
	language = {en},
	number = {12},
	urldate = {2025-07-21},
	journal = {Nature Astronomy},
	author = {Kankare, E. and Kotak, R. and Mattila, S. and Lundqvist, P. and Ward, M. J. and Fraser, M. and Lawrence, A. and Smartt, S. J. and Meikle, W. P. S. and Bruce, A. and Harmanen, J. and Hutton, S. J. and Inserra, C. and Kangas, T. and Pastorello, A. and Reynolds, T. and Romero-Cañizales, C. and Smith, K. W. and Valenti, S. and Chambers, K. C. and Hodapp, K. W. and Huber, M. E. and Kaiser, N. and Kudritzki, R.-P. and Magnier, E. A. and Tonry, J. L. and Wainscoat, R. J. and Waters, C.},
	month = dec,
	year = {2017},
	pages = {865--871},
}

@misc{makrygianni_double_2025,
	title = {The {Double} {Tidal} {Disruption} {Event} {AT} 2022dbl {Implies} {That} at {Least} {Some} "{Standard}" {Optical} {TDEs} are {Partial} {Disruptions}},
	url = {http://arxiv.org/abs/2505.16867},
	doi = {10.48550/arXiv.2505.16867},
	urldate = {2025-05-23},
	publisher = {arXiv},
	author = {Makrygianni, Lydia and Arcavi, Iair and Newsome, Megan and Bandopadhyay, Ananya and Coughlin, Eric R. and Linial, Itai and Mockler, Brenna and Quataert, Eliot and Nixon, Chris and Godson, Benjamin and Pursiainen, Miika and Leloudas, Giorgos and French, K. Decker and Zitrin, Adi and Faris, Sara and Lam, Marco C. and Horesh, Assaf and Sfaradi, Itai and Fausnaugh, Michael and Ackley, Kendall and Andrews, Moira and Charalampopoulos, Panos and Davies, Benjamin D. R. and Dgany, Yael and Dyer, Martin J. and Farah, Joseph and Fender, Rob and Green, David A. and Howell, D. Andrew and Killestein, Thomas and Koivisto, Niilo and Lyman, Joseph and McCully, Curtis and Mitchell, Morgan A. and Gonzalez, Estefania Padilla and Rhodes, Lauren and Sahu, Anwesha and Terreran, Giacomo and Warwick, Ben},
	month = may,
	year = {2025},
	note = {arXiv:2505.16867},
}

@article{ho_search_2023,
	title = {A {Search} for {Extragalactic} {Fast} {Blue} {Optical} {Transients} in {ZTF} and the {Rate} of {AT2018cow}-like {Transients}},
	volume = {949},
	issn = {0004-637X},
	url = {https://ui.adsabs.harvard.edu/abs/2023ApJ...949..120H/abstract},
	doi = {10.3847/1538-4357/acc533},
	language = {en},
	number = {2},
	urldate = {2025-07-21},
	journal = {The Astrophysical Journal},
	author = {Ho, Anna Y. Q. and Perley, Daniel A. and Gal-Yam, Avishay and Lunnan, Ragnhild and Sollerman, Jesper and Schulze, Steve and Das, Kaustav K. and Dobie, Dougal and Yao, Yuhan and Fremling, Christoffer and Adams, Scott and Anand, Shreya and Andreoni, Igor and Bellm, Eric C. and Bruch, Rachel J. and Burdge, Kevin B. and Castro-Tirado, Alberto J. and Dahiwale, Aishwarya and De, Kishalay and Dekany, Richard and Drake, Andrew J. and Duev, Dmitry A. and Graham, Matthew J. and Helou, George and Kaplan, David L. and Karambelkar, Viraj and Kasliwal, Mansi M. and Kool, Erik C. and Kulkarni, S. R. and Mahabal, Ashish A. and Medford, Michael S. and Miller, A. A. and Nordin, Jakob and Ofek, Eran and Petitpas, Glen and Riddle, Reed and Sharma, Yashvi and Smith, Roger and Stewart, Adam J. and Taggart, Kirsty and Tartaglia, Leonardo and Tzanidakis, Anastasios and Winters, Jan Martin},
	month = jun,
	year = {2023},
	pages = {120},
}

@article{wiseman_systematically_2025,
	title = {A systematically selected sample of luminous, long-duration, ambiguous nuclear transients},
	volume = {537},
	issn = {0035-8711},
	url = {https://ui.adsabs.harvard.edu/abs/2025MNRAS.537.2024W/abstract},
	doi = {10.1093/mnras/staf116},
	language = {en},
	number = {2},
	urldate = {2025-07-21},
	journal = {Monthly Notices of the Royal Astronomical Society},
	author = {Wiseman, P. and Williams, R. D. and Arcavi, I. and Galbany, L. and Graham, M. J. and Hönig, S. and Newsome, M. and Subrayan, B. and Sullivan, M. and Wang, Y. and Ilić, D. and Nicholl, M. and Oates, S. and Petrushevska, T. and Smith, K. W.},
	month = feb,
	year = {2025},
	pages = {2024--2045},
}

@article{malyali_at_2021,
	title = {{AT} 2019avd: a novel addition to the diverse population of nuclear transients},
	volume = {647},
	copyright = {© A. Malyali et al. 2021},
	issn = {0004-6361, 1432-0746},
	shorttitle = {{AT} 2019avd},
	url = {https://www.aanda.org/articles/aa/abs/2021/03/aa39681-20/aa39681-20.html},
	doi = {10.1051/0004-6361/202039681},
	language = {en},
	urldate = {2025-07-18},
	journal = {Astronomy \& Astrophysics},
	author = {Malyali, A. and Rau, A. and Merloni, A. and Nandra, K. and Buchner, J. and Liu, Z. and Gezari, S. and Sollerman, J. and Shappee, B. and Trakhtenbrot, B. and Arcavi, I. and Ricci, C. and Velzen, S. van and Goobar, A. and Frederick, S. and Kawka, A. and Tartaglia, L. and Burke, J. and Hiramatsu, D. and Schramm, M. and Boom, D. van der and Anderson, G. and Miller-Jones, J. C. A. and Bellm, E. and Drake, A. and Duev, D. and Fremling, C. and Graham, M. and Masci, F. and Rusholme, B. and Soumagnac, M. and Walters, R.},
	month = mar,
	year = {2021},
	pages = {A9},
}

@article{campana_multiple_2015,
	title = {Multiple tidal disruption flares in the active galaxy {IC} 3599},
	volume = {581},
	copyright = {© ESO, 2015},
	issn = {0004-6361, 1432-0746},
	url = {https://www.aanda.org/articles/aa/abs/2015/09/aa25965-15/aa25965-15.html},
	doi = {10.1051/0004-6361/201525965},
	language = {en},
	urldate = {2025-07-18},
	journal = {Astronomy \& Astrophysics},
	author = {Campana, S. and Mainetti, D. and Colpi, M. and Lodato, G. and D’Avanzo, P. and Evans, P. A. and Moretti, A.},
	month = sep,
	year = {2015},
	pages = {A17},
}

@article{sillanpaa_oj_1988,
	title = {{OJ} 287: {Binary} {Pair} of {Supermassive} {Black} {Holes}},
	volume = {325},
	issn = {0004-637X},
	shorttitle = {{OJ} 287},
	url = {https://ui.adsabs.harvard.edu/abs/1988ApJ...325..628S/abstract},
	doi = {10.1086/166033},
	language = {en},
	urldate = {2025-07-18},
	journal = {The Astrophysical Journal},
	author = {Sillanpaa, A. and Haarala, S. and Valtonen, M. J. and Sundelius, B. and Byrd, G. G.},
	month = feb,
	year = {1988},
	pages = {628},
}

@article{hinkle_discovery_2021,
	title = {Discovery and follow-up of {ASASSN}-19dj: an {X}-ray and {UV} luminous {TDE} in an extreme post-starburst galaxy},
	volume = {500},
	issn = {0035-8711},
	shorttitle = {Discovery and follow-up of {ASASSN}-19dj},
	url = {https://ui.adsabs.harvard.edu/abs/2021MNRAS.500.1673H/abstract},
	doi = {10.1093/mnras/staa3170},
	language = {en},
	number = {2},
	urldate = {2025-07-17},
	journal = {Monthly Notices of the Royal Astronomical Society},
	author = {Hinkle, Jason T. and Holoien, T. W.-S. and Auchettl, K. and Shappee, B. J. and Neustadt, J. M. M. and Payne, A. V. and Brown, J. S. and Kochanek, C. S. and Stanek, K. Z. and Graham, M. J. and Tucker, M. A. and Do, A. and Anderson, J. P. and Bose, S. and Chen, P. and Coulter, D. A. and Dimitriadis, G. and Dong, Subo and Foley, R. J. and Huber, M. E. and Hung, T. and Kilpatrick, C. D. and Pignata, G. and Piro, A. L. and Rojas-Bravo, C. and Siebert, M. R. and Stalder, B. and Thompson, Todd A. and Tonry, J. L. and Vallely, P. J. and Wisniewski, J. P.},
	month = jan,
	year = {2021},
	pages = {1673--1696},
}

@article{yao_tidal_2023,
	title = {Tidal {Disruption} {Event} {Demographics} with the {Zwicky} {Transient} {Facility}: {Volumetric} {Rates}, {Luminosity} {Function}, and {Implications} for the {Local} {Black} {Hole} {Mass} {Function}},
	volume = {955},
	issn = {0004-637X},
	shorttitle = {Tidal {Disruption} {Event} {Demographics} with the {Zwicky} {Transient} {Facility}},
	url = {https://ui.adsabs.harvard.edu/abs/2023ApJ...955L...6Y/abstract},
	doi = {10.3847/2041-8213/acf216},
	language = {en},
	number = {1},
	urldate = {2025-06-11},
	journal = {The Astrophysical Journal},
	author = {Yao, Yuhan and Ravi, Vikram and Gezari, Suvi and van Velzen, Sjoert and Lu, Wenbin and Schulze, Steve and Somalwar, Jean J. and Kulkarni, S. R. and Hammerstein, Erica and Nicholl, Matt and Graham, Matthew J. and Perley, Daniel A. and Cenko, S. Bradley and Stein, Robert and Ricarte, Angelo and Chadayammuri, Urmila and Quataert, Eliot and Bellm, Eric C. and Bloom, Joshua S. and Dekany, Richard and Drake, Andrew J. and Groom, Steven L. and Mahabal, Ashish A. and Prince, Thomas A. and Riddle, Reed and Rusholme, Ben and Sharma, Yashvi and Sollerman, Jesper and Yan, Lin},
	month = sep,
	year = {2023},
	pages = {L6},
}

@misc{langis_repeating_2025,
	title = {Repeating {Flares}, {X}-ray {Outbursts} and {Delayed} {Infrared} {Emission}: {A} {Comprehensive} {Compilation} of {Optical} {Tidal} {Disruption} {Events}},
	shorttitle = {Repeating {Flares}, {X}-ray {Outbursts} and {Delayed} {Infrared} {Emission}},
	url = {https://arxiv.org/abs/2506.05476v1},
	language = {en},
	urldate = {2025-06-09},
	journal = {arXiv.org},
	author = {Langis, D. A. and Liodakis, I. and Koljonen, K. I. I. and Paggi, A. and Globus, N. and Wyrzykowski, L. and Mikołajczyk, P. J. and Kotysz, K. and Zieliński, P. and Ihanec, N. and Ding, J. and Morshed, D. and Torres, Z.},
	month = jun,
	year = {2025},
}

@article{hinkle_extreme_2024,
	title = {Extreme {Nuclear} {Transients} {Resulting} from the {Tidal} {Disruption} of {Intermediate} {Mass} {Stars}},
	url = {https://ui.adsabs.harvard.edu/abs/2024arXiv240508855H/abstract},
	doi = {10.48550/arXiv.2405.08855},
	language = {en},
	urldate = {2025-06-05},
	journal = {arXiv e-prints},
	author = {Hinkle, Jason T. and Shappee, Benjamin J. and Auchettl, Katie and Kochanek, Christopher S. and Neustadt, Jack M. M. and Polin, Abigail and Strader, Jay and Holoien, Thomas W.-S. and Huber, Mark E. and Tucker, Michael A. and Ashall, Christopher and de Jaeger, Thomas and Desai, Dhvanil D. and Do, Aaron and Hoogendam, Willem B. and Payne, Anna V.},
	month = may,
	year = {2024},
	pages = {arXiv:2405.08855},
}

@article{sun_recurring_2024,
	title = {Recurring tidal disruption events a decade apart in {IRAS} {F01004}-2237},
	volume = {692},
	issn = {0004-6361},
	url = {https://ui.adsabs.harvard.edu/abs/2024A&A...692A.262S/abstract},
	doi = {10.1051/0004-6361/202452380},
	language = {en},
	urldate = {2025-06-04},
	journal = {Astronomy and Astrophysics},
	author = {Sun, Luming and Jiang, Ning and Dou, Liming and Shu, Xinwen and Zhu, Jiazheng and Dong, Subo and Buckley, David and Bradley Cenko, S. and Fan, Xiaohui and Gromadzki, Mariusz and Liu, Zhu and Wang, Jianguo and Wang, Tinggui and Wang, Yibo and Wu, Tao and Yang, Lei and Zhang, Fabao and Zhang, Wenjie and Zhang, Xiaer},
	month = dec,
	year = {2024},
	pages = {A262},
}

@article{milan_veres_back_2024,
	title = {Back from the dead: {AT2019aalc} as a candidate repeating {TDE} in an {AGN}},
	shorttitle = {Back from the dead},
	url = {https://ui.adsabs.harvard.edu/abs/2024arXiv240817419M/abstract},
	doi = {10.48550/arXiv.2408.17419},
	language = {en},
	urldate = {2025-06-04},
	journal = {arXiv e-prints},
	author = {Milán Veres, Patrik and Franckowiak, Anna and van Velzen, Sjoert and Adebahr, Bjoern and Taziaux, Sam and Necker, Jannis and Stein, Robert and Kier, Alexander and Mueller, Ancla and Bomans, Dominik J. and Jordana-Mitjans, Nuria and Kowalski, Marek and Hammerstein, Erica and Marci-Boehncke, Elena and Reusch, Simeon and Garrappa, Simone and Rose, Sam and Kashyap Das, Kaustav},
	month = aug,
	year = {2024},
	pages = {arXiv:2408.17419},
}

@article{sun_at2021aeuk_2025,
	title = {{AT2021aeuk}: {A} {Repeating} {Partial} {Tidal} {Disruption} {Event} {Candidate} in a {Narrow}-line {Seyfert} 1 {Galaxy}},
	volume = {982},
	issn = {0004-637X},
	shorttitle = {{AT2021aeuk}},
	url = {https://ui.adsabs.harvard.edu/abs/2025ApJ...982..150S/abstract},
	doi = {10.3847/1538-4357/adb724},
	language = {en},
	number = {2},
	urldate = {2025-06-04},
	journal = {The Astrophysical Journal},
	author = {Sun, Jingbo and Guo, Hengxiao and Gu, Minfeng and Li, Ya-Ping and Chen, Yongjun and González-Buitrago, D. and Wang, Jian-Guo and Li, Sha-Sha and Feng, Hai-Cheng and Xiong, Dingrong and Wang, Yanan and Yuan, Qi and Jin, Jun-jie and Zhang, Wenda and Deng, Hongping and Zhang, Minghao},
	month = apr,
	year = {2025},
	pages = {150},
}

@article{pessi_spectroscopic_2024,
	title = {Spectroscopic classification of {ZTF} transients},
	volume = {250},
	url = {https://ui.adsabs.harvard.edu/abs/2024TNSAN.250....1P/abstract},
	language = {en},
	urldate = {2025-06-02},
	journal = {Transient Name Server AstroNote},
	author = {Pessi, P. and Lunnan, R. and Sollerman, J. and Schulze, S. and Gkini, A. and Gangopadhyay, A. and Yan, L. and Gal-Yam, A. and Perley, D. A. and Chen, T. W. and Hinds, K. R. and Brennan, S. J. and Hu, Y. and Singh, A. and Andreoni, I. and Cook, D. O. and Fremling, C. and Ho, A. Y. Q. and Sharma, Y. and Velzen, S. V. and Wold, A. and Bellm, E. C. and Bloom, J. S. and Graham, M. J. and Kasliwal, M. M. and Kulkarni, S. R. and Riddle, R. and Rusholme, B.},
	month = sep,
	year = {2024},
	pages = {1},
}

@article{pessi_sample_2025,
	title = {Sample of hydrogen-rich superluminous supernovae from the {Zwicky} {Transient} {Facility}},
	volume = {695},
	copyright = {© The Authors 2025},
	issn = {0004-6361, 1432-0746},
	url = {https://www.aanda.org/articles/aa/abs/2025/03/aa52014-24/aa52014-24.html},
	doi = {10.1051/0004-6361/202452014},
	language = {en},
	urldate = {2025-06-02},
	journal = {Astronomy \& Astrophysics},
	author = {Pessi, P. J. and Lunnan, R. and Sollerman, J. and Schulze, S. and Gkini, A. and Gangopadhyay, A. and Yan, L. and Gal-Yam, A. and Perley, D. A. and Chen, T.-W. and Hinds, K. R. and Brennan, S. J. and Hu, Y. and Singh, A. and Andreoni, I. and Cook, D. O. and Fremling, C. and Ho, A. Y. Q. and Sharma, Y. and Velzen, S. van and Kangas, T. and Wold, A. and Bellm, E. C. and Bloom, J. S. and Graham, M. J. and Kasliwal, M. M. and Kulkarni, S. R. and Riddle, R. and Rusholme, B.},
	month = mar,
	year = {2025},
	pages = {A142},
}

@article{pavez-herrera_alerce_2025,
	title = {{ALeRCE} light curve classifier: {Tidal} disruption event expansion pack},
	volume = {696},
	copyright = {© The Authors 2025},
	issn = {0004-6361, 1432-0746},
	shorttitle = {{ALeRCE} light curve classifier},
	url = {https://www.aanda.org/articles/aa/abs/2025/04/aa51951-24/aa51951-24.html},
	doi = {10.1051/0004-6361/202451951},
	language = {en},
	urldate = {2025-06-02},
	journal = {Astronomy \& Astrophysics},
	author = {Pavez-Herrera, M. and Sánchez-Sáez, P. and Hernández-García, L. and Bauer, F. E. and Förster, F. and Catelan, M. and Arancibia, A. Muñoz and Ricci, C. and Reyes-Jainaga, I. and Bayo, A. and Huijse, P. and Cabrera-Vives, G.},
	month = apr,
	year = {2025},
	pages = {A153},
}

@article{duncan_all-purpose_2022,
	title = {All-purpose, all-sky photometric redshifts for the {Legacy} {Imaging} {Surveys} {Data} {Release} 8},
	volume = {512},
	issn = {0035-8711},
	url = {https://ui.adsabs.harvard.edu/abs/2022MNRAS.512.3662D/abstract},
	doi = {10.1093/mnras/stac608},
	language = {en},
	number = {3},
	urldate = {2025-06-02},
	journal = {Monthly Notices of the Royal Astronomical Society},
	author = {Duncan, Kenneth J.},
	month = may,
	year = {2022},
	pages = {3662--3683},
}

@article{dalen_epessto_2024,
	title = {{ePESSTO}+ {Transient} {Classification} {Report} for 2024-04-10},
	volume = {2024-1012},
	url = {https://ui.adsabs.harvard.edu/abs/2024TNSCR1012....1D/abstract},
	language = {en},
	urldate = {2025-05-30},
	journal = {Transient Name Server Classification Report},
	author = {Dalen, J. V. and Hoof, A. V. and Shlentsova, A. and Fraser, M. and Yaron, O.},
	month = apr,
	year = {2024},
	pages = {1},
}

@article{chen_fate_2024,
	title = {Fate of the {Remnant} in {Tidal} {Stripping} {Event}: {Repeating} and {Nonrepeating}},
	volume = {977},
	issn = {0004-637X},
	shorttitle = {Fate of the {Remnant} in {Tidal} {Stripping} {Event}},
	url = {https://ui.adsabs.harvard.edu/abs/2024ApJ...977...80C/abstract},
	doi = {10.3847/1538-4357/ad8b24},
	language = {en},
	number = {1},
	urldate = {2025-05-30},
	journal = {The Astrophysical Journal},
	author = {Chen, Jin-Hong and Dai, Lixin and Liu, Shang-Fei and Ou, Jian-Wen},
	month = dec,
	year = {2024},
	pages = {80},
}

@article{lin_unluckiest_2024,
	title = {The {Unluckiest} {Star}: {A} {Spectroscopically} {Confirmed} {Repeated} {Partial} {Tidal} {Disruption} {Event} {AT} 2022dbl},
	volume = {971},
	issn = {0004-637X},
	shorttitle = {The {Unluckiest} {Star}},
	url = {https://ui.adsabs.harvard.edu/abs/2024ApJ...971L..26L/abstract},
	doi = {10.3847/2041-8213/ad638e},
	language = {en},
	number = {1},
	urldate = {2025-05-30},
	journal = {The Astrophysical Journal},
	author = {Lin, Zheyu and Jiang, Ning and Wang, Tinggui and Kong, Xu and Li, Dongyue and He, Han and Wang, Yibo and Zhu, Jiazheng and Li, Wentao and Jiang, Ji-an and Singh, Avinash and Teja, Rishabh Singh and Sahu, D. K. and Jin, Chichuan and Maeda, Keiichi and Huang, Shifeng},
	month = aug,
	year = {2024},
	pages = {L26},
}

@article{liu_repeating_2025,
	title = {Repeating {Partial} {Tidal} {Encounters} of {Sun}-like {Stars} {Leading} to {Their} {Complete} {Disruption}},
	volume = {979},
	issn = {0004-637X},
	url = {https://ui.adsabs.harvard.edu/abs/2025ApJ...979...40L/abstract},
	doi = {10.3847/1538-4357/ad9b0b},
	language = {en},
	number = {1},
	urldate = {2025-05-30},
	journal = {The Astrophysical Journal},
	author = {Liu, Chang and Yarza, Ricardo and Ramirez-Ruiz, Enrico},
	month = jan,
	year = {2025},
	pages = {40},
}

@article{hernandez-garcia_discovery_2025,
	title = {Discovery of extreme quasi-periodic eruptions in a newly accreting massive black hole},
	copyright = {2025 The Author(s), under exclusive licence to Springer Nature Limited},
	issn = {2397-3366},
	url = {https://arxiv.org/abs/2504.07169},
	doi = {10.1038/s41550-025-02523-9},
	language = {en},
	urldate = {2025-05-23},
	journal = {Nature Astronomy},
	author = {Hernández-García, Lorena and Chakraborty, Joheen and Sánchez-Sáez, Paula and Ricci, Claudio and Cuadra, Jorge and McKernan, Barry and Ford, K. E. Saavik and Arévalo, Patricia and Rau, Arne and Arcodia, Riccardo and Kara, Erin and Liu, Zhu and Merloni, Andrea and Bruni, Gabriele and Goodwin, Adelle and Arzoumanian, Zaven and Assef, Roberto J. and Baldini, Pietro and Bayo, Amelia and Bauer, Franz E. and Bernal, Santiago and Brightman, Murray and Calistro Rivera, Gabriela and Gendreau, Keith and Homan, David and Krumpe, Mirko and Lira, Paulina and Martínez-Aldama, Mary Loli and Salvato, Mara and Sotomayor, Belén},
	month = apr,
	year = {2025},
	pages = {1--12},
}

@article{sanchez-saez_sdss13350728_2024,
	title = {{SDSS1335}+0728: {The} awakening of a ∼106 {M}⊙ black hole},
	volume = {688},
	copyright = {© The Authors 2024},
	issn = {0004-6361, 1432-0746},
	shorttitle = {{SDSS1335}+0728},
	url = {https://www.aanda.org/articles/aa/abs/2024/08/aa47957-23/aa47957-23.html},
	doi = {10.1051/0004-6361/202347957},
	language = {en},
	urldate = {2025-04-25},
	journal = {Astronomy \& Astrophysics},
	author = {Sánchez-Sáez, P. and Hernández-García, L. and Bernal, S. and Bayo, A. and Rivera, G. Calistro and Bauer, F. E. and Ricci, C. and Merloni, A. and Graham, M. J. and Cartier, R. and Arévalo, P. and Assef, R. J. and Concas, A. and Homan, D. and Krumpe, M. and Lira, P. and Malyali, A. and Martínez-Aldama, M. L. and Arancibia, A. M. Muñoz and Rau, A. and Bruni, G. and Förster, F. and Pavez-Herrera, M. and Tubín-Arenas, D. and Brightman, M.},
	month = aug,
	year = {2024},
	pages = {A157},
}

@article{magee_search_2023,
	title = {A search for gravitationally lensed supernovae within the {Zwicky} transient facility public survey},
	volume = {525},
	issn = {0035-8711},
	url = {https://ui.adsabs.harvard.edu/abs/2023MNRAS.525..542M/abstract},
	doi = {10.1093/mnras/stad2263},
	language = {en},
	number = {1},
	urldate = {2025-04-22},
	journal = {Monthly Notices of the Royal Astronomical Society},
	author = {Magee, M. R. and Sainz de Murieta, A. and Collett, T. E. and Enzi, W.},
	month = oct,
	year = {2023},
	pages = {542--560},
}

@article{van_velzen_late-time_2019,
	title = {Late-time {UV} {Observations} of {Tidal} {Disruption} {Flares} {Reveal} {Unobscured}, {Compact} {Accretion} {Disks}},
	volume = {878},
	issn = {0004-637X},
	url = {https://ui.adsabs.harvard.edu/abs/2019ApJ...878...82V/abstract},
	doi = {10.3847/1538-4357/ab1844},
	language = {en},
	number = {2},
	urldate = {2025-04-22},
	journal = {The Astrophysical Journal},
	author = {van Velzen, Sjoert and Stone, Nicholas C. and Metzger, Brian D. and Gezari, Suvi and Brown, Thomas M. and Fruchter, Andrew S.},
	month = jun,
	year = {2019},
	pages = {82},
}

@article{trakhtenbrot_new_2019,
	title = {A new class of flares from accreting supermassive black holes},
	volume = {3},
	issn = {2397-3366},
	url = {https://ui.adsabs.harvard.edu/abs/2019NatAs...3..242T/abstract},
	doi = {10.1038/s41550-018-0661-3},
	language = {en},
	urldate = {2025-04-15},
	journal = {Nature Astronomy},
	author = {Trakhtenbrot, Benny and Arcavi, Iair and Ricci, Claudio and Tacchella, Sandro and Stern, Daniel and Netzer, Hagai and Jonker, Peter G. and Horesh, Assaf and Mejía-Restrepo, Julián Esteban and Hosseinzadeh, Griffin and Hallefors, Valentina and Howell, D. Andrew and McCully, Curtis and Baloković, Mislav and Heida, Marianne and Kamraj, Nikita and Lansbury, George Benjamin and Wyrzykowski, Łukasz and Gromadzki, Mariusz and Hamanowicz, Aleksandra and Cenko, S. Bradley and Sand, David J. and Hsiao, Eric Y. and Phillips, Mark M. and Diamond, Tiara R. and Kara, Erin and Gendreau, Keith C. and Arzoumanian, Zaven and Remillard, Ron},
	month = jan,
	year = {2019},
	pages = {242--250},
}

@article{condon_nrao_1998,
	title = {The {NRAO} {VLA} {Sky} {Survey}},
	volume = {115},
	issn = {0004-6256},
	url = {https://ui.adsabs.harvard.edu/abs/1998AJ....115.1693C/abstract},
	doi = {10.1086/300337},
	language = {en},
	number = {5},
	urldate = {2025-04-15},
	journal = {The Astronomical Journal},
	author = {Condon, J. J. and Cotton, W. D. and Greisen, E. W. and Yin, Q. F. and Perley, R. A. and Taylor, G. B. and Broderick, J. J.},
	month = may,
	year = {1998},
	pages = {1693--1716},
}

@article{adrian-martinez_letter_2016,
	title = {Letter of intent for {KM3NeT} 2.0},
	volume = {43},
	issn = {0954-3899},
	url = {https://ui.adsabs.harvard.edu/abs/2016JPhG...43h4001A},
	doi = {10.1088/0954-3899/43/8/084001},
	urldate = {2025-01-08},
	journal = {Journal of Physics G Nuclear Physics},
	author = {Adrián-Martínez, S. and Ageron, M. and Aharonian, F. and Aiello, S. and Albert, A. and Ameli, F. and Anassontzis, E. and Andre, M. and Androulakis, G. and Anghinolfi, M. and Anton, G. and Ardid, M. and Avgitas, T. and Barbarino, G. and Barbarito, E. and Baret, B. and Barrios-Martí, J. and Belhorma, B. and Belias, A. and Berbee, E. and van den Berg, A. and Bertin, V. and Beurthey, S. and van Beveren, V. and Beverini, N. and Biagi, S. and Biagioni, A. and Billault, M. and Bondì, M. and Bormuth, R. and Bouhadef, B. and Bourlis, G. and Bourret, S. and Boutonnet, C. and Bouwhuis, M. and Bozza, C. and Bruijn, R. and Brunner, J. and Buis, E. and Busto, J. and Cacopardo, G. and Caillat, L. and Calamai, M. and Calvo, D. and Capone, A. and Caramete, L. and Cecchini, S. and Celli, S. and Champion, C. and Cherkaoui El Moursli, R. and Cherubini, S. and Chiarusi, T. and Circella, M. and Classen, L. and Cocimano, R. and Coelho, J. A. B. and Coleiro, A. and Colonges, S. and Coniglione, R. and Cordelli, M. and Cosquer, A. and Coyle, P. and Creusot, A. and Cuttone, G. and D'Amico, A. and De Bonis, G. and De Rosa, G. and De Sio, C. and Di Capua, F. and Di Palma, I. and Díaz García, A. F. and Distefano, C. and Donzaud, C. and Dornic, D. and Dorosti-Hasankiadeh, Q. and Drakopoulou, E. and Drouhin, D. and Drury, L. and Durocher, M. and Eberl, T. and Eichie, S. and van Eijk, D. and El Bojaddaini, I. and El Khayati, N. and Elsaesser, D. and Enzenhöfer, A. and Fassi, F. and Favali, P. and Fermani, P. and Ferrara, G. and Filippidis, C. and Frascadore, G. and Fusco, L. A. and Gal, T. and Galatà, S. and Garufi, F. and Gay, P. and Gebyehu, M. and Giordano, V. and Gizani, N. and Gracia, R. and Graf, K. and Grégoire, T. and Grella, G. and Habel, R. and Hallmann, S. and van Haren, H. and Harissopulos, S. and Heid, T. and Heijboer, A. and Heine, E. and Henry, S. and Hernández-Rey, J. J. and Hevinga, M. and Hofestädt, J. and Hugon, C. M. F. and Illuminati, G. and James, C. W. and Jansweijer, P. and Jongen, M. and de Jong, M. and Kadler, M. and Kalekin, O. and Kappes, A. and Katz, U. F. and Keller, P. and Kieft, G. and Kießling, D. and Koffeman, E. N. and Kooijman, P. and Kouchner, A. and Kulikovskiy, V. and Lahmann, R. and Lamare, P. and Leisos, A. and Leonora, E. and Clark, M. Lindsey and Liolios, A. and Llorens Alvarez, C. D. and Lo Presti, D. and Löhner, H. and Lonardo, A. and Lotze, M. and Loucatos, S. and Maccioni, E. and Mannheim, K. and Margiotta, A. and Marinelli, A. and Mariş, O. and Markou, C. and Martínez-Mora, J. A. and Martini, A. and Mele, R. and Melis, K. W. and Michael, T. and Migliozzi, P. and Migneco, E. and Mijakowski, P. and Miraglia, A. and Mollo, C. M. and Mongelli, M. and Morganti, M. and Moussa, A. and Musico, P. and Musumeci, M. and Navas, S. and Nicolau, C. A. and Olcina, I. and Olivetto, C. and Orlando, A. and Papaikonomou, A. and Papaleo, R. and Păvălaş, G. E. and Peek, H. and Pellegrino, C. and Perrina, C. and Pfutzner, M. and Piattelli, P. and Pikounis, K. and Poma, G. E. and Popa, V. and Pradier, T. and Pratolongo, F. and Pühlhofer, G. and Pulvirenti, S. and Quinn, L. and Racca, C. and Raffaelli, F. and Randazzo, N. and Rapidis, P. and Razis, P. and Real, D. and Resvanis, L. and Reubelt, J. and Riccobene, G. and Rossi, C. and Rovelli, A. and Saldaña, M. and Salvadori, I. and Samtleben, D. F. E.},
	month = aug,
	year = {2016},
	note = {ADS Bibcode: 2016JPhG...43h4001A},
	pages = {084001},
}

@article{russeil_identification_2024,
	title = {Identification of {AT} 2020ukj as an uniquely slow {TDE}},
	volume = {379},
	url = {https://ui.adsabs.harvard.edu/abs/2024TNSAN.379....1R},
	urldate = {2025-01-07},
	journal = {Transient Name Server AstroNote},
	author = {Russeil, E. and Quintin, E. and Lanza, M. L. and Ishida, E. and Peloton, J. and Karpov, S. and Pruzhinskaya, M. V. and S, Z. and A, D. and A, B.},
	month = dec,
	year = {2024},
	note = {ADS Bibcode: 2024TNSAN.379....1R},
	pages = {1},
}

@article{mummery_fundamental_2024,
	title = {Fundamental scaling relationships revealed in the optical light curves of tidal disruption events},
	volume = {527},
	issn = {0035-8711},
	url = {https://ui.adsabs.harvard.edu/abs/2024MNRAS.527.2452M},
	doi = {10.1093/mnras/stad3001},
	urldate = {2024-12-11},
	journal = {Monthly Notices of the Royal Astronomical Society},
	author = {Mummery, Andrew and van Velzen, Sjoert and Nathan, Edward and Ingram, Adam and Hammerstein, Erica and Fraser-Taliente, Ludovic and Balbus, Steven},
	month = jan,
	year = {2024},
	note = {ADS Bibcode: 2024MNRAS.527.2452M},
	pages = {2452--2489},
}

@article{quintin_type_2024,
	title = {Type {Ia} {Supernova} detected by {Fink} {TDE} module},
	volume = {176},
	url = {https://ui.adsabs.harvard.edu/abs/2024TNSAN.176....1Q},
	urldate = {2024-11-21},
	journal = {Transient Name Server AstroNote},
	author = {Quintin, E. and Lanza, M. L. and Russeil, E. and Ishida, E. and Peloton, J. and Karpov, S. and Pruzhinskaya, M. V. and Tatarnikov, A. A. and Dodin, A. V. and Belinski, A. A. and Shatski, N. I.},
	month = jul,
	year = {2024},
	note = {ADS Bibcode: 2024TNSAN.176....1Q},
	pages = {1},
}

@article{llamas_lanza_identification_2024,
	title = {Identification of {AT} 2023adr as a candidate repeated partial {TDE}},
	volume = {178},
	url = {https://ui.adsabs.harvard.edu/abs/2024TNSAN.178....1L},
	urldate = {2024-11-21},
	journal = {Transient Name Server AstroNote},
	author = {Llamas Lanza, M. and Quintin, E. and Russeil, E. and Ishida, E. and Peloton, J. and Karpov, S. and Pruzhinskaya, M. V.},
	month = jul,
	year = {2024},
	note = {ADS Bibcode: 2024TNSAN.178....1L},
	pages = {1},
}

@article{quintin_stonks_2024,
	title = {{STONKS}: {Quasi}-real time {XMM}-{Newton} transient detection system},
	volume = {687},
	issn = {0004-6361},
	shorttitle = {{STONKS}},
	url = {https://ui.adsabs.harvard.edu/abs/2024A&A...687A.250Q},
	doi = {10.1051/0004-6361/202348317},
	urldate = {2024-10-28},
	journal = {Astronomy and Astrophysics},
	author = {Quintin, E. and Webb, N. A. and Georgantopoulos, I. and Gupta, M. and Kammoun, E. and Michel, L. and Schwope, A. and Tranin, H. and Traulsen, I.},
	month = jul,
	year = {2024},
	note = {ADS Bibcode: 2024A\&A...687A.250Q},
	pages = {A250},
}

@misc{russeil_rainbow_2023,
	title = {Rainbow: a colorful approach on multi-passband light curve estimation},
	shorttitle = {Rainbow},
	url = {http://arxiv.org/abs/2310.02916},
	doi = {10.48550/arXiv.2310.02916},
	urldate = {2024-09-18},
	publisher = {arXiv},
	author = {Russeil, E. and Malanchev, K. L. and Aleo, P. D. and Ishida, E. E. O. and Pruzhinskaya, M. V. and Gangler, E. and Lavrukhina, A. D. and Volnova, A. A. and Voloshina, A. and Semenikhin, T. and Sreejith, S. and Kornilov, M. V. and Korolev, V. S.},
	month = oct,
	year = {2023},
	note = {arXiv:2310.02916 [astro-ph, physics:physics]},
}

@article{malyali_erasst_2023,
	title = {{eRASSt} {J074426}.3 + 291606: prompt accretion disc formation in a 'faint and slow' tidal disruption event},
	volume = {520},
	issn = {0035-8711},
	shorttitle = {{eRASSt} {J074426}.3 + 291606},
	url = {https://ui.adsabs.harvard.edu/abs/2023MNRAS.520.4209M},
	doi = {10.1093/mnras/stad046},
	urldate = {2024-05-17},
	journal = {Monthly Notices of the Royal Astronomical Society},
	author = {Malyali, A. and Liu, Z. and Merloni, A. and Rau, A. and Buchner, J. and Ciroi, S. and Di Mille, F. and Grotova, I. and Dwelly, T. and Nandra, K. and Salvato, M. and Homan, D. and Krumpe, M.},
	month = apr,
	year = {2023},
	note = {Publisher: OUP
ADS Bibcode: 2023MNRAS.520.4209M},
	pages = {4209--4225},
}

@article{evans_real-time_2023,
	title = {A real-time transient detector and the living {Swift}-{XRT} point source catalogue},
	volume = {518},
	issn = {0035-8711},
	url = {https://ui.adsabs.harvard.edu/abs/2023MNRAS.518..174E},
	doi = {10.1093/mnras/stac2937},
	urldate = {2024-05-17},
	journal = {Monthly Notices of the Royal Astronomical Society},
	author = {Evans, P. A. and Page, K. L. and Beardmore, A. P. and Eyles-Ferris, R. A. J. and Osborne, J. P. and Campana, S. and Kennea, J. A. and Cenko, S. B.},
	month = jan,
	year = {2023},
	note = {Publisher: OUP
ADS Bibcode: 2023MNRAS.518..174E},
	pages = {174--184},
}

@article{ricci_changing-look_2023,
	title = {Changing-look active galactic nuclei},
	volume = {7},
	copyright = {2023 Springer Nature Limited},
	issn = {2397-3366},
	url = {https://www.nature.com/articles/s41550-023-02108-4},
	doi = {10.1038/s41550-023-02108-4},
	language = {en},
	number = {11},
	urldate = {2024-02-05},
	journal = {Nature Astronomy},
	author = {Ricci, Claudio and Trakhtenbrot, Benny},
	month = nov,
	year = {2023},
	note = {Number: 11
Publisher: Nature Publishing Group},
	pages = {1282--1294},
}

@misc{masterson_new_2024,
	title = {A {New} {Population} of {Mid}-{Infrared}-{Selected} {Tidal} {Disruption} {Events}: {Implications} for {Tidal} {Disruption} {Event} {Rates} and {Host} {Galaxy} {Properties}},
	shorttitle = {A {New} {Population} of {Mid}-{Infrared}-{Selected} {Tidal} {Disruption} {Events}},
	url = {http://arxiv.org/abs/2401.01403},
	urldate = {2024-01-04},
	publisher = {arXiv},
	author = {Masterson, Megan and De, Kishalay and Panagiotou, Christos and Kara, Erin and Arcavi, Iair and Eilers, Anna-Christina and Frostig, Danielle and Gezari, Suvi and Grotova, Iuliia and Liu, Zhu and Malyali, Adam and Meisner, Aaron M. and Merloni, Andrea and Newsome, Megan and Rau, Arne and Simcoe, Robert A. and van Velzen, Sjoert},
	month = jan,
	year = {2024},
	note = {arXiv:2401.01403 [astro-ph]},
}

@article{bianchi_revised_2017,
	title = {Revised {Catalog} of {GALEX} {Ultraviolet} {Sources}. {I}. {The} {All}-{Sky} {Survey}: {GUVcat}\_AIS},
	volume = {230},
	issn = {0067-0049},
	shorttitle = {Revised {Catalog} of {GALEX} {Ultraviolet} {Sources}. {I}. {The} {All}-{Sky} {Survey}},
	url = {https://ui.adsabs.harvard.edu/abs/2017ApJS..230...24B},
	doi = {10.3847/1538-4365/aa7053},
	urldate = {2024-01-03},
	journal = {The Astrophysical Journal Supplement Series},
	author = {Bianchi, Luciana and Shiao, Bernie and Thilker, David},
	month = jun,
	year = {2017},
	note = {ADS Bibcode: 2017ApJS..230...24B},
	pages = {24},
}

@misc{somalwar_first_2023,
	title = {The first systematically identified repeating partial tidal disruption event},
	url = {http://arxiv.org/abs/2310.03782},
	language = {en},
	urldate = {2023-10-09},
	publisher = {arXiv},
	author = {Somalwar, Jean J. and Ravi, Vikram and Yao, Yuhan and Guolo, Muriel and Graham, Matthew and Hammerstein, Erica and Lu, Wenbin and Nicholl, Matt and Sharma, Yashvi and Stein, Robert and van Velzen, Sjoert and Bellm, Eric C. and Coughlin, Michael W. and Groom, Steven L. and Masci, Frank J. and Riddle, Reed},
	month = oct,
	year = {2023},
	note = {arXiv:2310.03782 null},
}

@article{webb_xmm2athena_2023,
	title = {{XMM2ATHENA}, the {H2020} project to improve {XMM}-{Newton} analysis software and prepare for {Athena}},
	volume = {344},
	copyright = {© 2023 Wiley-VCH GmbH},
	issn = {1521-3994},
	url = {https://onlinelibrary.wiley.com/doi/abs/10.1002/asna.20220102},
	doi = {10.1002/asna.20220102},
	language = {en},
	number = {7},
	urldate = {2023-09-28},
	journal = {Astronomische Nachrichten},
	author = {Webb, Natalie A. and Carrera, Francisco J. and Schwope, Axel and Motch, Christian and Ballet, Jean and Watson, Mike and Page, Mat and Freyberg, Michael and Georgantopoulos, Ioannis and Coriat, Mickael and Barret, Didier and Massida, Zoe and Gupta, Maitrayee and Tranin, Hugo and Quintin, Erwan and Teresa Ceballos, M. and Mateos, Silvia and Corral, Amalia and Dominguez, Rosa and Stiele, Holger and Traulsen, Iris and Pires, Adriana and Nebot, Ada and Michel, Laurent and Xavier Pineau, François and Kuuttila, Jere and Maggi, Pierre and Chakroborty, Sudip and Birchall, Keir and Kuin, Paul and Akylas, Athanassios and Ruiz, Angel and Pouliasis, Ektoras and Georgakakis, Antonis},
	year = {2023},
	note = {\_eprint: https://onlinelibrary.wiley.com/doi/pdf/10.1002/asna.20220102},
	pages = {e220102},
}

@article{gaia_collaboration_gaia_2023,
	title = {Gaia {Data} {Release} 3. {The} extragalactic content},
	volume = {674},
	issn = {0004-6361},
	url = {https://ui.adsabs.harvard.edu/abs/2023A&A...674A..41G},
	doi = {10.1051/0004-6361/202243232},
	urldate = {2023-09-12},
	journal = {Astronomy and Astrophysics},
	author = {{Gaia Collaboration} and Bailer-Jones, C. A. L. and Teyssier, D. and Delchambre, L. and Ducourant, C. and Garabato, D. and Hatzidimitriou, D. and Klioner, S. A. and Rimoldini, L. and Bellas-Velidis, I. and Carballo, R. and Carnerero, M. I. and Diener, C. and Fouesneau, M. and Galluccio, L. and Gavras, P. and Krone-Martins, A. and Raiteri, C. M. and Teixeira, R. and Brown, A. G. A. and Vallenari, A. and Prusti, T. and de Bruijne, J. H. J. and Arenou, F. and Babusiaux, C. and Biermann, M. and Creevey, O. L. and Evans, D. W. and Eyer, L. and Guerra, R. and Hutton, A. and Jordi, C. and Lammers, U. L. and Lindegren, L. and Luri, X. and Mignard, F. and Panem, C. and Pourbaix, D. and Randich, S. and Sartoretti, P. and Soubiran, C. and Tanga, P. and Walton, N. A. and Bastian, U. and Drimmel, R. and Jansen, F. and Katz, D. and Lattanzi, M. G. and van Leeuwen, F. and Bakker, J. and Cacciari, C. and Castañeda, J. and De Angeli, F. and Fabricius, C. and Frémat, Y. and Guerrier, A. and Heiter, U. and Masana, E. and Messineo, R. and Mowlavi, N. and Nicolas, C. and Nienartowicz, K. and Pailler, F. and Panuzzo, P. and Riclet, F. and Roux, W. and Seabroke, G. M. and Sordo, R. and Thévenin, F. and Gracia-Abril, G. and Portell, J. and Altmann, M. and Andrae, R. and Audard, M. and Benson, K. and Berthier, J. and Blomme, R. and Burgess, P. W. and Busonero, D. and Busso, G. and Cánovas, H. and Carry, B. and Cellino, A. and Cheek, N. and Clementini, G. and Damerdji, Y. and Davidson, M. and de Teodoro, P. and Nuñez Campos, M. and Dell'Oro, A. and Esquej, P. and Fernández-Hernández, J. and Fraile, E. and García-Lario, P. and Gosset, E. and Haigron, R. and Halbwachs, J. -L. and Hambly, N. C. and Harrison, D. L. and Hernández, J. and Hestroffer, D. and Hodgkin, S. T. and Holl, B. and Janßen, K. and Jevardat de Fombelle, G. and Jordan, S. and Lanzafame, A. C. and Löffler, W. and Marchal, O. and Marrese, P. M. and Moitinho, A. and Muinonen, K. and Osborne, P. and Pancino, E. and Pauwels, T. and Recio-Blanco, A. and Reylé, C. and Riello, M. and Roegiers, T. and Rybizki, J. and Sarro, L. M. and Siopis, C. and Smith, M. and Sozzetti, A. and Utrilla, E. and van Leeuwen, M. and Abbas, U. and Ábrahám, P. and Abreu Aramburu, A. and Aerts, C. and Aguado, J. J. and Ajaj, M. and Aldea-Montero, F. and Altavilla, G. and Álvarez, M. A. and Alves, J. and Anderson, R. I. and Anglada Varela, E. and Antoja, T. and Baines, D. and Baker, S. G. and Balaguer-Núñez, L. and Balbinot, E. and Balog, Z. and Barache, C. and Barbato, D. and Barros, M. and Barstow, M. A. and Bartolomé, S. and Bassilana, J. -L. and Bauchet, N. and Becciani, U. and Bellazzini, M. and Berihuete, A. and Bernet, M. and Bertone, S. and Bianchi, L. and Binnenfeld, A. and Blanco-Cuaresma, S. and Boch, T. and Bombrun, A. and Bossini, D. and Bouquillon, S. and Bragaglia, A. and Bramante, L. and Breedt, E. and Bressan, A. and Brouillet, N. and Brugaletta, E. and Bucciarelli, B. and Burlacu, A. and Butkevich, A. G. and Buzzi, R. and Caffau, E. and Cancelliere, R. and Cantat-Gaudin, T. and Carlucci, T. and Carrasco, J. M. and Casamiquela, L. and Castellani, M. and Castro-Ginard, A. and Chaoul, L. and Charlot, P. and Chemin, L. and Chiaramida, V. and Chiavassa, A. and Chornay, N. and Comoretto, G. and Contursi, G. and Cooper, W. J. and Cornez, T. and Cowell, S. and Crifo, F. and Cropper, M. and Crosta, M. and Crowley, C. and Dafonte, C. and Dapergolas, A. and David, P. and de Laverny, P. and De Luise, F. and De March, R. and De Ridder, J. and de Souza, R. and de Torres, A. and del Peloso, E. F. and del Pozo, E. and Delbo, M. and Delgado, A. and Delisle, J. -B. and Demouchy, C. and Dharmawardena, T. E. and Diakite, S. and Distefano, E. and Dolding, C. and Enke, H. and Fabre, C. and Fabrizio, M. and Faigler, S. and Fedorets, G. and Fernique, P. and Figueras, F. and Fournier, Y. and Fouron, C. and Fragkoudi, F. and Gai, M. and Garcia-Gutierrez, A. and Garcia-Reinaldos, M. and García-Torres, M. and Garofalo, A. and Gavel, A. and Gerlach, E. and Geyer, R. and Giacobbe, P. and Gilmore, G. and Girona, S. and Giuffrida, G. and Gomel, R. and Gomez, A. and González-Núñez, J. and González-Santamaría, I. and González-Vidal, J. J. and Granvik, M. and Guillout, P. and Guiraud, J. and Gutiérrez-Sánchez, R. and Guy, L. P. and Hauser, M. and Haywood, M. and Helmer, A. and Helmi, A. and Sarmiento, M. H. and Hidalgo, S. L. and Hilger, T. and Hładczuk, N. and Hobbs, D. and Holland, G. and Huckle, H. E. and Jardine, K. and Jasniewicz, G. and Jean-Antoine Piccolo, A. and Jiménez-Arranz, Ó. and Juaristi Campillo, J. and Julbe, F. and Karbevska, L. and Kervella, P. and Khanna, S. and Kontizas, M. and Kordopatis, G. and Korn, A. J. and Kóspál, Á. and Kostrzewa-Rutkowska, Z. and Kruszyńska, K. and Kun, M. and Laizeau, P. and Lambert, S. and Lanza, A. F. and Lasne, Y. and Le Campion, J. -F. and Lebreton, Y. and Lebzelter, T. and Leccia, S. and Leclerc, N. and Lecoeur-Taibi, I. and Liao, S. and Licata, E. L. and Lindstrøm, H. E. P. and Lister, T. A. and Livanou, E. and Lobel, A. and Lorca, A. and Loup, C. and Madrero Pardo, P. and Magdaleno Romeo, A. and Managau, S. and Mann, R. G. and Manteiga, M. and Marchant, J. M. and Marconi, M. and Marcos, J. and Marcos Santos, M. M. S. and Marín Pina, D. and Marinoni, S. and Marocco, F. and Marshall, D. J. and Martin Polo, L. and Martín-Fleitas, J. M. and Marton, G. and Mary, N. and Masip, A. and Massari, D. and Mastrobuono-Battisti, A. and Mazeh, T. and McMillan, P. J. and Messina, S. and Michalik, D. and Millar, N. R. and Mints, A. and Molina, D. and Molinaro, R. and Molnár, L. and Monari, G. and Monguió, M. and Montegriffo, P. and Montero, A. and Mor, R. and Mora, A. and Morbidelli, R. and Morel, T. and Morris, D. and Muraveva, T. and Murphy, C. P. and Musella, I. and Nagy, Z. and Noval, L. and Ocaña, F. and Ogden, A. and Ordenovic, C. and Osinde, J. O. and Pagani, C. and Pagano, I. and Palaversa, L. and Palicio, P. A. and Pallas-Quintela, L. and Panahi, A. and Payne-Wardenaar, S. and Peñalosa Esteller, X. and Penttilä, A. and Pichon, B. and Piersimoni, A. M. and Pineau, F. -X. and Plachy, E. and Plum, G. and Poggio, E. and Prša, A. and Pulone, L. and Racero, E. and Ragaini, S. and Rainer, M. and Ramos, P. and Ramos-Lerate, M. and Re Fiorentin, P. and Regibo, S. and Richards, P. J. and Rios Diaz, C. and Ripepi, V. and Riva, A. and Rix, H. -W. and Rixon, G. and Robichon, N. and Robin, A. C. and Robin, C. and Roelens, M. and Rogues, H. R. O. and Rohrbasser, L. and Romero-Gómez, M. and Rowell, N. and Royer, F. and Ruz Mieres, D. and Rybicki, K. A. and Sadowski, G. and Sáez Núñez, A. and Sagristà Sellés, A. and Sahlmann, J. and Salguero, E. and Samaras, N. and Sanchez Gimenez, V. and Sanna, N. and Santoveña, R. and Sarasso, M. and Schultheis, M. and Sciacca, E. and Segol, M. and Segovia, J. C. and Ségransan, D. and Semeux, D. and Shahaf, S. and Siddiqui, H. I. and Siebert, A. and Siltala, L. and Silvelo, A. and Slezak, E. and Slezak, I. and Smart, R. L. and Snaith, O. N. and Solano, E. and Solitro, F. and Souami, D. and Souchay, J. and Spagna, A. and Spina, L. and Spoto, F. and Steele, I. A. and Steidelmüller, H. and Stephenson, C. A. and Süveges, M. and Surdej, J. and Szabados, L. and Szegedi-Elek, E. and Taris, F. and Taylor, M. B. and Tolomei, L. and Tonello, N. and Torra, F. and Torra, J. and Torralba Elipe, G. and Trabucchi, M. and Tsounis, A. T. and Turon, C. and Ulla, A. and Unger, N. and Vaillant, M. V. and van Dillen, E. and van Reeven, W. and Vanel, O. and Vecchiato, A. and Viala, Y. and Vicente, D. and Voutsinas, S. and Weiler, M. and Wevers, T. and Wyrzykowski, Ł. and Yoldas, A. and Yvard, P. and Zhao, H. and Zorec, J. and Zucker, S. and Zwitter, T.},
	month = jun,
	year = {2023},
	note = {ADS Bibcode: 2023A\&A...674A..41G},
	pages = {A41},
}

@article{2019ApJ...873..111I,
	title = {{LSST}: {From} {Science} {Drivers} to {Reference} {Design} and {Anticipated} {Data} {Products}},
	volume = {873},
	issn = {0004-637X},
	shorttitle = {{LSST}},
	url = {https://ui.adsabs.harvard.edu/abs/2019ApJ...873..111I},
	doi = {10.3847/1538-4357/ab042c},
	urldate = {2021-08-31},
	journal = {The Astrophysical Journal},
	author = {Ivezić, Zeljko and Kahn, Steven M. and Tyson, J. Anthony and Abel, Bob and Acosta, Emily and Allsman, Robyn and Alonso, David and AlSayyad, Yusra and Anderson, Scott F. and Andrew, John and Angel, James Roger P. and Angeli, George Z. and Ansari, Reza and Antilogus, Pierre and Araujo, Constanza and Armstrong, Robert and Arndt, Kirk T. and Astier, Pierre and Aubourg, Éric and Auza, Nicole and Axelrod, Tim S. and Bard, Deborah J. and Barr, Jeff D. and Barrau, Aurelian and Bartlett, James G. and Bauer, Amanda E. and Bauman, Brian J. and Baumont, Sylvain and Bechtol, Ellen and Bechtol, Keith and Becker, Andrew C. and Becla, Jacek and Beldica, Cristina and Bellavia, Steve and Bianco, Federica B. and Biswas, Rahul and Blanc, Guillaume and Blazek, Jonathan and Blandford, Roger D. and Bloom, Josh S. and Bogart, Joanne and Bond, Tim W. and Booth, Michael T. and Borgland, Anders W. and Borne, Kirk and Bosch, James F. and Boutigny, Dominique and Brackett, Craig A. and Bradshaw, Andrew and Brandt, William Nielsen and Brown, Michael E. and Bullock, James S. and Burchat, Patricia and Burke, David L. and Cagnoli, Gianpietro and Calabrese, Daniel and Callahan, Shawn and Callen, Alice L. and Carlin, Jeffrey L. and Carlson, Erin L. and Chandrasekharan, Srinivasan and Charles-Emerson, Glenaver and Chesley, Steve and Cheu, Elliott C. and Chiang, Hsin-Fang and Chiang, James and Chirino, Carol and Chow, Derek and Ciardi, David R. and Claver, Charles F. and Cohen-Tanugi, Johann and Cockrum, Joseph J. and Coles, Rebecca and Connolly, Andrew J. and Cook, Kem H. and Cooray, Asantha and Covey, Kevin R. and Cribbs, Chris and Cui, Wei and Cutri, Roc and Daly, Philip N. and Daniel, Scott F. and Daruich, Felipe and Daubard, Guillaume and Daues, Greg and Dawson, William and Delgado, Francisco and Dellapenna, Alfred and de Peyster, Robert and de Val-Borro, Miguel and Digel, Seth W. and Doherty, Peter and Dubois, Richard and Dubois-Felsmann, Gregory P. and Durech, Josef and Economou, Frossie and Eifler, Tim and Eracleous, Michael and Emmons, Benjamin L. and Fausti Neto, Angelo and Ferguson, Henry and Figueroa, Enrique and Fisher-Levine, Merlin and Focke, Warren and Foss, Michael D. and Frank, James and Freemon, Michael D. and Gangler, Emmanuel and Gawiser, Eric and Geary, John C. and Gee, Perry and Geha, Marla and Gessner, Charles J. B. and Gibson, Robert R. and Gilmore, D. Kirk and Glanzman, Thomas and Glick, William and Goldina, Tatiana and Goldstein, Daniel A. and Goodenow, Iain and Graham, Melissa L. and Gressler, William J. and Gris, Philippe and Guy, Leanne P. and Guyonnet, Augustin and Haller, Gunther and Harris, Ron and Hascall, Patrick A. and Haupt, Justine and Hernandez, Fabio and Herrmann, Sven and Hileman, Edward and Hoblitt, Joshua and Hodgson, John A. and Hogan, Craig and Howard, James D. and Huang, Dajun and Huffer, Michael E. and Ingraham, Patrick and Innes, Walter R. and Jacoby, Suzanne H. and Jain, Bhuvnesh and Jammes, Fabrice and Jee, M. James and Jenness, Tim and Jernigan, Garrett and Jevremović, Darko and Johns, Kenneth and Johnson, Anthony S. and Johnson, Margaret W. G. and Jones, R. Lynne and Juramy-Gilles, Claire and Jurić, Mario and Kalirai, Jason S. and Kallivayalil, Nitya J. and Kalmbach, Bryce and Kantor, Jeffrey P. and Karst, Pierre and Kasliwal, Mansi M. and Kelly, Heather and Kessler, Richard and Kinnison, Veronica and Kirkby, David and Knox, Lloyd and Kotov, Ivan V. and Krabbendam, Victor L. and Krughoff, K. Simon and Kubánek, Petr and Kuczewski, John and Kulkarni, Shri and Ku, John and Kurita, Nadine R. and Lage, Craig S. and Lambert, Ron and Lange, Travis and Langton, J. Brian and Le Guillou, Laurent and Levine, Deborah and Liang, Ming and Lim, Kian-Tat and Lintott, Chris J. and Long, Kevin E. and Lopez, Margaux and Lotz, Paul J. and Lupton, Robert H. and Lust, Nate B. and MacArthur, Lauren A. and Mahabal, Ashish and Mandelbaum, Rachel and Markiewicz, Thomas W. and Marsh, Darren S. and Marshall, Philip J. and Marshall, Stuart and May, Morgan and McKercher, Robert and McQueen, Michelle and Meyers, Joshua and Migliore, Myriam and Miller, Michelle and Mills, David J. and Miraval, Connor and Moeyens, Joachim and Moolekamp, Fred E. and Monet, David G. and Moniez, Marc and Monkewitz, Serge and Montgomery, Christopher and Morrison, Christopher B. and Mueller, Fritz and Muller, Gary P. and Muñoz Arancibia, Freddy and Neill, Douglas R. and Newbry, Scott P. and Nief, Jean-Yves and Nomerotski, Andrei and Nordby, Martin and O'Connor, Paul and Oliver, John and Olivier, Scot S. and Olsen, Knut and O'Mullane, William and Ortiz, Sandra and Osier, Shawn and Owen, Russell E. and Pain, Reynald and Palecek, Paul E. and Parejko, John K. and Parsons, James B. and Pease, Nathan M. and Peterson, J. Matt and Peterson, John R. and Petravick, Donald L. and Libby Petrick, M. E. and Petry, Cathy E. and Pierfederici, Francesco and Pietrowicz, Stephen and Pike, Rob and Pinto, Philip A. and Plante, Raymond and Plate, Stephen and Plutchak, Joel P. and Price, Paul A. and Prouza, Michael and Radeka, Veljko and Rajagopal, Jayadev and Rasmussen, Andrew P. and Regnault, Nicolas and Reil, Kevin A. and Reiss, David J. and Reuter, Michael A. and Ridgway, Stephen T. and Riot, Vincent J. and Ritz, Steve and Robinson, Sean and Roby, William and Roodman, Aaron and Rosing, Wayne and Roucelle, Cecille and Rumore, Matthew R. and Russo, Stefano and Saha, Abhijit and Sassolas, Benoit and Schalk, Terry L. and Schellart, Pim and Schindler, Rafe H. and Schmidt, Samuel and Schneider, Donald P. and Schneider, Michael D. and Schoening, William and Schumacher, German and Schwamb, Megan E. and Sebag, Jacques and Selvy, Brian and Sembroski, Glenn H. and Seppala, Lynn G. and Serio, Andrew and Serrano, Eduardo and Shaw, Richard A. and Shipsey, Ian and Sick, Jonathan and Silvestri, Nicole and Slater, Colin T. and Smith, J. Allyn and Smith, R. Chris and Sobhani, Shahram and Soldahl, Christine and Storrie-Lombardi, Lisa and Stover, Edward and Strauss, Michael A. and Street, Rachel A. and Stubbs, Christopher W. and Sullivan, Ian S. and Sweeney, Donald and Swinbank, John D. and Szalay, Alexander and Takacs, Peter and Tether, Stephen A. and Thaler, Jon J. and Thayer, John Gregg and Thomas, Sandrine and Thornton, Adam J. and Thukral, Vaikunth and Tice, Jeffrey and Trilling, David E. and Turri, Max and Van Berg, Richard and Vanden Berk, Daniel and Vetter, Kurt and Virieux, Francoise and Vucina, Tomislav and Wahl, William and Walkowicz, Lucianne and Walsh, Brian and Walter, Christopher W. and Wang, Daniel L. and Wang, Shin-Yawn and Warner, Michael and Wiecha, Oliver and Willman, Beth and Winters, Scott E. and Wittman, David and Wolff, Sidney C. and Wood-Vasey, W. Michael and Wu, Xiuqin and Xin, Bo and Yoachim, Peter and Zhan, Hu},
	month = mar,
	year = {2019},
	note = {ADS Bibcode: 2019ApJ...873..111I},
	pages = {111},
}

@article{van_velzen_seventeen_2021,
	title = {Seventeen {Tidal} {Disruption} {Events} from the {First} {Half} of {ZTF} {Survey} {Observations}: {Entering} a {New} {Era} of {Population} {Studies}},
	volume = {908},
	issn = {0004-637X},
	shorttitle = {Seventeen {Tidal} {Disruption} {Events} from the {First} {Half} of {ZTF} {Survey} {Observations}},
	url = {https://ui.adsabs.harvard.edu/abs/2021ApJ...908....4V},
	doi = {10.3847/1538-4357/abc258},
	urldate = {2022-07-22},
	journal = {The Astrophysical Journal},
	author = {van Velzen, Sjoert and Gezari, Suvi and Hammerstein, Erica and Roth, Nathaniel and Frederick, Sara and Ward, Charlotte and Hung, Tiara and Cenko, S. Bradley and Stein, Robert and Perley, Daniel A. and Taggart, Kirsty and Foley, Ryan J. and Sollerman, Jesper and Blagorodnova, Nadejda and Andreoni, Igor and Bellm, Eric C. and Brinnel, Valery and De, Kishalay and Dekany, Richard and Feeney, Michael and Fremling, Christoffer and Giomi, Matteo and Golkhou, V. Zach and Graham, Matthew J. and Ho, Anna. Y. Q. and Kasliwal, Mansi M. and Kilpatrick, Charles D. and Kulkarni, Shrinivas R. and Kupfer, Thomas and Laher, Russ R. and Mahabal, Ashish and Masci, Frank J. and Miller, Adam A. and Nordin, Jakob and Riddle, Reed and Rusholme, Ben and van Santen, Jakob and Sharma, Yashvi and Shupe, David L. and Soumagnac, Maayane T.},
	month = feb,
	year = {2021},
	note = {ADS Bibcode: 2021ApJ...908....4V},
	pages = {4},
}

@article{chan_tidal_2019,
	title = {Tidal {Disruption} {Events} in {Active} {Galactic} {Nuclei}},
	volume = {881},
	issn = {0004-637X},
	url = {https://dx.doi.org/10.3847/1538-4357/ab2b40},
	doi = {10.3847/1538-4357/ab2b40},
	language = {en},
	number = {2},
	urldate = {2023-07-25},
	journal = {The Astrophysical Journal},
	author = {Chan, Chi-Ho and Piran, Tsvi and Krolik, Julian H. and Saban, Dekel},
	month = aug,
	year = {2019},
	note = {Publisher: The American Astronomical Society},
	pages = {113},
}

@article{ginsburg_astroquery_2019,
	title = {astroquery: {An} {Astronomical} {Web}-querying {Package} in {Python}},
	volume = {157},
	issn = {0004-6256},
	shorttitle = {astroquery},
	url = {https://ui.adsabs.harvard.edu/abs/2019AJ....157...98G},
	doi = {10.3847/1538-3881/aafc33},
	urldate = {2023-07-11},
	journal = {The Astronomical Journal},
	author = {Ginsburg, Adam and Sipőcz, Brigitta M. and Brasseur, C. E. and Cowperthwaite, Philip S. and Craig, Matthew W. and Deil, Christoph and Guillochon, James and Guzman, Giannina and Liedtke, Simon and Lian Lim, Pey and Lockhart, Kelly E. and Mommert, Michael and Morris, Brett M. and Norman, Henrik and Parikh, Madhura and Persson, Magnus V. and Robitaille, Thomas P. and Segovia, Juan-Carlos and Singer, Leo P. and Tollerud, Erik J. and de Val-Borro, Miguel and Valtchanov, Ivan and Woillez, Julien and {Astroquery Collaboration} and {a subset of astropy Collaboration}},
	month = mar,
	year = {2019},
	note = {ADS Bibcode: 2019AJ....157...98G},
	pages = {98},
}

@article{moriya_superluminous_2018,
	title = {Superluminous supernovae},
	volume = {214},
	issn = {0038-6308, 1572-9672},
	url = {http://arxiv.org/abs/1803.01875},
	doi = {10.1007/s11214-018-0493-6},
	number = {2},
	urldate = {2023-06-28},
	journal = {Space Science Reviews},
	author = {Moriya, Takashi J. and Sorokina, Elena I. and Chevalier, Roger A.},
	month = mar,
	year = {2018},
	note = {arXiv:1803.01875 [astro-ph]},
	pages = {59},
}

@article{trakhtenbrot_1es_2019,
	title = {{1ES} 1927+654: {An} {AGN} {Caught} {Changing} {Look} on a {Timescale} of {Months}},
	volume = {883},
	issn = {0004-637X},
	shorttitle = {{1ES} 1927+654},
	url = {https://dx.doi.org/10.3847/1538-4357/ab39e4},
	doi = {10.3847/1538-4357/ab39e4},
	language = {en},
	number = {1},
	urldate = {2023-06-21},
	journal = {The Astrophysical Journal},
	author = {Trakhtenbrot, Benny and Arcavi, Iair and MacLeod, Chelsea L. and Ricci, Claudio and Kara, Erin and Graham, Melissa L. and Stern, Daniel and Harrison, Fiona A. and Burke, Jamison and Hiramatsu, Daichi and Hosseinzadeh, Griffin and Howell, D. Andrew and Smartt, Stephen J. and Rest, Armin and Prieto, Jose L. and Shappee, Benjamin J. and Holoien, Thomas W.-S. and Bersier, David and Filippenko, Alexei V. and Brink, Thomas G. and Zheng, WeiKang and Li, Ruancun and Remillard, Ronald A. and Loewenstein, Michael},
	month = sep,
	year = {2019},
	note = {Publisher: The American Astronomical Society},
	pages = {94},
}

@article{gezari_tidal_2021,
	title = {Tidal {Disruption} {Events}},
	volume = {59},
	issn = {0066-4146},
	url = {https://ui.adsabs.harvard.edu/abs/2021ARA&A..59...21G},
	doi = {10.1146/annurev-astro-111720-030029},
	urldate = {2023-05-09},
	journal = {Annual Review of Astronomy and Astrophysics},
	author = {Gezari, Suvi},
	month = sep,
	year = {2021},
	note = {ADS Bibcode: 2021ARA\&A..59...21G},
	pages = {21--58},
}

@article{antonucci_unified_1993,
	title = {Unified models for active galactic nuclei and quasars.},
	volume = {31},
	issn = {0066-4146},
	url = {https://ui.adsabs.harvard.edu/abs/1993ARA&A..31..473A},
	doi = {10.1146/annurev.aa.31.090193.002353},
	urldate = {2023-06-15},
	journal = {Annual Review of Astronomy and Astrophysics},
	author = {Antonucci, Robert},
	month = jan,
	year = {1993},
	note = {ADS Bibcode: 1993ARA\&A..31..473A},
	pages = {473--521},
}

@article{moller_fink_2021,
	title = {Fink, a new generation of broker for the {LSST} community},
	volume = {501},
	issn = {0035-8711, 1365-2966},
	url = {http://arxiv.org/abs/2009.10185},
	doi = {10.1093/mnras/staa3602},
	number = {3},
	urldate = {2023-06-14},
	journal = {Monthly Notices of the Royal Astronomical Society},
	author = {Möller, Anais and Peloton, Julien and Ishida, Emille E. O. and Arnault, Chris and Bachelet, Etienne and Blaineau, Tristan and Boutigny, Dominique and Chauhan, Abhishek and Gangler, Emmanuel and Hernandez, Fabio and Hrivnac, Julius and Leoni, Marco and Leroy, Nicolas and Moniez, Marc and Pateyron, Sacha and Ramparison, Adrien and Turpin, Damien and Ansari, Réza and Allam Jr., Tarek and Bajat, Armelle and Biswas, Biswajit and Boucaud, Alexandre and Bregeon, Johan and Campagne, Jean-Eric and Cohen-Tanugi, Johann and Coleiro, Alexis and Dornic, Damien and Fouchez, Dominique and Godet, Olivier and Gris, Philippe and Karpov, Sergey and Gomez-Moran, Ada Nebot and Neveu, Jérémy and Plaszczynski, Stephane and Savchenko, Volodymyr and Webb, Natalie},
	month = jan,
	year = {2021},
	note = {arXiv:2009.10185 [astro-ph]},
	pages = {3272--3288},
}

@article{hammerstein_final_2022,
	title = {The {Final} {Season} {Reimagined}: 30 {Tidal} {Disruption} {Events} from the {ZTF}-{I} {Survey}},
	volume = {942},
	issn = {0004-637X},
	shorttitle = {The {Final} {Season} {Reimagined}},
	url = {https://dx.doi.org/10.3847/1538-4357/aca283},
	doi = {10.3847/1538-4357/aca283},
	language = {en},
	number = {1},
	urldate = {2023-04-03},
	journal = {The Astrophysical Journal},
	author = {Hammerstein, Erica and Velzen, Sjoert van and Gezari, Suvi and Cenko, S. Bradley and Yao, Yuhan and Ward, Charlotte and Frederick, Sara and Villanueva, Natalia and Somalwar, Jean J. and Graham, Matthew J. and Kulkarni, Shrinivas R. and Stern, Daniel and Andreoni, Igor and Bellm, Eric C. and Dekany, Richard and Dhawan, Suhail and Drake, Andrew J. and Fremling, Christoffer and Gatkine, Pradip and Groom, Steven L. and Ho, Anna Y. Q. and Kasliwal, Mansi M. and Karambelkar, Viraj and Kool, Erik C. and Masci, Frank J. and Medford, Michael S. and Perley, Daniel A. and Purdum, Josiah and Roestel, Jan van and Sharma, Yashvi and Sollerman, Jesper and Taggart, Kirsty and Yan, Lin},
	month = dec,
	year = {2022},
	note = {Publisher: The American Astronomical Society},
	pages = {9},
}

@article{astropy_collaboration_astropy_2013,
	title = {Astropy: {A} community {Python} package for astronomy},
	volume = {558},
	issn = {0004-6361},
	shorttitle = {Astropy},
	url = {https://ui.adsabs.harvard.edu/abs/2013A&A...558A..33A/abstract},
	doi = {10.1051/0004-6361/201322068},
	language = {en},
	urldate = {2023-03-23},
	journal = {Astronomy and Astrophysics},
	author = {{Astropy Collaboration} and Robitaille, Thomas P. and Tollerud, Erik J. and Greenfield, Perry and Droettboom, Michael and Bray, Erik and Aldcroft, Tom and Davis, Matt and Ginsburg, Adam and Price-Whelan, Adrian M. and Kerzendorf, Wolfgang E. and Conley, Alexander and Crighton, Neil and Barbary, Kyle and Muna, Demitri and Ferguson, Henry and Grollier, Frédéric and Parikh, Madhura M. and Nair, Prasanth H. and Unther, Hans M. and Deil, Christoph and Woillez, Julien and Conseil, Simon and Kramer, Roban and Turner, James E. H. and Singer, Leo and Fox, Ryan and Weaver, Benjamin A. and Zabalza, Victor and Edwards, Zachary I. and Azalee Bostroem, K. and Burke, D. J. and Casey, Andrew R. and Crawford, Steven M. and Dencheva, Nadia and Ely, Justin and Jenness, Tim and Labrie, Kathleen and Lim, Pey Lian and Pierfederici, Francesco and Pontzen, Andrew and Ptak, Andy and Refsdal, Brian and Servillat, Mathieu and Streicher, Ole},
	month = oct,
	year = {2013},
	pages = {A33},
}

@article{astropy_collaboration_astropy_2018,
	title = {The {Astropy} {Project}: {Building} an {Open}-science {Project} and {Status} of the v2.0 {Core} {Package}},
	volume = {156},
	issn = {0004-6256},
	shorttitle = {The {Astropy} {Project}},
	url = {https://ui.adsabs.harvard.edu/abs/2018AJ....156..123A},
	doi = {10.3847/1538-3881/aabc4f},
	urldate = {2023-03-23},
	journal = {The Astronomical Journal},
	author = {{Astropy Collaboration} and Price-Whelan, A. M. and Sipőcz, B. M. and Günther, H. M. and Lim, P. L. and Crawford, S. M. and Conseil, S. and Shupe, D. L. and Craig, M. W. and Dencheva, N. and Ginsburg, A. and VanderPlas, J. T. and Bradley, L. D. and Pérez-Suárez, D. and de Val-Borro, M. and Aldcroft, T. L. and Cruz, K. L. and Robitaille, T. P. and Tollerud, E. J. and Ardelean, C. and Babej, T. and Bach, Y. P. and Bachetti, M. and Bakanov, A. V. and Bamford, S. P. and Barentsen, G. and Barmby, P. and Baumbach, A. and Berry, K. L. and Biscani, F. and Boquien, M. and Bostroem, K. A. and Bouma, L. G. and Brammer, G. B. and Bray, E. M. and Breytenbach, H. and Buddelmeijer, H. and Burke, D. J. and Calderone, G. and Cano Rodríguez, J. L. and Cara, M. and Cardoso, J. V. M. and Cheedella, S. and Copin, Y. and Corrales, L. and Crichton, D. and D'Avella, D. and Deil, C. and Depagne, É. and Dietrich, J. P. and Donath, A. and Droettboom, M. and Earl, N. and Erben, T. and Fabbro, S. and Ferreira, L. A. and Finethy, T. and Fox, R. T. and Garrison, L. H. and Gibbons, S. L. J. and Goldstein, D. A. and Gommers, R. and Greco, J. P. and Greenfield, P. and Groener, A. M. and Grollier, F. and Hagen, A. and Hirst, P. and Homeier, D. and Horton, A. J. and Hosseinzadeh, G. and Hu, L. and Hunkeler, J. S. and Ivezić, Ž. and Jain, A. and Jenness, T. and Kanarek, G. and Kendrew, S. and Kern, N. S. and Kerzendorf, W. E. and Khvalko, A. and King, J. and Kirkby, D. and Kulkarni, A. M. and Kumar, A. and Lee, A. and Lenz, D. and Littlefair, S. P. and Ma, Z. and Macleod, D. M. and Mastropietro, M. and McCully, C. and Montagnac, S. and Morris, B. M. and Mueller, M. and Mumford, S. J. and Muna, D. and Murphy, N. A. and Nelson, S. and Nguyen, G. H. and Ninan, J. P. and Nöthe, M. and Ogaz, S. and Oh, S. and Parejko, J. K. and Parley, N. and Pascual, S. and Patil, R. and Patil, A. A. and Plunkett, A. L. and Prochaska, J. X. and Rastogi, T. and Reddy Janga, V. and Sabater, J. and Sakurikar, P. and Seifert, M. and Sherbert, L. E. and Sherwood-Taylor, H. and Shih, A. Y. and Sick, J. and Silbiger, M. T. and Singanamalla, S. and Singer, L. P. and Sladen, P. H. and Sooley, K. A. and Sornarajah, S. and Streicher, O. and Teuben, P. and Thomas, S. W. and Tremblay, G. R. and Turner, J. E. H. and Terrón, V. and van Kerkwijk, M. H. and de la Vega, A. and Watkins, L. L. and Weaver, B. A. and Whitmore, J. B. and Woillez, J. and Zabalza, V. and {Astropy Contributors}},
	month = sep,
	year = {2018},
	note = {ADS Bibcode: 2018AJ....156..123A},
	pages = {123},
}

@article{astropy_collaboration_astropy_2022,
	title = {The {Astropy} {Project}: {Sustaining} and {Growing} a {Community}-oriented {Open}-source {Project} and the {Latest} {Major} {Release} (v5.0) of the {Core} {Package}},
	volume = {935},
	issn = {0004-637X},
	shorttitle = {The {Astropy} {Project}},
	url = {https://ui.adsabs.harvard.edu/abs/2022ApJ...935..167A},
	doi = {10.3847/1538-4357/ac7c74},
	urldate = {2023-03-23},
	journal = {The Astrophysical Journal},
	author = {{Astropy Collaboration} and Price-Whelan, Adrian M. and Lim, Pey Lian and Earl, Nicholas and Starkman, Nathaniel and Bradley, Larry and Shupe, David L. and Patil, Aarya A. and Corrales, Lia and Brasseur, C. E. and Nöthe, Maximilian and Donath, Axel and Tollerud, Erik and Morris, Brett M. and Ginsburg, Adam and Vaher, Eero and Weaver, Benjamin A. and Tocknell, James and Jamieson, William and van Kerkwijk, Marten H. and Robitaille, Thomas P. and Merry, Bruce and Bachetti, Matteo and Günther, H. Moritz and Aldcroft, Thomas L. and Alvarado-Montes, Jaime A. and Archibald, Anne M. and Bódi, Attila and Bapat, Shreyas and Barentsen, Geert and Bazán, Juanjo and Biswas, Manish and Boquien, Médéric and Burke, D. J. and Cara, Daria and Cara, Mihai and Conroy, Kyle E. and Conseil, Simon and Craig, Matthew W. and Cross, Robert M. and Cruz, Kelle L. and D'Eugenio, Francesco and Dencheva, Nadia and Devillepoix, Hadrien A. R. and Dietrich, Jörg P. and Eigenbrot, Arthur Davis and Erben, Thomas and Ferreira, Leonardo and Foreman-Mackey, Daniel and Fox, Ryan and Freij, Nabil and Garg, Suyog and Geda, Robel and Glattly, Lauren and Gondhalekar, Yash and Gordon, Karl D. and Grant, David and Greenfield, Perry and Groener, Austen M. and Guest, Steve and Gurovich, Sebastian and Handberg, Rasmus and Hart, Akeem and Hatfield-Dodds, Zac and Homeier, Derek and Hosseinzadeh, Griffin and Jenness, Tim and Jones, Craig K. and Joseph, Prajwel and Kalmbach, J. Bryce and Karamehmetoglu, Emir and Kałuszyński, Mikołaj and Kelley, Michael S. P. and Kern, Nicholas and Kerzendorf, Wolfgang E. and Koch, Eric W. and Kulumani, Shankar and Lee, Antony and Ly, Chun and Ma, Zhiyuan and MacBride, Conor and Maljaars, Jakob M. and Muna, Demitri and Murphy, N. A. and Norman, Henrik and O'Steen, Richard and Oman, Kyle A. and Pacifici, Camilla and Pascual, Sergio and Pascual-Granado, J. and Patil, Rohit R. and Perren, Gabriel I. and Pickering, Timothy E. and Rastogi, Tanuj and Roulston, Benjamin R. and Ryan, Daniel F. and Rykoff, Eli S. and Sabater, Jose and Sakurikar, Parikshit and Salgado, Jesús and Sanghi, Aniket and Saunders, Nicholas and Savchenko, Volodymyr and Schwardt, Ludwig and Seifert-Eckert, Michael and Shih, Albert Y. and Jain, Anany Shrey and Shukla, Gyanendra and Sick, Jonathan and Simpson, Chris and Singanamalla, Sudheesh and Singer, Leo P. and Singhal, Jaladh and Sinha, Manodeep and Sipőcz, Brigitta M. and Spitler, Lee R. and Stansby, David and Streicher, Ole and Šumak, Jani and Swinbank, John D. and Taranu, Dan S. and Tewary, Nikita and Tremblay, Grant R. and de Val-Borro, Miguel and Van Kooten, Samuel J. and Vasović, Zlatan and Verma, Shresth and de Miranda Cardoso, José Vinícius and Williams, Peter K. G. and Wilson, Tom J. and Winkel, Benjamin and Wood-Vasey, W. M. and Xue, Rui and Yoachim, Peter and Zhang, Chen and Zonca, Andrea and {Astropy Project Contributors}},
	month = aug,
	year = {2022},
	note = {ADS Bibcode: 2022ApJ...935..167A},
	pages = {167},
}

@article{hunter_matplotlib_2007,
	title = {Matplotlib: {A} {2D} {Graphics} {Environment}},
	volume = {9},
	issn = {1558-366X},
	shorttitle = {Matplotlib},
	doi = {10.1109/MCSE.2007.55},
	number = {3},
	journal = {Computing in Science \& Engineering},
	author = {Hunter, John D.},
	month = may,
	year = {2007},
	note = {Conference Name: Computing in Science \& Engineering},
	pages = {90--95},
}

@article{harris_array_2020,
	title = {Array programming with {NumPy}},
	volume = {585},
	copyright = {2020 The Author(s)},
	issn = {1476-4687},
	url = {https://www.nature.com/articles/s41586-020-2649-2},
	doi = {10.1038/s41586-020-2649-2},
	language = {en},
	number = {7825},
	urldate = {2023-03-23},
	journal = {Nature},
	author = {Harris, Charles R. and Millman, K. Jarrod and van der Walt, Stéfan J. and Gommers, Ralf and Virtanen, Pauli and Cournapeau, David and Wieser, Eric and Taylor, Julian and Berg, Sebastian and Smith, Nathaniel J. and Kern, Robert and Picus, Matti and Hoyer, Stephan and van Kerkwijk, Marten H. and Brett, Matthew and Haldane, Allan and del Río, Jaime Fernández and Wiebe, Mark and Peterson, Pearu and Gérard-Marchant, Pierre and Sheppard, Kevin and Reddy, Tyler and Weckesser, Warren and Abbasi, Hameer and Gohlke, Christoph and Oliphant, Travis E.},
	month = sep,
	year = {2020},
	note = {Number: 7825
Publisher: Nature Publishing Group},
	pages = {357--362},
}

@inproceedings{bellm_zwicky_2014,
	address = {eprint: arXiv:1410.8185},
	title = {The {Zwicky} {Transient} {Facility}},
	url = {https://ui.adsabs.harvard.edu/abs/2014htu..conf...27B},
	urldate = {2022-05-10},
	booktitle = {The {Third} {Hot}-wiring the {Transient} {Universe} {Workshop}},
	author = {Bellm, E.},
	month = jan,
	year = {2014},
	note = {Conference Name: The Third Hot-wiring the Transient Universe Workshop
Pages: 27-33
ADS Bibcode: 2014htu..conf...27B},
	pages = {27--33},
}

@misc{gomez_search_2023,
	title = {The {Search} for {Thermonuclear} {Transients} from the {Tidal} {Disruption} of a {White} {Dwarf} by an {Intermediate} {Mass} {Black} {Hole}},
	url = {http://arxiv.org/abs/2302.14070},
	urldate = {2023-03-01},
	publisher = {arXiv},
	author = {Gomez, Sebastian and Gezari, Suvi},
	month = feb,
	year = {2023},
	note = {arXiv:2302.14070 [astro-ph]},
}

@article{baldwin_classification_1981,
	title = {Classification parameters for the emission-line spectra of extragalactic objects.},
	volume = {93},
	issn = {0004-6280},
	url = {https://ui.adsabs.harvard.edu/abs/1981PASP...93....5B},
	doi = {10.1086/130766},
	urldate = {2023-02-27},
	journal = {Publications of the Astronomical Society of the Pacific},
	author = {Baldwin, J. A. and Phillips, M. M. and Terlevich, R.},
	month = feb,
	year = {1981},
	note = {ADS Bibcode: 1981PASP...93....5B},
	pages = {5--19},
}

@article{masci_zwicky_2018,
	title = {The {Zwicky} {Transient} {Facility}: {Data} {Processing}, {Products}, and {Archive}},
	volume = {131},
	issn = {1538-3873},
	shorttitle = {The {Zwicky} {Transient} {Facility}},
	url = {https://dx.doi.org/10.1088/1538-3873/aae8ac},
	doi = {10.1088/1538-3873/aae8ac},
	language = {en},
	number = {995},
	urldate = {2022-12-21},
	journal = {Publications of the Astronomical Society of the Pacific},
	author = {Masci, Frank J. and Laher, Russ R. and Rusholme, Ben and Shupe, David L. and Groom, Steven and Surace, Jason and Jackson, Edward and Monkewitz, Serge and Beck, Ron and Flynn, David and Terek, Scott and Landry, Walter and Hacopians, Eugean and Desai, Vandana and Howell, Justin and Brooke, Tim and Imel, David and Wachter, Stefanie and Ye, Quan-Zhi and Lin, Hsing-Wen and Cenko, S. Bradley and Cunningham, Virginia and Rebbapragada, Umaa and Bue, Brian and Miller, Adam A. and Mahabal, Ashish and Bellm, Eric C. and Patterson, Maria T. and Jurić, Mario and Golkhou, V. Zach and Ofek, Eran O. and Walters, Richard and Graham, Matthew and Kasliwal, Mansi M. and Dekany, Richard G. and Kupfer, Thomas and Burdge, Kevin and Cannella, Christopher B. and Barlow, Tom and Sistine, Angela Van and Giomi, Matteo and Fremling, Christoffer and Blagorodnova, Nadejda and Levitan, David and Riddle, Reed and Smith, Roger M. and Helou, George and Prince, Thomas A. and Kulkarni, Shrinivas R.},
	month = dec,
	year = {2018},
	note = {Publisher: The Astronomical Society of the Pacific},
	pages = {018003},
}

@article{virtanen_scipy_2020,
	title = {{SciPy} 1.0: fundamental algorithms for scientific computing in {Python}},
	volume = {17},
	copyright = {2020 The Author(s)},
	issn = {1548-7105},
	shorttitle = {{SciPy} 1.0},
	url = {https://www.nature.com/articles/s41592-019-0686-2},
	doi = {10.1038/s41592-019-0686-2},
	language = {en},
	number = {3},
	urldate = {2022-12-02},
	journal = {Nature Methods},
	author = {Virtanen, Pauli and Gommers, Ralf and Oliphant, Travis E. and Haberland, Matt and Reddy, Tyler and Cournapeau, David and Burovski, Evgeni and Peterson, Pearu and Weckesser, Warren and Bright, Jonathan and van der Walt, Stéfan J. and Brett, Matthew and Wilson, Joshua and Millman, K. Jarrod and Mayorov, Nikolay and Nelson, Andrew R. J. and Jones, Eric and Kern, Robert and Larson, Eric and Carey, C. J. and Polat, İlhan and Feng, Yu and Moore, Eric W. and VanderPlas, Jake and Laxalde, Denis and Perktold, Josef and Cimrman, Robert and Henriksen, Ian and Quintero, E. A. and Harris, Charles R. and Archibald, Anne M. and Ribeiro, Antônio H. and Pedregosa, Fabian and van Mulbregt, Paul},
	month = mar,
	year = {2020},
	note = {Number: 3
Publisher: Nature Publishing Group},
	pages = {261--272},
}

@article{gaia_collaboration_gaia_2022,
	title = {Gaia {Data} {Release} 3. {The} extragalactic content},
	issn = {0004-6361, 1432-0746},
	url = {https://www.aanda.org/10.1051/0004-6361/202243232},
	doi = {10.1051/0004-6361/202243232},
	language = {en},
	urldate = {2022-10-19},
	journal = {Astronomy \& Astrophysics},
	author = {{Gaia Collaboration} and Bailer-Jones, Coryn A. L. and {al.}},
	month = jun,
	year = {2022},
}

@misc{wevers_rebrightening_2022,
	title = {The rebrightening of {AT2018fyk} as a repeating partial tidal disruption event},
	url = {http://arxiv.org/abs/2209.07538},
	urldate = {2022-09-19},
	publisher = {arXiv},
	author = {Wevers, T. and Coughlin, E. R. and Pasham, D. R. and Guolo, M. and Sun, Y. and Wen, S. and Jonker, P. G. and Zabludoff, A. and Malyali, A. and Arcodia, R. and Liu, Z. and Merloni, A. and Rau, A. and Grotova, I. and Short, P. and Cao, Z.},
	month = sep,
	year = {2022},
	note = {arXiv:2209.07538 [astro-ph]},
}

@article{guillochon_mosfit_2018,
	title = {{MOSFiT}: {Modular} {Open} {Source} {Fitter} for {Transients}},
	volume = {236},
	issn = {0067-0049},
	shorttitle = {{MOSFiT}},
	url = {https://doi.org/10.3847/1538-4365/aab761},
	doi = {10.3847/1538-4365/aab761},
	language = {en},
	number = {1},
	urldate = {2022-09-07},
	journal = {The Astrophysical Journal Supplement Series},
	author = {Guillochon, James and Nicholl, Matt and Villar, V. Ashley and Mockler, Brenna and Narayan, Gautham and Mandel, Kaisey S. and Berger, Edo and Williams, Peter K. G.},
	month = may,
	year = {2018},
	note = {Publisher: American Astronomical Society},
	pages = {6},
}

@article{miniutti_nine-hour_2019,
	title = {Nine-hour {X}-ray quasi-periodic eruptions from a low-mass black hole galactic nucleus},
	volume = {573},
	issn = {0028-0836},
	url = {https://ui.adsabs.harvard.edu/abs/2019Natur.573..381M},
	doi = {10.1038/s41586-019-1556-x},
	urldate = {2022-06-28},
	journal = {Nature},
	author = {Miniutti, G. and Saxton, R. D. and Giustini, M. and Alexander, K. D. and Fender, R. P. and Heywood, I. and Monageng, I. and Coriat, M. and Tzioumis, A. K. and Read, A. M. and Knigge, C. and Gandhi, P. and Pretorius, M. L. and Agís-González, B.},
	month = sep,
	year = {2019},
	note = {ADS Bibcode: 2019Natur.573..381M},
	pages = {381--384},
}

@article{sanchez-saez_searching_2021,
	title = {Searching for changing-state {AGNs} in massive datasets -- {I}: applying deep learning and anomaly detection techniques to find {AGNs} with anomalous variability behaviours},
	volume = {162},
	issn = {0004-6256, 1538-3881},
	shorttitle = {Searching for changing-state {AGNs} in massive datasets -- {I}},
	url = {http://arxiv.org/abs/2106.07660},
	doi = {10.3847/1538-3881/ac1426},
	number = {5},
	urldate = {2022-05-16},
	journal = {The Astronomical Journal},
	author = {Sánchez-Sáez, P. and Lira, H. and Martí, L. and Sánchez-Pi, N. and Arredondo, J. and Bauer, F. E. and Bayo, A. and Cabrera-Vives, G. and Donoso-Oliva, C. and Estévez, P. A. and Eyheramendy, S. and Förster, F. and Hernández-García, L. and Arancibia, A. M. Muñoz and Pérez-Carrasco, M. and Sepúlveda, M. and Vergara, J. R.},
	month = nov,
	year = {2021},
	note = {arXiv:2106.07660 [astro-ph]},
	pages = {206},
}

@article{payne_asassn-14ko_2021,
	title = {{ASASSN}-14ko is a {Periodic} {Nuclear} {Transient} in {ESO} 253-{G003}},
	volume = {910},
	issn = {0004-637X},
	url = {https://doi.org/10.3847/1538-4357/abe38d},
	doi = {10.3847/1538-4357/abe38d},
	language = {en},
	number = {2},
	urldate = {2022-03-17},
	journal = {The Astrophysical Journal},
	author = {Payne, Anna V. and Shappee, Benjamin J. and Hinkle, Jason T. and Vallely, Patrick J. and Kochanek, Christopher S. and Holoien, Thomas W.-S. and Auchettl, Katie and Stanek, K. Z. and Thompson, Todd A. and Neustadt, Jack M. M. and Tucker, Michael A. and Armstrong, James D. and Brimacombe, Joseph and Cacella, Paulo and Cornect, Robert and Denneau, Larry and Fausnaugh, Michael M. and Flewelling, Heather and Grupe, Dirk and Heinze, A. N. and Lopez, Laura A. and Monard, Berto and Prieto, Jose L. and Schneider, Adam C. and Sheppard, Scott S. and Tonry, John L. and Weiland, Henry},
	month = apr,
	year = {2021},
	note = {Publisher: American Astronomical Society},
	pages = {125},
}

@article{modiano_tuvopipe_2022,
	title = {{TUVOpipe}: a pipeline to search for {UV} transients with {Swift}-{UVOT}},
	shorttitle = {{TUVOpipe}},
	url = {http://arxiv.org/abs/2202.10143},
	language = {en},
	urldate = {2022-02-22},
	journal = {arXiv:2202.10143 [astro-ph]},
	author = {Modiano, David and Wijnands, Rudy and Parikh, Aastha and van Opijnen, Jari and Verberne, Sill and van Etten, Marieke},
	month = feb,
	year = {2022},
	note = {arXiv: 2202.10143},
}

@article{nicholl_systematic_2022,
	title = {Systematic light curve modelling of {TDEs}: statistical differences between the spectroscopic classes},
	shorttitle = {Systematic light curve modelling of {TDEs}},
	url = {http://arxiv.org/abs/2201.02649},
	urldate = {2022-01-11},
	journal = {arXiv:2201.02649 [astro-ph]},
	author = {Nicholl, Matt and Lanning, Daniel and Ramsden, Paige and Mockler, Brenna and Lawrence, Andy and Short, Phil and Ridley, Evan J.},
	month = jan,
	year = {2022},
	note = {arXiv: 2201.02649},
}

@article{zabludoff_distinguishing_2021,
	title = {Distinguishing {Tidal} {Disruption} {Events} from {Impostors}},
	url = {http://arxiv.org/abs/2103.12150},
	urldate = {2021-03-24},
	journal = {arXiv:2103.12150 [astro-ph]},
	author = {Zabludoff, Ann and Arcavi, Iair and La Massa, Stephanie and Perets, Hagai B. and Trakhtenbrot, Benny and Zauderer, B. Ashley and Auchettl, Katie and Dai, Jane L. and French, K. Decker and Hung, Tiara and Kara, Erin and Lodato, Giuseppe and Maksym, W. Peter and Qin, Yujing and Ramirez-Ruiz, Enrico and Roth, Nathaniel and Runnoe, Jessie C. and Wevers, Thomas},
	month = mar,
	year = {2021},
	note = {arXiv: 2103.12150},
}

\begin{appendix}

\onecolumn
 
\section{Fitted properties of the optical transients}

\begin{table}[H]
    \centering
    \begin{tabular}{lcccc}
        \hline\hline \\[-0.2cm]
        Name       & $t_{peak}$ (UTC) &$\tau_{rise}$ (days)& $\tau_{decay}$(days)\\[0.2cm] \hline\hline \\[-0.2cm]
        AT2020aexc & 2021-02-22 $\pm$2.9d. & $38.2\pm2.8$ & $198.1\pm11.0$\\
        AT2021lnu & 2021-06-25 $\pm$1.5d. & $34.2\pm2.3$ & $20.4\pm4.8$\\
        AT2022yhf & 2023-01-02 $\pm$3.4d. & $64.0\pm2.0$ & $668.2\pm34.3$\\
        AT2020ukj & 2020-12-12 $\pm$7.9d. & $68.5\pm8.8$ & $4984.9\pm497.2$\\
        AT2023adr \#1 & 2023-01-30 $\pm$1d. & $25.5\pm0.9$ & $64.3\pm1.6$\\
        AT2023adr \#2 & 2024-03-19 $\pm$1d. & $16.6\pm1.5$ & $40.5\pm3.7$\\
        AT2024ljd & 2024-06-26 $\pm$0.3d. & $7.2\pm0.4$ & $8.3\pm0.9$\\
        AT2023rav & 2023-09-02 $\pm$1.1d. & $13.6\pm0.9$ & $45.0\pm3.0$\\
        AT2024nxp & 2024-06-23 $\pm$1.1d. & $30.4\pm1.0$ & $137.1\pm5.9$\\
        AT2020mvg & 2020-06-12 $\pm$0.6d. & $12.9\pm0.5$ & $31.5\pm2.4$\\
        AT2021ovg & 2021-06-19 $\pm$2.1d. & $36.9\pm2.2$ & $59.5\pm5.7$\\
        AT2021wxd & 2021-09-16 $\pm$4.8d. & $24.8\pm3.4$ & $160.6\pm16.6$\\
        AT2023jag & 2023-06-12 $\pm$1.4d. & $28.1\pm1.8$ & $89.8\pm7.8$\\
        AT2024hhj & 2024-05-07 $\pm$4.5d. & $18.8\pm2.6$ & $82.8\pm11.1$\\
        AT2023szj & 2024-01-02 $\pm$13.1d. & $162.3\pm11.5$ & $375.4\pm50.6$\\
        ZTF23abjvojy \#1& 2008-01-22 $\pm$6.3d. & $16.1\pm8.4$ & $523.5\pm58.2$\\
        ZTF23abjvojy \#2& 2023-11-24 $\pm$3.8d. & $25.5\pm2.8$ & $186.1\pm24.4$\\
        ZTF20aatpzog & 2020-04-08 $\pm$2.6d. & $27.9\pm1.8$ & $31.7\pm5.5$\\
        AT2024gzn & 2024-06-02 $\pm$5.2d. & $138.6\pm6.4$ & $111.5\pm15.2$\\
        AT2020actc & 2021-01-25 $\pm$3.1d. & $29.5\pm2.8$ & $136.6\pm15.4$\\
        AT2020pno & 2020-08-01 $\pm$0.6d. & $19.0\pm0.7$ & $50.5\pm1.9$\\
        AT2020afap & 2021-03-01 $\pm$3.7d. & $59.2\pm5.0$ & $92.2\pm10.6$\\
        AT2023zaj & 2024-10-21 $\pm$9.1d. & $168.7\pm10.1$ & $150.4\pm33.4$\\

    \end{tabular}
    \caption{Timing properties of the TDE candidates, as obtained from the simple $g-$band phenomenological fit. Errors are $1\sigma$. }
    \label{tab:TDE_TimingProperties}
\end{table}

\begin{table}[H]
\centering
\resizebox{\textwidth}{!}{
\begin{tabular}{lcccccccccc}\\
        \hline\hline \\[-0.2cm]
Name & log($M_{\rm BH}$) & $M_{\rm star}$ & log($t_{\rm visc}$) & $\beta^{*}$ & log($\eta$) & log($R_{ph-0}$) & $L^{*}_{photo}$ & redshift & $t_{0}$ \\[0.2cm]
\tiny{\textit{Units}} & \tiny{\textit{log($M_\odot$)}} & \tiny{\textit{$M_\odot$}} & \tiny{\textit{log(days)}} & \tiny{\textit{--}} & \tiny{\textit{--}} & \tiny{\textit{--}} & \tiny{\textit{--}} & \tiny{\textit{--}} & \tiny{\textit{days}} \\[0.2cm] 
 \hline\hline \\[-0.2cm]
ZTF23abjvojy & $6.79^{+0.06}_{-0.06}$ & $0.10^{+0.01}_{-0.01}$ & $-0.71^{+1.13}_{-1.55}$ & $0.96^{+0.05}_{-0.07}$ & $-1.25^{+0.06}_{-0.07}$ & $1.67^{+0.33}_{-0.25}$ & $0.89^{+0.23}_{-0.18}$ & $0.27$ & $-17.05^{+1.55}_{-2.15}$ \\
AT2023adr (1st peak) & $6.79^{+0.06}_{-0.08}$ & $0.76^{+0.14}_{-0.14}$ & $-1.15^{+1.28}_{-1.23}$ & $1.00^{+0.10}_{-0.12}$ & $-3.47^{+0.18}_{-0.25}$ & $1.41^{+0.25}_{-0.24}$ & $0.71^{+0.13}_{-0.12}$ & $0.06^{+0.01}_{-0.01}$ & $-8.38^{+1.73}_{-1.68}$ \\
ZTF20aatpzog & $6.75^{+0.10}_{-0.13}$ & $0.96^{+0.13}_{-0.50}$ & $-0.68^{+1.38}_{-1.62}$ & $0.81^{+0.33}_{-0.20}$ & $-3.00^{+0.30}_{-0.22}$ & $2.74^{+0.83}_{-0.80}$ & $1.24^{+0.77}_{-0.75}$ & $0.18^{+0.03}_{-0.03}$ & $-21.04^{+2.69}_{-3.08}$ \\
AT2020mvg & $6.06^{+0.05}_{-0.05}$ & $0.87^{+0.09}_{-0.14}$ & $-1.12^{+1.30}_{-1.16}$ & $0.10^{+0.04}_{-0.02}$ & $-1.46^{+0.19}_{-0.27}$ & $2.99^{+0.69}_{-0.91}$ & $0.66^{+0.46}_{-0.44}$ & $0.06^{+0.00}_{-0.00}$ & $-6.39^{+0.76}_{-0.87}$ \\
AT2022yhf & $7.13^{+0.04}_{-0.05}$ & $1.14^{+0.07}_{-0.05}$ & $0.97^{+0.05}_{-0.05}$ & $1.00^{+0.00}_{-0.00}$ & $-0.46^{+0.04}_{-0.06}$ & $0.47^{+0.02}_{-0.02}$ & $1.13^{+0.06}_{-0.06}$ & $0.32^{+0.01}_{-0.01}$ & $15.92^{+1.35}_{-1.25}$ \\
AT2024gzn & $7.29^{+0.13}_{-0.11}$ & $0.24^{+0.15}_{-0.13}$ & $-0.74^{+1.20}_{-1.46}$ & $1.18^{+0.07}_{-0.06}$ & $-1.90^{+0.30}_{-0.39}$ & $2.33^{+0.31}_{-0.27}$ & $1.29^{+0.34}_{-0.21}$ & $0.22^{+0.06}_{-0.05}$ & $-39.83^{+4.67}_{-5.15}$ \\
AT2020ukj & $5.75^{+0.08}_{-0.09}$ & $0.98^{+0.02}_{-0.04}$ & $3.21^{+0.02}_{-0.02}$ & $1.72^{+0.06}_{-0.10}$ & $-3.98^{+0.03}_{-0.01}$ & $0.37^{+0.13}_{-0.12}$ & $0.29^{+0.05}_{-0.04}$ & $0.01^{+0.00}_{-0.00}$ & $-18.4^{+1.8}_{-2.0}$ \\
AT2023npi & $5.66^{+0.16}_{-0.06}$ & $0.01^{+0.00}_{-0.00}$ & $-1.29^{+1.09}_{-1.13}$ & $0.98^{+0.02}_{-0.04}$ & $-1.89^{+0.03}_{-0.08}$ & $0.03^{+0.05}_{-0.09}$ & $0.08^{+0.02}_{-0.02}$ & $0.03$ & $-16.30^{+0.96}_{-1.13}$ \\
AT2020afap & $5.39^{+0.16}_{-0.09}$ & $0.62^{+0.29}_{-0.20}$ & $1.57^{+0.17}_{-0.21}$ & $0.63^{+0.29}_{-0.27}$ & $-1.42^{+0.43}_{-0.83}$ & $1.40^{+0.26}_{-0.13}$ & $2.95^{+0.68}_{-0.84}$ & $0.10$ & $-31.31^{+8.02}_{-6.23}$ \\
AT2023szj & $7.95^{+0.03}_{-0.06}$ & $0.28^{+0.04}_{-0.05}$ & $-1.29^{+1.01}_{-1.15}$ & $0.93^{+0.14}_{-0.07}$ & $-0.51^{+0.08}_{-0.13}$ & $-0.19^{+0.08}_{-0.08}$ & $0.11^{+0.04}_{-0.04}$ & $0.46^{+0.04}_{-0.04}$ & $-48.62^{+1.96}_{-1.02}$ \\
AT2020pno & $7.03^{+0.02}_{-0.02}$ & $0.99^{+0.01}_{-0.02}$ & $0.56^{+0.09}_{-0.11}$ & $0.72^{+0.01}_{-0.01}$ & $-2.26^{+0.02}_{-0.01}$ & $3.95^{+0.04}_{-0.07}$ & $3.51^{+0.06}_{-0.06}$ & $0.28$ & $-43.59^{+1.58}_{-1.71}$ \\
AT2024hhj & $7.30^{+0.06}_{-0.07}$ & $0.10^{+0.01}_{-0.00}$ & $0.52^{+0.35}_{-2.08}$ & $0.85^{+0.10}_{-0.10}$ & $-0.76^{+0.07}_{-0.07}$ & $3.51^{+0.28}_{-0.34}$ & $3.15^{+0.16}_{-0.17}$ & $0.22$ & $-31.13^{+3.57}_{-3.83}$ \\
AT2023zaj & $7.35^{+0.21}_{-0.27}$ & $0.33^{+0.24}_{-0.14}$ & $-0.58^{+1.52}_{-1.45}$ & $1.00^{+0.28}_{-0.26}$ & $-0.85^{+0.28}_{-0.33}$ & $2.16^{+1.26}_{-1.05}$ & $1.38^{+1.31}_{-0.86}$ & $0.41^{+0.05}_{-0.06}$ & $-44.22^{+6.61}_{-4.03}$ \\
AT2021wxd & $6.90^{+0.35}_{-0.18}$ & $0.34^{+0.38}_{-0.17}$ & $-0.12^{+1.35}_{-1.32}$ & $0.84^{+0.60}_{-0.55}$ & $-1.68^{+0.72}_{-0.66}$ & $3.04^{+0.70}_{-0.82}$ & $1.80^{+0.53}_{-0.52}$ & $0.23$ & $-30.89^{+7.18}_{-6.51}$ \\
AT2021ovg & $6.92^{+0.24}_{-0.30}$ & $0.46^{+0.29}_{-0.30}$ & $-0.64^{+1.13}_{-1.59}$ & $1.01^{+0.31}_{-0.45}$ & $-2.19^{+0.35}_{-0.42}$ & $2.83^{+0.71}_{-0.65}$ & $1.90^{+0.66}_{-0.57}$ & $0.25^{+0.05}_{-0.05}$ & $-26.74^{+8.71}_{-12.83}$ \\
AT2023adr (2nd peak)& $5.84^{+0.32}_{-0.31}$ & $0.32^{+0.23}_{-0.12}$ & $1.87^{+0.09}_{-0.10}$ & $0.55^{+0.10}_{-0.10}$ & $-0.81^{+0.26}_{-0.41}$ & $2.17^{+0.33}_{-0.35}$ & $3.74^{+0.19}_{-0.36}$ & $0.14^{+0.05}_{-0.04}$ & $-38.29^{+8.14}_{-8.05}$ \\
AT2020act & $6.83^{+0.08}_{-0.09}$ & $0.12^{+0.12}_{-0.02}$ & $1.35^{+0.04}_{-0.05}$ & $0.90^{+0.07}_{-0.07}$ & $-1.84^{+0.07}_{-0.28}$ & $1.84^{+0.16}_{-0.15}$ & $1.02^{+0.12}_{-0.11}$ & $0.14$ & $-35.92^{+2.88}_{-2.75}$ \\
AT2023jag & $6.96^{+0.08}_{-0.09}$ & $0.75^{+0.19}_{-0.16}$ & $-0.17^{+1.20}_{-1.18}$ & $0.99^{+0.23}_{-0.26}$ & $-3.38^{+0.27}_{-0.22}$ & $1.58^{+1.14}_{-0.34}$ & $0.63^{+0.62}_{-0.16}$ & $0.11^{+0.03}_{-0.03}$ & $-24.76^{+2.94}_{-3.38}$ \\
AT2020act &	$6.8^{+0.07}_{-0.07}$ &	$0.20^{+0.09}_{-0.08}$ & $1.34^{+0.05}_{-0.05}$ & $0.90^{+0.08}_{-0.07}$ & $-2.04^{+0.19}_{-0.16}$ & $1.83^{+0.15}_{-0.15}$ & $1.03^{+0.12}_{-0.11}$ & $0.15^{+0.0}_{-0.0}$ & $-36.8^{+2.6}_{-2.7}$ \\
AT2020aex & $6.92^{+0.10}_{-0.07}$ & $0.35^{+0.17}_{-0.17}$ & $0.85^{+0.25}_{-2.60}$ & $0.74^{+0.13}_{-0.12}$ & $-1.12^{+0.29}_{-0.21}$ & $0.63^{+0.09}_{-0.11}$ & $0.72^{+0.19}_{-0.16}$ & $0.26^{+0.02}_{-0.01}$ & $-18.13^{+4.69}_{-3.24}$ \\
Ansky & $6.52^{+0.05}_{-0.05}$ & $1.00^{+0.01}_{-0.01}$ & $3.28^{+0.01}_{-0.01}$&$0.32^{+0.02}_{-0.01}$&$-1.77^{+0.05}_{-0.07}$&$-0.52^{+0.04}_{0.04}$&$0.007^{+0.01}_{-0.005}$&$0.024$&$72.67^{+1.63}_{-1.76}$

\\
\hline
\end{tabular}
}
\caption{Fitted parameter values from the \textit{tde\_fallback} model using \textsc{Redback}. The central values correspond to the median of the posterior distribution. The upper and lower respectively correspond to the 84th and 16th percentile. Redshift values with no error displayed indicate that they were fixed to their known spectroscopic redshift value. Because the fit failed to converge, AT2023rav is missing from this sample.}
\label{tab:Redback_values}
\end{table}

\newpage

\section{Skyview and lightcurves of the optical transients}

\begin{figure}[H]
    \centering
    \includegraphics[height=.92\textheight]{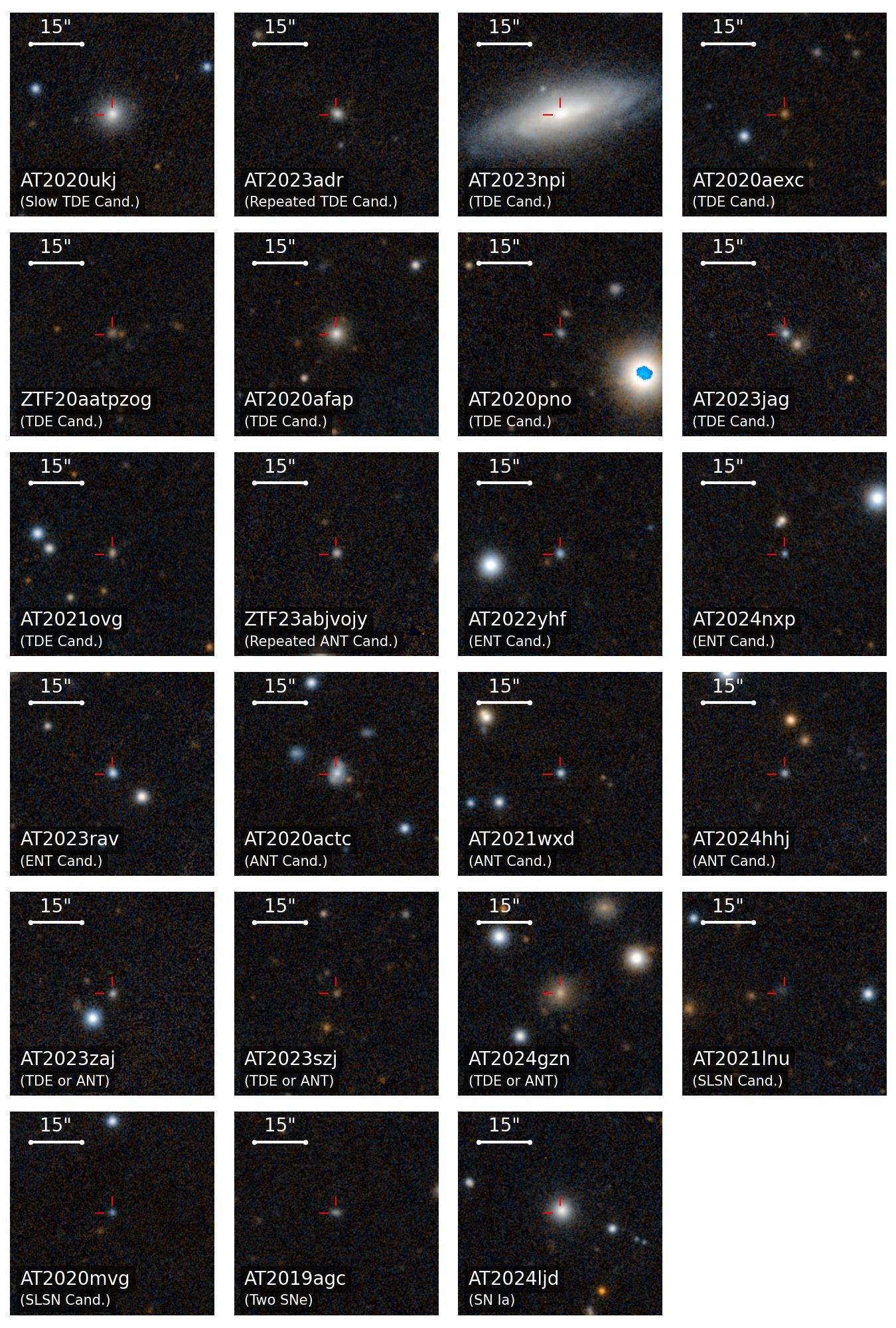}
    \caption{Optical view of the hosts of the optical transients presented in this paper, from PanSTARRS, with a $1'\times1'$ field of view.}
    \label{fig:ImageTDEs}
\end{figure}



\begin{figure}
    \centering
    \includegraphics[height=.95\textheight]{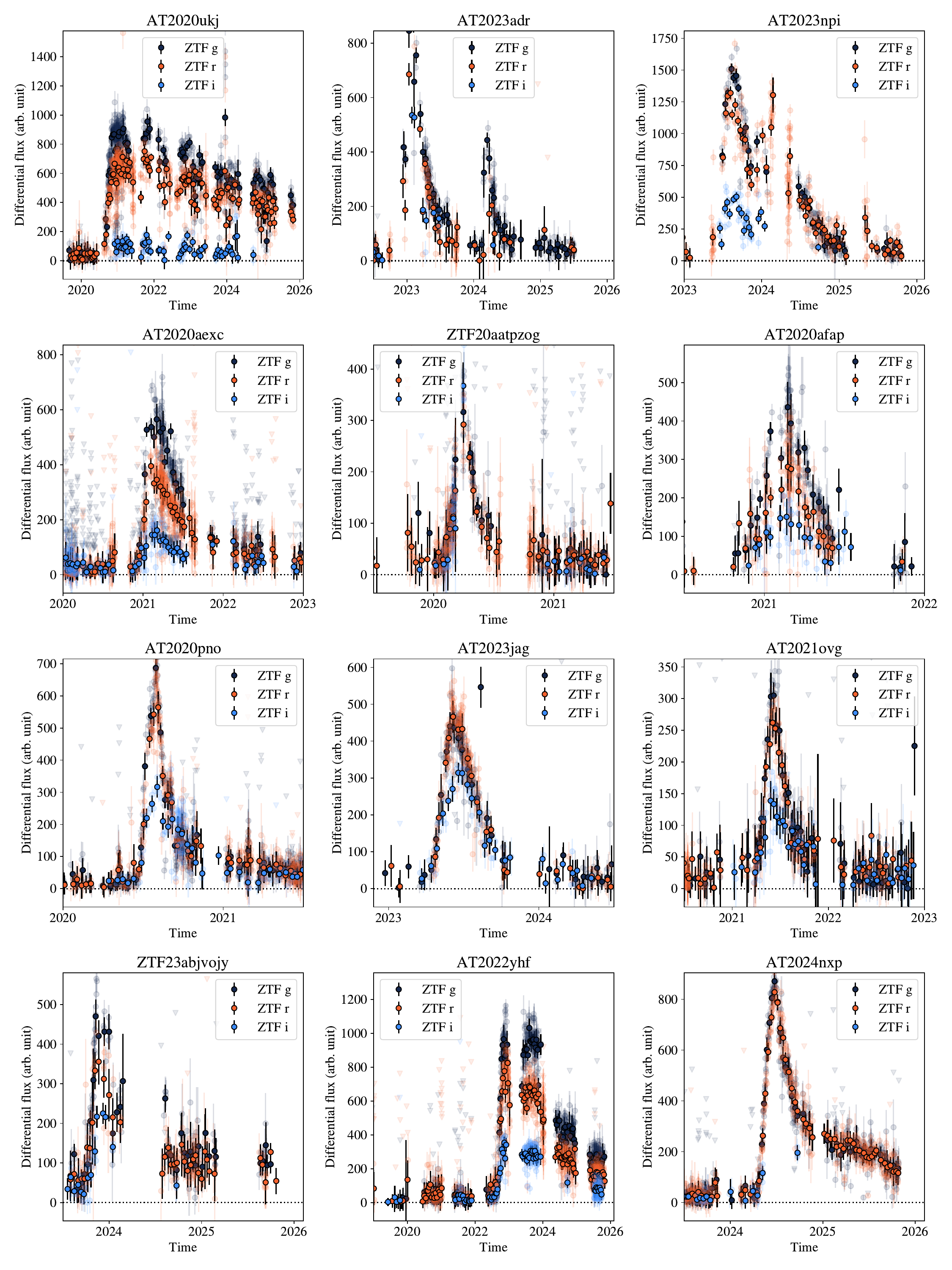}
    \caption{ZTF lightcurves of the optical transients in this paper, expressed in the instrumental ZTF units ('Digital Number'), and zoomed in on the transient. The translucent points correspond to the unbinned lightcurve, and the opaque points are binned in bins of 10 days. }
    \label{fig:ZTF_TDEs1}
\end{figure}

\begin{figure}
    \centering
    \includegraphics[height=.95\textheight]{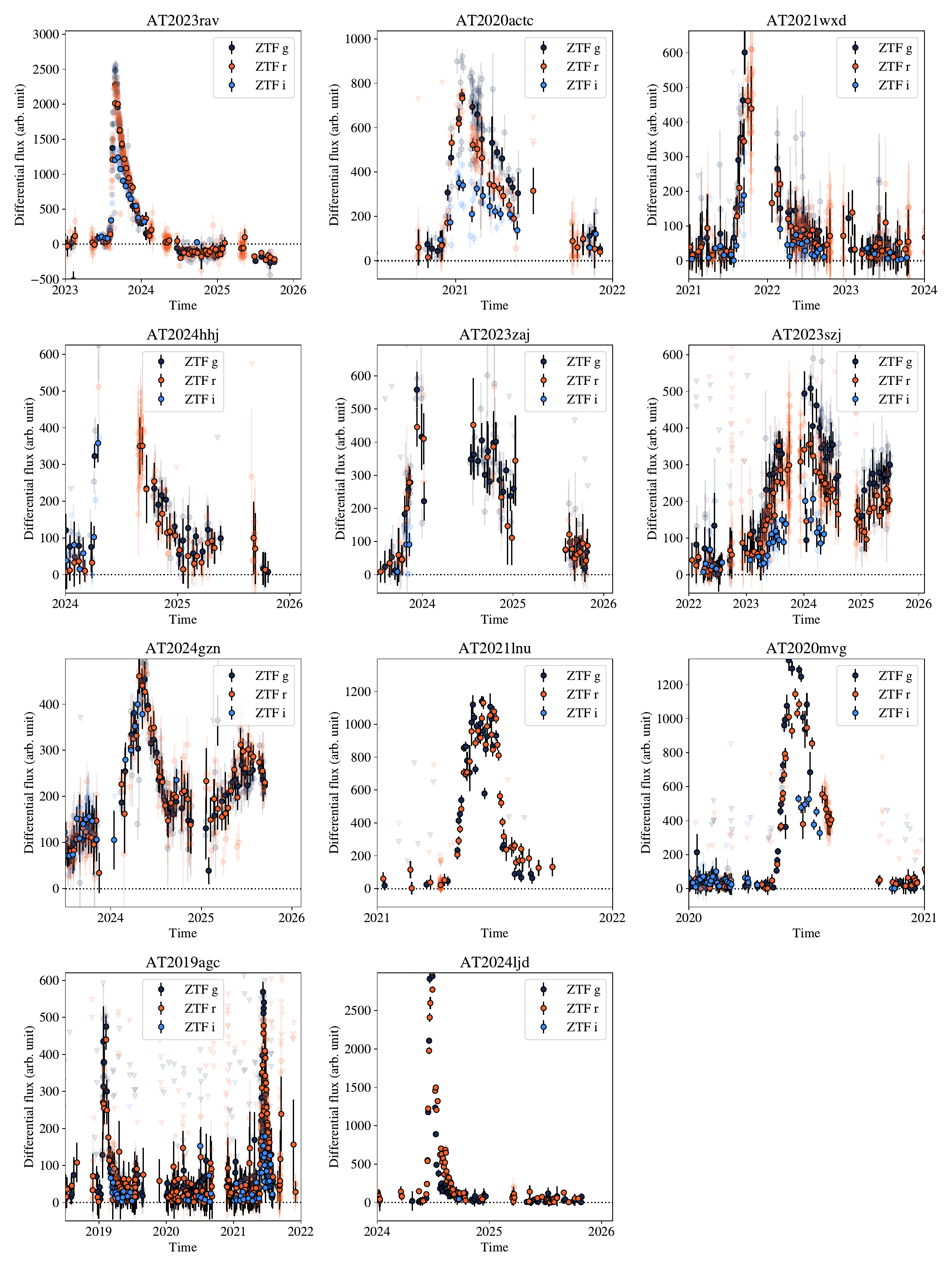}
    \caption{Continuation of Fig. \ref{fig:ZTF_TDEs1}, except for the last four objects, which as SNe or SLSNe candidates are plotted in bins of 1 day instead of 10 days.}
    \label{fig:ZTF_TDEs2}
\end{figure}

\begin{figure}
    \centering
    \includegraphics[height=.94\textheight]{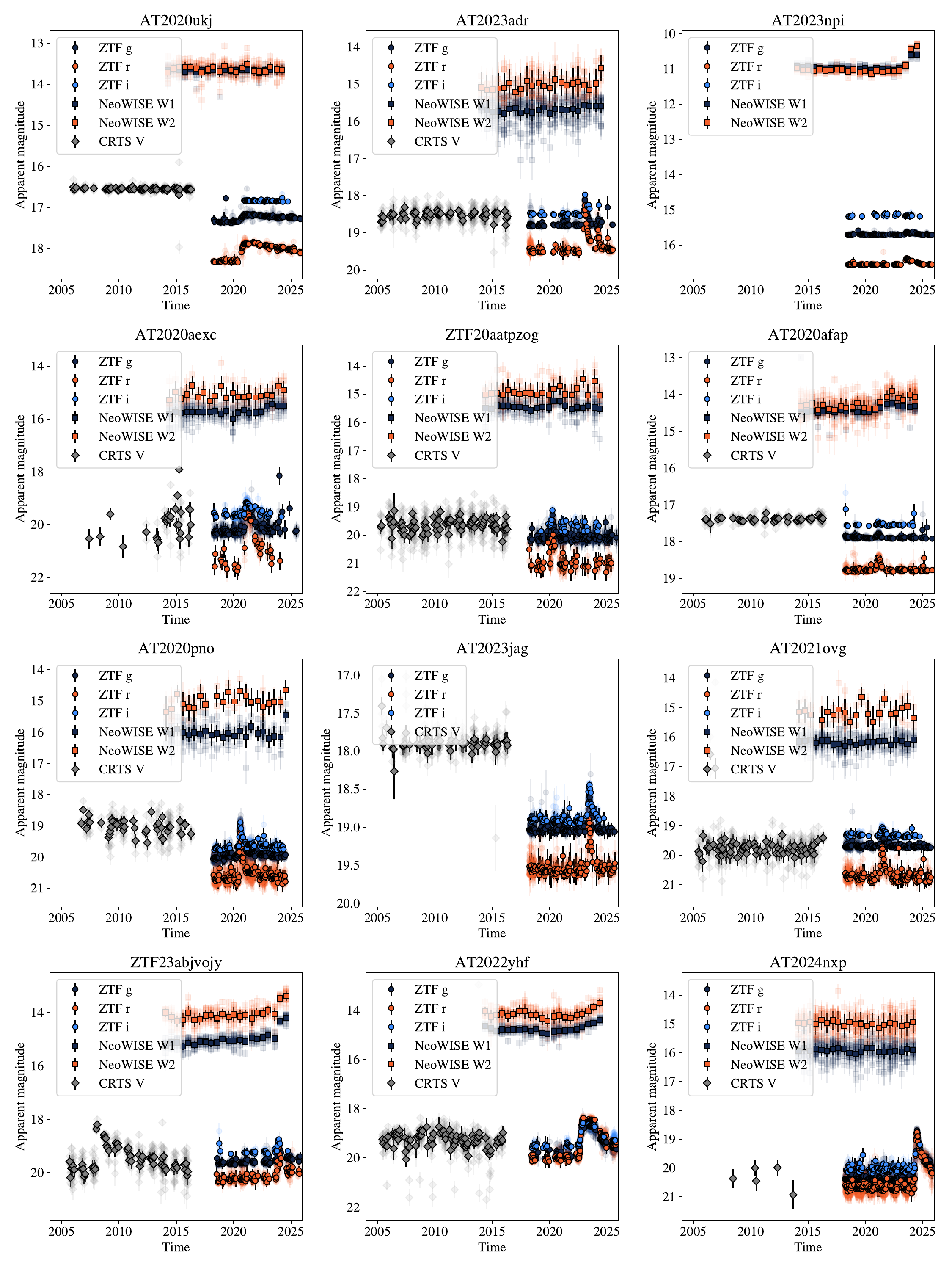}
    \caption{Multi-epoch, multi-wavelenght lightcurves of the optical transients in this paper, showing optical (CRTS and ZTF) and infrared (WISE) evolution. The optical points are binned at 15 days. The CRTS counterpart to AT2023jag is a combination of the two central sources, hence the much larger magnitude than ZTF.}
    \label{fig:MultiL_TDEs1}
\end{figure}

\begin{figure}
    \centering
    \includegraphics[height=.95\textheight]{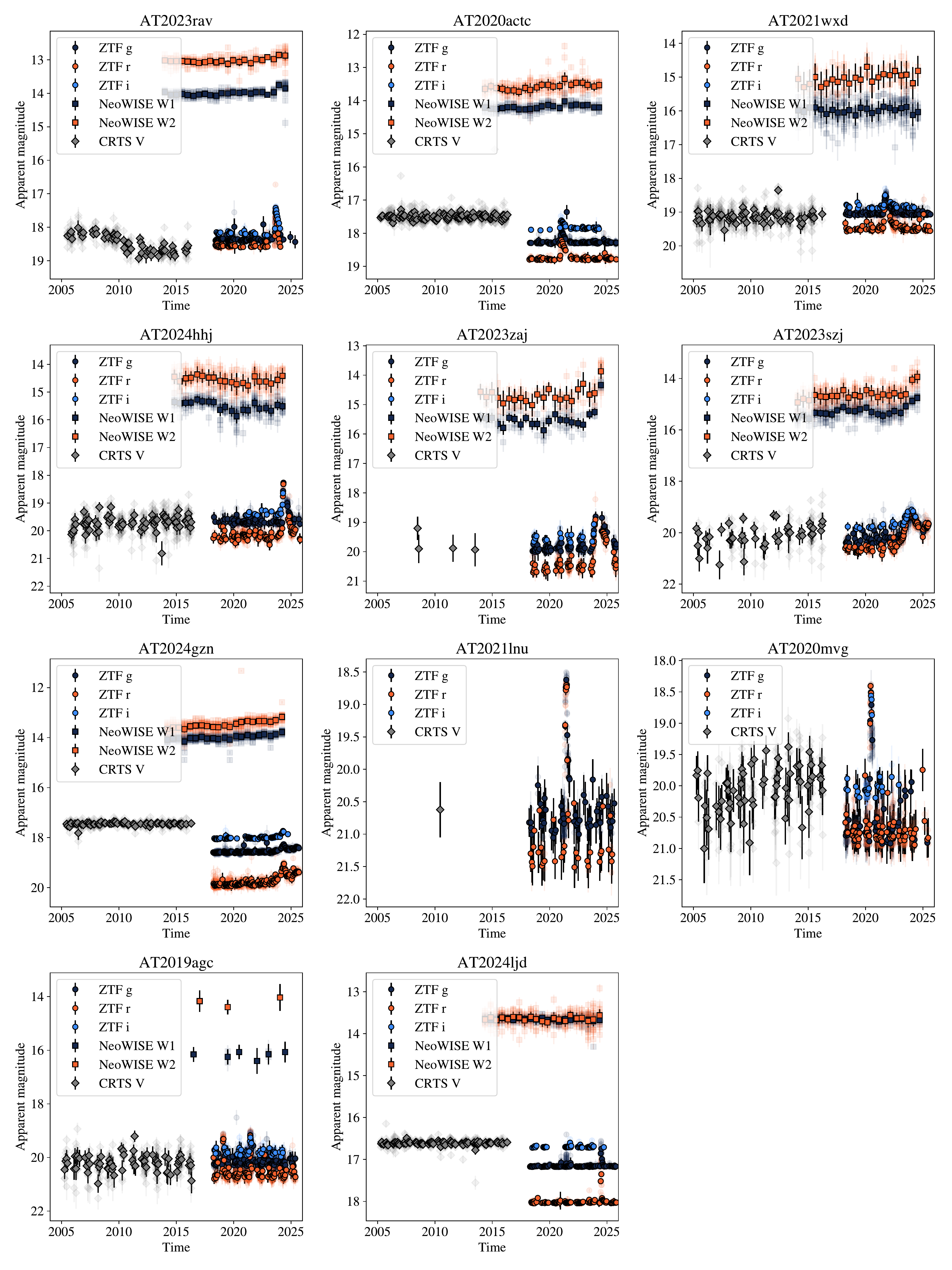}
    \caption{Continuation of Fig. \ref{fig:MultiL_TDEs1}}
    \label{fig:MultiL_TDEs2}
\end{figure}

\twocolumn
\section{Supernova modeling}
\label{app:sn}

We used the \textsc{Python} library \textsc{sncosmo}\footnote{\url{https://sncosmo.readthedocs.io/}}, version 2.12.1 to obtain a preliminary photometric classification for the identified supernovae candidates. Their lightcurves were fitted with Peter Nugent's supernova models\fnsref{foot:sn_models}  which cover the main SN types (Ia, Ib/c, IIP, IIL, IIn) \emi{and  the SALT2 model~\citep{guy2007}}. Nugent's models are simple spectral time series which can be scaled up and down; the model parameters are the redshift $z$, the observer-frame time corresponding to the source's zero phase, $t_0$, and the amplitude. 
\emi{In case of SALT2 fit, the free parameters are $z$, $t_0$, the normalization of the SED sequence $x_0$, stretch-factor $x_1$, and color $c$.}

For this work, we mostly fit the alert data from \fink\ (consequently, $g$ and $r$ bands), except in a few cases where significantly more photometric points were available in ZTF DR23 (which enabled fit also using the $i$ band). \emi{To obtain the transient lightcurve from the DR photometry, we accounted for the host galaxy contribution using the ZTF reference magnitudes}. The reference magnitudes were retrieved from the SNAD viewer\footnote{\url{https://ztf.snad.space/}} \citep{malanchev2023}. We also corrected \emi{all lightcurves} for a line-of-sight reddening in the Milky Way galaxy using \cite{2011ApJ...737..103S} estimates. For sources holding \emi{spectroscopic redshift of the host galaxy}, we fixed the redshift to this value.  If this was not available, we adopted $[-15; -22]$ as an acceptable range for the supernova absolute magnitude \citep{2014AJ....147..118R} and then, using the maximum apparent magnitude, roughly transformed it to the corresponding redshift range. We applied a $\chi^2$ criterion to choose the best-fit model for each object. 

\begin{figure}
    \centering
    \includegraphics[scale=0.5, width=0.5\textwidth]{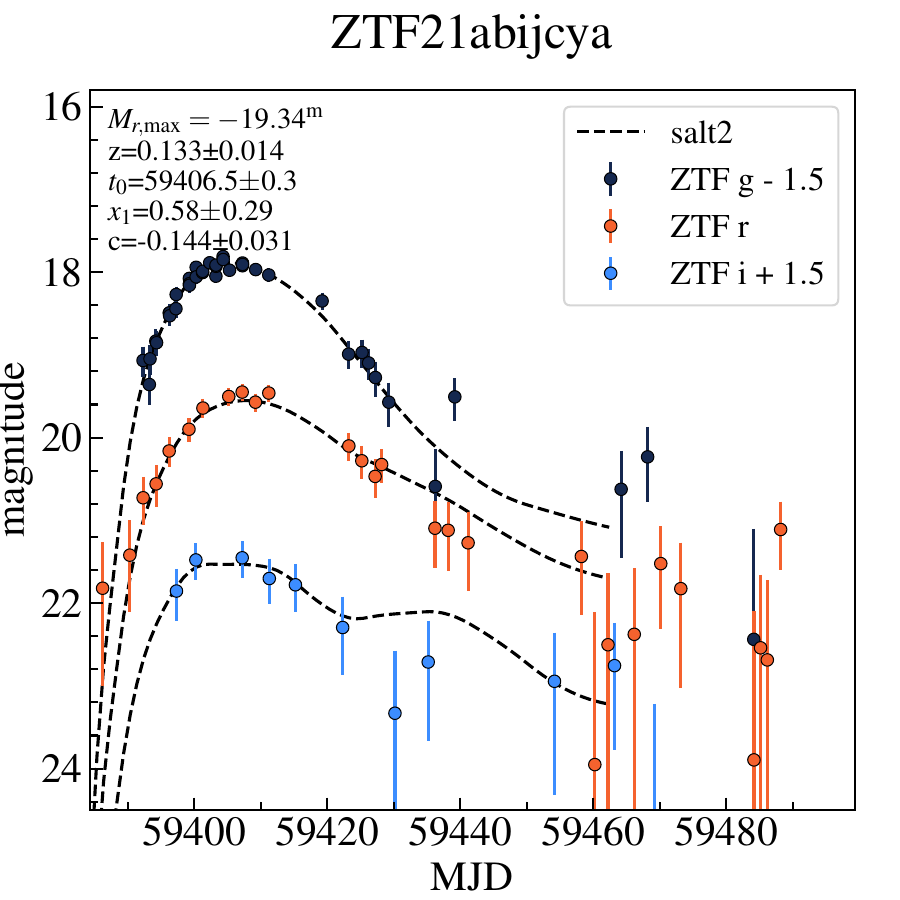}
    \vspace{1cm}
    \includegraphics[scale=0.5, width=0.5\textwidth, trim=0 70 0 0]{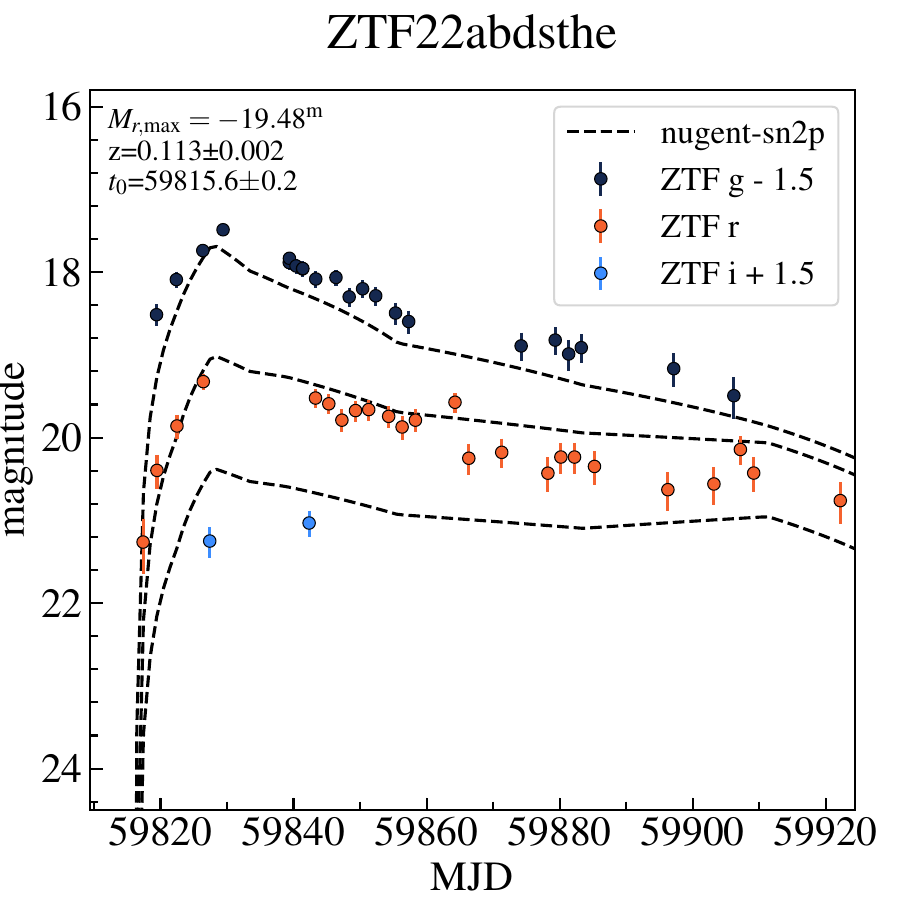}
    \caption{Example of best-fitted \emi{ \textsc{sncosmo} built-in models}. Different colors correspond to different ZTF filters. lightcurves in $g$ and $i$ band were shifted by 1.5 mag to improve visualization. The inset shows best-fit parameter values. \textbf{Top}: SN~Ia candidate. \textbf{Bottom}: SN Type IIP candidate.}
    \label{fig:sn_lc}
\end{figure}

It should be noted that we did not intend to make a detailed fit, but rather to show that the candidate lightcurves, selected initially by eye, can be satisfactorily fitted by different supernova models.  Also, we did not take into account the possible extinction in host galaxies, therefore our fit is less accurate for highly reddened objects. \emi{Another potential limitation arises from the fact that the models employed are limited in both wavelength and time range}. However, this simple fit is enough to show that a few transients have significant agreement with expected supernova lightcurves, as shown in Section \ref{subsec:sn_candidates} and Fig.  \ref{fig:sn_lc}.

\end{appendix}

\end{document}